\documentclass[useAMS, usenatbib]{mn2e}

\usepackage{times} 
\usepackage{comment}
\usepackage{graphicx}
\usepackage{epsfig}
\usepackage{lscape}
\usepackage{longtable}
\usepackage{amsmath,amssymb,multirow,epsfig,color,rotating,placeins}
\usepackage{color}

\newcommand{\samplesize}{245}
\newcommand{\aosamplesize}{196}
\newcommand{\archivesamplesize}{181}

\title[Stellar multiplicity of nearby low-mass stars]{The M-dwarfs in Multiples ({\sc MinMs}) survey - I. Stellar multiplicity among low-mass stars within 15 pc\thanks{Based on observations collected at the European Organisation for Astronomical Research in the Southern Hemisphere, Chile (programme 091.D-0804), and observations obtained at the MMT Observatory, a joint facility of the University of Arizona and the Smithsonian Institution.}}
\author[K. Ward-Duong et al.]{K. Ward-Duong$^{1}$\thanks{E-mail: kwardduo@asu.edu}, J. Patience$^{1}$, R. J. De Rosa$^{1, 2, 3}$, J. Bulger$^{1}$, A. Rajan$^{1}$, \newauthor S. P. Goodwin$^{4}$, Richard J. Parker$^{5}$, D. W. McCarthy$^{6}$ \& C. Kulesa$^{6}$ \\
$^{1}$School of Earth and Space Exploration, Arizona State University, Tempe, AZ 85287, USA \\
$^{2}$Astrophysics Group, School of Physics, University of Exeter, Stocker Road, Exeter EX4 4QL, UK\\
$^{3}$Department of Astronomy, University of California, Berkeley, CA 94720, USA \\
$^{4}$Department of Physics and Astronomy, University of Sheffield, Hounsfield Road, Sheffield, S3 7RH, UK \\
$^{5}$Astrophysics Research Institute, Liverpool John Moores University, 146 Brownlow Hill, L3 5RF, UK \\ 
$^{6}$Steward Observatory, University of Arizona, 933 N. Cherry Ave., Tucson, AZ 85721, USA}

\begin{document}

\date{Accepted 20 February 2015. Received 19 February 2015; in original form 9 June 2014}

\pagerange{\pageref{firstpage}--\pageref{lastpage}} \pubyear{2015}

\maketitle

\label{firstpage}

\begin{abstract}
We present a large-scale, volume-limited companion survey of \samplesize\ late-K to mid-M (K7-M6) dwarfs within 15~pc. Infrared adaptive optics (AO) data were analysed from the Very Large Telescope, Subaru Telescope, Canada-France-Hawaii Telescope, and MMT Observatory to detect close companions to the sample from $\sim$1~au to 100~au, while digitised wide-field archival plates were searched for wide companions from $\sim$100~au to 10,000~au. With sensitivity to the bottom of the main sequence over a separation range of 3~au to 10,000~au, multiple AO and wide-field epochs allow us to confirm candidates with common proper motions, minimize background contamination, and enable a measurement of comprehensive binary statistics. We detected 65 co-moving stellar companions and find a companion star fraction of $23.5 \pm 3.2$~per~cent over the 3~au to 10,000~au separation range. The companion separation distribution is observed to rise to a higher frequency at smaller separations, peaking at closer separations than measured for more massive primaries. The mass ratio distribution across the $q = 0.2 - 1.0$ range is flat, similar to that of multiple systems with solar-type primaries. The characterisation of binary and multiple star frequency for low-mass field stars can provide crucial comparisons with star forming environments and hold implications for the frequency and evolutionary histories of their associated disks and planets.
\end{abstract}

\begin{keywords}
techniques: high angular resolution - binaries: close
- binaries: general - binaries: visual - stars: late-type - stars: low-mass
\end{keywords}


\section{Introduction}
Among the nearest stars, the large majority are M-dwarfs (e.g. \citealp{Reid1997}), with a recent accounting indicating that M-dwarfs outnumber higher mass stars by a factor of $\sim$3 \citep{lepine2011}. Ongoing parallax programs designed to survey the Solar Neighborhood continue to discover additional nearby M-dwarfs, increasing the proportion of the lowest mass members among the nearest stars (e.g. \citealp{Henry2006}). The nearest star-forming regions are also dominated by low-mass stars; for example, $\sim$50~per~cent of the known members of Taurus have spectral types later than M3, the spectral type that will correspond to M0 and later after contraction onto the main sequence \citep{Luhman2010}. The preponderance of the low-mass population of stars highlights the importance of understanding their properties, including the statistics of their companions. As they represent a common outcome of star formation, multiple star systems provide key signatures of the physical processes which affect both star and planet formation. 

One of the key scientific questions of star-formation is the universality of the process, and the distribution of binary stars and the initial mass function (IMF) are among the main observational products that can be used to investigate this area \citep{king2}. There is no evidence that the IMF varies systematically between different environments (e.g. \citealp{Luhman2003,Bastian2010}), indicating that determinations of the IMF are unable to probe the universality or diversity of star-formation. An alternate approach is the analysis of binary populations, since observed differences in binary populations in independent regions suggest differences in the star-formation process (e.g. \citealp{Goodwin2010,king2,king1,Parker2014}). A well-characterised field M-dwarf binary sample is essential for comparison with the full membership of star-forming regions. 

In addition to the importance for understanding the products and process of star-formation, companion stars may critically impact planet formation and evolution. A companion star is expected to gravitationally truncate a protoplanetary disk to a radius of approximately one-third the binary star orbital semi-major axis \citep{Artymowicz1994}, thereby limiting the amount of material and the region over which planet formation can occur. Determining the population of companions with separations comparable to or less than typical disk sizes of $\sim$100~au \citep{Andrews:2009} are particularly important in considering dynamical effects on the disk, and nearby stars are ideal targets to probe the disk-sized separation range with imaging. 

Once planets form, the presence of even a distant companion can alter the dynamics of a planetary system through the Kozai mechanism \citep{Kozai1962,Lidov1962}. There are indications that some exoplanet systems have been impacted by this dynamical effect (e.g. \citealp{Wu2003}), and simulations of early dynamical interactions in star-forming regions have shown that the Kozai mechanism could be induced in up to 20 per cent of field binaries \citep{parkergoodwin}, subsequently affecting a sizable fraction of the exoplanet population. Determining the binarity of the field population and comparing with primordial binary distributions are critical to determining the fraction of stable stellar systems amenable to hosting planets, as demonstrated recently for G-dwarfs \citep{parkerquanz}. As the population statistics of exoplanets around low-mass stellar hosts are explored, an understanding of the stellar companions to M-dwarfs represents an important comparison and environmental factor.

Existing M-dwarf binary surveys have sample sizes or selection criteria that impact the interpretation of the population statistics of the companions. The benchmark survey of nearby M-dwarfs by \citet{fischermarcy} remains the only survey with complete separation coverage by combining radial velocity, speckle, and direct imaging, however the sample size is restricted to fewer than 65 stars per technique, resulting in large uncertainties on the distributions of mass ratio and separations relative to surveys of more massive primaries (e.g. \citealp{raghavan,2014MNRAS.437.1216D}). While more recent M-dwarf surveys include larger samples, the selection criteria include a mixed collection of distances (often based on photometry) or activity indicators, and only the single technique of adaptive optics (AO) or lucky imaging was employed, so the separation range coverage was limited (\citealp{Bergfors,janson}).

To develop comprehensive population statistics on the companions to a large-scale volume-limited sample of nearby low-mass stars, we have conducted a binary survey of 245 K7-M6 dwarfs within 15~pc based on \emph{Hipparcos} parallaxes. By combining archive and new observations with high resolution and wide-field imaging, the study spans three and a half orders of magnitude in separation. In Section~2, the definition and characteristics of the M-dwarfs in Multiples ({\sc MinMs}) sample are presented. In Section~3, the new and archival observations are described. The data reduction and analysis method are explained in Section~4. The results and discussion are explored in Section~5. Finally, Section~6 reports a summary of the results.


\section[]{Sample}
\label{sec:Sample}

\subsection{Sample selection}
\begin{figure}
\includegraphics[width=0.50\textwidth]{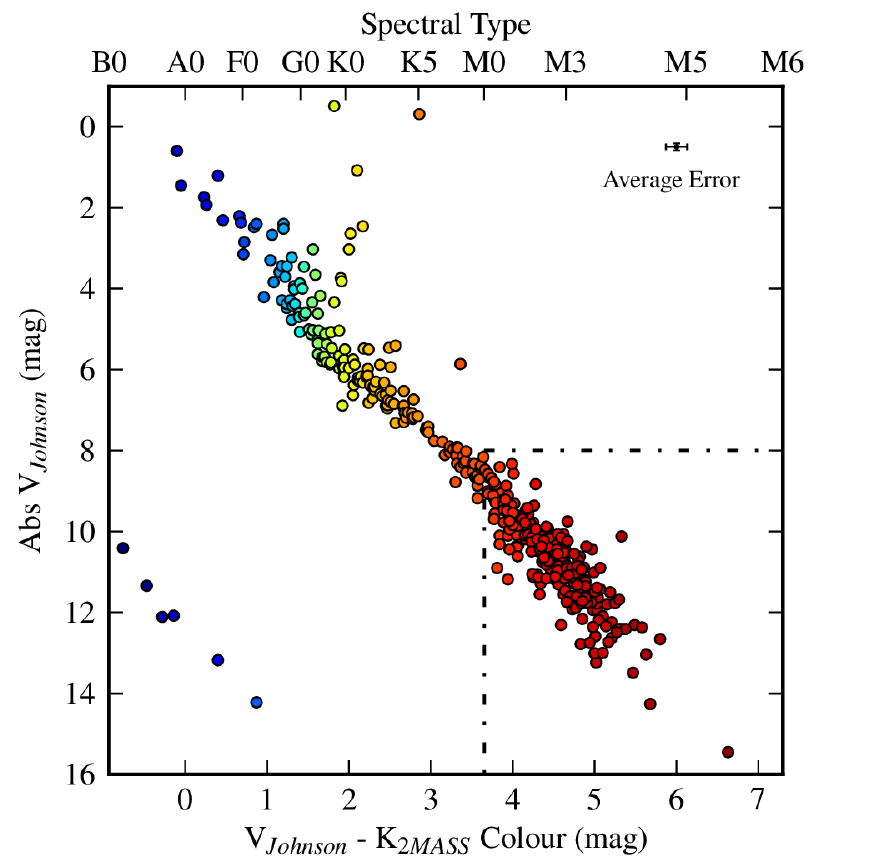}
\caption{Colour-magnitude diagram of all 449 stars within 15 pc meeting our \emph{Hipparcos} parallax error ($\sigma_{\pi}/\pi \leq 10$~per~cent) and photometry criteria. The colour and magnitude criteria used to select our sample of 245 M-dwarfs ($V-K_{\rm S} \geq 3.65$, $M_{V} > 8$) are shown with dashed-dotted lines enclosing our sample space on the CMD. The stars are colour-coded by spectral type, with the paucity of lowest-mass/reddest M-dwarfs at far distances indicative of the sensitivity limit of the \emph{Hipparcos} catalog. An example of the typical errorbar size is shown in the upper right corner of the figure.}
\label{fig:cmd15pc}
\end{figure}

The \textsc{MinMs} sample is derived from the new reduction of the \emph{Hipparcos} catalogue \citep{vanleeuwen}. We selected all stars with parallaxes greater than $\pi \geq 66.67$~mas, corresponding to stars located within a distance limit of $D\le15$~pc. In order to obtain precise distances and absolute magnitudes, stars with parallax errors larger than $\sigma_{\pi}/\pi \geq 0.10$ were excluded from the sample. Johnson $V$-band magnitudes were obtained from the original \emph{Hipparcos} catalogue \citep{oldhip}, which comprises ground and space-based photometry with uncertainties $\leq 0.08$~mag, and $K_{\rm S}$-band magnitudes were obtained from the Two Micron All Sky Survey \citep[2MASS;][]{cutri}, providing $V-K_{\rm S}$ colours. 

Stars with $K_{\rm S}$ magnitudes with significant errors and/or poor quality flags (``X'', ``U'', or ``F'') were excluded from the sample selection. We also excluded five stars identified as companions to earlier spectral types, as given in the Washington Double Star catalogue (WDS; \citealp{mason}) and confirmed by common proper motion (see Section~\ref{subsection:astro_confirmation}): HIP83599, HIP26801, HIP42762, HIP45343/HIP120005. The parallax and photometric quality criteria provided an all-sky, volume-limited sample of 449 stars of over a wide range of spectral types, as shown in the colour-magnitude diagram (CMD) in Figure~\ref{fig:cmd15pc}. M-dwarfs were then selected from this sample by adopting a colour cut of $V-K_{\rm S} > 3.65$, corresponding to spectral class M0 and later \citep{kenyonhartmann}. Additionally, only stars with $M_{V} > 8$ were included to remove any possibility of contamination from evolved stars. The combined parallax, colour, and magnitude criteria define a total sample of \samplesize\ K and M-dwarfs, of which over 95 per cent are M spectral types.

\subsection{Sample properties}
\begin{figure}
\includegraphics[width=0.50\textwidth]{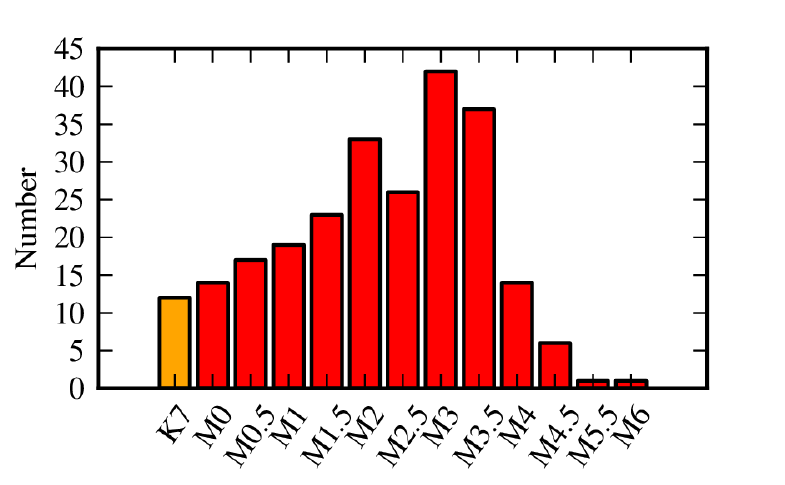}
\caption{The spectral type distribution of the 15~pc volume-limited 245 M-dwarf sample. The sample has a higher proportion of early-type M-dwarfs, owing to the sensitivity limit of the \emph{Hipparcos} instrument. One star in the sample, HIP~117828, has an early spectral type of ``Ma'' as classified by \citet{houk} but lacks a more recent classification, and is not included in this histogram.}
\label{fig:sptydist}
\end{figure}
\begin{figure}
\includegraphics[width=0.50\textwidth]{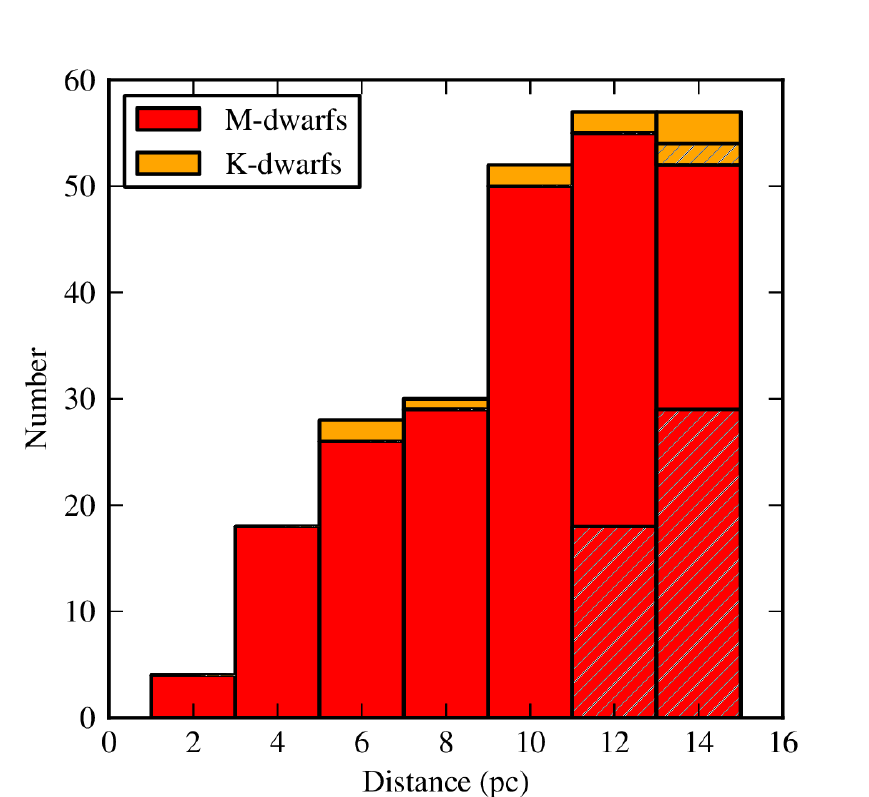}
\caption{The distance distribution of the 15~pc volume-limited 245 star sample. The darker, hatched portion of the histogram represents stars without new and/or archival adaptive optics data. All of the K and M-dwarfs in our sample up to 11~pc (189 targets) have high-resolution imaging data, covering projected separations of $\sim$$1-100$~au. Archival digitised photographic plates were analysed for the entire sample (both darker and lighter solid regions), covering projected separations of $\sim$$100-10,000$~au.}
\label{fig:distdistn}
\end{figure}
\begin{figure}
\includegraphics[width=0.50\textwidth]{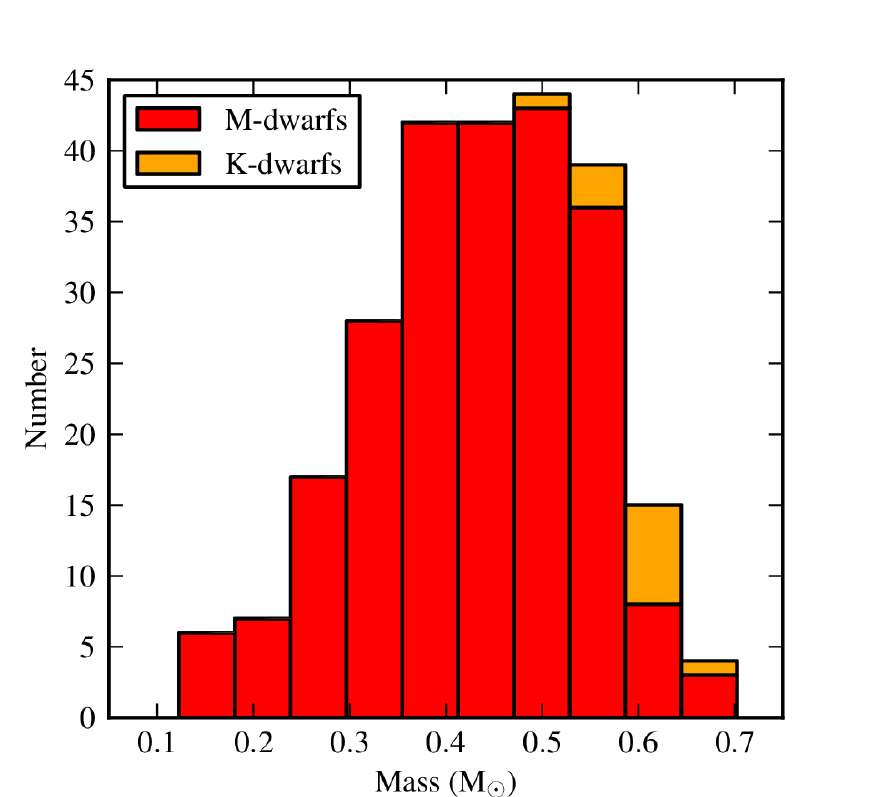}
\caption{The distribution of primary masses for the targets within the sample. The masses were estimated by comparing the absolute $K_{\rm S}$ magnitude of each target with theoretical solar-metallicity mass magnitude relations \citep{baraffe98}.}
\label{fig:prim_mass_dist}
\end{figure}

The spectral type distribution of the sample is shown in Figure~\ref{fig:sptydist}, with the majority (94~per cent) of the sample spectral type classifications obtained from the Palomar/MSU Nearby Star Spectroscopic Survey \citep[PMSU, ][]{pmsu1,pmsu2}. The spectral types of the remaining stars without PMSU classifications were obtained from the SIMBAD database (individual references listed in Table~\ref{table:sample}). Uncertainties on the colours and a spread in metallicity led to the inclusion of 12 K7 spectral types, as shown in Figure~\ref{fig:distdistn} (discussed in further detail in Section~\ref{sec:malm}). Given the limiting magnitudes of the \emph{Hipparcos} catalogue of $V = 12.4$ \citep{esaref}, the latest M spectral types were too faint to be detected, leading to the larger frequency of early M-dwarfs in the sample. The distance distribution of the sample is shown in Figure~\ref{fig:distdistn}. The mass of each primary was estimated from theoretical mass-magnitude relations \citep{baraffe98}, with the resulting distribution shown in Figure \ref{fig:prim_mass_dist}.

{\scriptsize 
\onecolumn 
\begin{center}
\begin{longtable}{lccccccccccccc}
\caption{The 245 K7-M6 dwarf sample within 15 pc}\\
\label{table:sample}\\
\hline
HIP           &  $\alpha$  &  $\delta$  &  $\mu_{\alpha}$ &  $\mu_{\delta}$ &  $V$ & $J$ &  $H$ &  $K_{\rm S}$ &  $\pi$ & Distance & SpTy. &   [Fe/H] & Ref. \\

  &  (J2000.0) &  (J2000.0) &  (mas yr$^{-1}$)       &  (mas yr$^{-1}$)       &  (mag) &     (mag) &  (mag)   &   (mag)   &   (mas) &   (pc)   &   & (dex)           \\
\hline
\endfirsthead
\caption{continued.}\\
\hline
HIP           &  $\alpha$  &  $\delta$  &  $\mu_{\alpha}$ &  $\mu_{\delta}$ &  $V$ & $J$ &  $H$ &  $K_{\rm S}$ &  $\pi$ & Distance & SpTy. &   [Fe/H] & Ref. \\

  &  (J2000.0) &  (J2000.0) &  (mas yr$^{-1}$)       &  (mas yr$^{-1}$)       &  (mag) &     (mag) &  (mag)   &   (mag)   &   (mas) &   (pc)   &  & (dex)       \\
\hline
\endhead
\hline
\endfoot
\hline
\multicolumn{11}{l}{References (metallicities). -- $^{1}$ \citet{neves2014}; $^{2}$ \citet{gaidosmann2014}; $^{3}$ \citet{rojasayala}; $^{4}$ \citet{newton2014}} \\
\multicolumn{14}{l}{Spectral type references for targets without PMSU classifications. -- $^{a}$ \citet{gray2006}; $^{b}$ \citet{browning}; $^{c}$ \citet{torres}; $^{d}$ \citet{poveda}; $^{e}$ \cite{koen}; $^{f}$ \citet{houk}}\\
\endlastfoot

428	&	1.295368	&	45.786567	&	869.7	&	-151	&	9.95	&	  6.704$\pm$0.021 	&	  6.098$\pm$0.024 	&	  5.853$\pm$0.018 	&	    88.88$\pm$1.42 	&	  11.25$\pm$0.18 	&	M1	&	-0.01	&	2	\\
439	&	1.351784	&	-37.357361	&	5634.4	&	-2336.4	&	8.56	&	  5.328$\pm$0.019 	&	  4.828$\pm$0.076 	&	  4.523$\pm$0.017 	&	   230.42$\pm$0.90 	&	   4.34$\pm$0.02 	&	M1.5	&	-0.45	&	1	\\
1242	&	3.873033	&	-16.133639	&	-32.1	&	-149.7	&	11.49	&	  7.215$\pm$0.019 	&	  6.712$\pm$0.034 	&	  6.390$\pm$0.016 	&	   200.53$\pm$9.41 	&	   4.99$\pm$0.23 	&	M4	&		&		\\
1475	&	4.595353	&	44.022949	&	2887.5	&	408.9	&	8.09	&	  5.252$\pm$0.264 	&	  4.476$\pm$0.200 	&	  4.018$\pm$0.020 	&	   278.76$\pm$0.77 	&	   3.59$\pm$0.01 	&	M3.5	&	-0.26	&	2	\\
2552	&	8.122664	&	67.235616	&	1741.1	&	-246.8	&	10.27	&	  6.844$\pm$0.023 	&	  6.268$\pm$0.021 	&	  6.037$\pm$0.023 	&	    99.35$\pm$2.17 	&	  10.07$\pm$0.22 	&	M2	&	-0.28	&	3	\\
3937	&	12.63845	&	24.816829	&	223.7	&	-40.9	&	12.01	&	  7.951$\pm$0.023 	&	  7.350$\pm$0.015 	&	  7.119$\pm$0.020 	&	    84.42$\pm$4.65 	&	  11.85$\pm$0.65 	&	M3.5	&		&		\\
4569	&	14.616383	&	-27.857032	&	1293.4	&	-298.9	&	11.77	&	  7.763$\pm$0.023 	&	  7.203$\pm$0.051 	&	  6.892$\pm$0.021 	&	    80.95$\pm$3.21 	&	  12.35$\pm$0.49 	&	M3.5	&		&		\\
4856	&	15.6343	&	71.679816	&	1745.3	&	-380.9	&	9.98	&	  6.301$\pm$0.034 	&	  5.699$\pm$0.049 	&	  5.449$\pm$0.017 	&	   121.41$\pm$1.25 	&	   8.24$\pm$0.08 	&	M3	&		&		\\
\hline
\end{longtable}
\end{center}
}
\twocolumn


\section{Data Sources and Acquisition}
\label{sec:obs}
The \textsc{MinMs} survey made use of new and archival AO observations and archival digitised photographic plates to detect companions to M-dwarfs. The AO data provide the capacity to search for companions at high angular resolution, while the plate data provide a complementary widefield search space. Combined, the two detection techniques provide nearly continuous angular separation coverage for companions from $\sim$0.1 arcsec to $\sim$10 arcmin. In this section, we describe the archival data origins and the newly-obtained AO observations.

\subsection{Archival high-resolution imaging data}
To search for companions at projected separations of $\sim$$1 - 100$~au from their host stars, we queried all publicly-available high-resolution and adaptive optics (AO) archives. Of the \samplesize\ star total sample, \archivesamplesize\ stars had previous high-resolution AO imaging. We obtained archival near-infrared AO imaging data from the following instruments/facilities: NAOS-CONICA (NaCo: Nasmyth Adaptive Optics System Near-Infrared Imager and Spectrograph, \citealt{lenzen2003,rousset2003}) on the Very Large Telescope (VLT); the AOB/PUEO KIR camera \citep{doyon} on the Canada-France-Hawaii Telescope (CFHT); and CIAO (Coronagraphic Imager; \citealp{Murakawa2003,Murakawa2004}), HiCIAO (High Contrast Instrument for the Subaru Next Generation Adaptive Optics; \citealp{Hodapp2008}), and IRCS (Infrared Camera and Spectropgrah; \citealp{Tokunaga1998,Kobayashi2000}) on Subaru. Table~\ref{table:ao_obs} provides a summary of the telescopes and instruments used to obtain the first epoch data. Archive AO data from three sources -- CFHT, VLT, and Subaru -- combined with new AO images from MMT, covered 196 of the 245-star \textsc{MinMs} sample. Some of the targets have additional archive data, and the data used in this study were selected based on the availability of unsaturated images, with preference given to the three instruments that observed the largest number of targets in archives. Existing observations and their corresponding calibration files were downloaded from the Canadian Astronomy Data Centre, the European Southern Observatory Science Archive Facility, and the Subaru-Mitaka-Okayama-Kiso Archive System. Detailed information on the observations, including observation, filter, exposure time, date, programme ID and project PI for each target are provided in Table~\ref{table:ao_obs}.

\onecolumn 
\begin{center}
\begin{longtable}{lcccclcc}
\caption{\label{table:ao_obs}High-resolution imaging data} \\ 
\hline
 HIP & Telescope & Instrument  & Filter &  Exp. time & Programme & PI &  UT \\
   &  &   &  &  (s) & ID &  &  Date \\
\hline
\endfirsthead
\caption{continued.}\\
\hline
 HIP & Telescope & Instrument  & Filter &  Exp. time & Programme & PI &  UT \\
   &  &   &  &  (s) & ID &  &  Date \\
\hline
\endhead
\hline
\endfoot
\hline
\noalign{\smallskip}
\multicolumn{8}{l}{$^{a}$ Observatory technical time.} \\
\endlastfoot
428	&	CFHT	&	PUEO/KIR	&	Br$\gamma$	&	6	&	00BF25	&	Perrier-Bellet	&	2000-08-21	\\
439	&	VLT	&	NaCo/S13	&	NB1.64	&	8	&	072.C-0570(A)	&	Beuzit	&	2003-12-08	\\
1242	&	VLT	&	NaCo/S13	&	NB1.64	&	0.7	&	072.C-0570(A)	&	Beuzit	&	2003-12-10	\\
1475	&	CFHT	&	PUEO/KIR	&	Fe~{\scriptsize II}	&	0.5	&	99IID409	&	M{\'e}nard	&	1999-08-25	\\
2552	&	CFHT	&	PUEO/KIR	&	Br$\gamma$	&	10	&	07AF12	&	Forveille	&	2007-01-28	\\
3937	&	CFHT	&	PUEO/KIR	&	$H$	&	10	&	h8	&	Jewitt	&	2000-08-19	\\
4856	&	CFHT	&	PUEO/KIR	&	$H_{2}(\nu=1-0)$ 	&	4	&	03BD2	&	Forveille	&	2003-10-16	\\
4872	&	CFHT	&	PUEO/KIR	&	Fe~{\scriptsize II}	&	2	&	97IIF28	&	Beuzit	&	1997-12-27	\\
5496	&	VLT	&	NaCo/S13	&	$H$	&	5	&	382.D-0754(A)	&	Bean	&	2008-10-17	\\
5643	&	CFHT	&	PUEO/KIR	&	$K$	&	1	&	00BH12	&	Roddier	&	2000-12-12	\\
8051	&	CFHT	&	PUEO/KIR	&	Fe~{\scriptsize II}	&	3	&	97IIF28	&	Beuzit	&	1997-12-28	\\
8768	&	CFHT	&	PUEO/KIR	&	Br$\gamma$	&	2	&	01BF50	&	Perrier-Bellet	&	2001-08-06	\\
9291	&	CFHT	&	PUEO/KIR	&	Fe~{\scriptsize II}	&	15	&	04BF8	&	Catala	&	2004-09-28	\\
9724	&	VLT	&	NaCo/S13	&	NB1.64	&	0.345	&	072.C-0570(A)	&	Beuzit	&	2003-12-09	\\
9786	&	VLT	&	NaCo/S13	&	NB1.64	&	3.5	&	072.C-0570(A)	&	Beuzit	&	2003-12-08	\\
10279	&	CFHT	&	PUEO/KIR	&	Br$\gamma$	&	4	&	01BF50	&	Perrier-Bellet	&	2001-08-06	\\
10395	&	CFHT	&	PUEO/KIR	&	Br$\gamma$	&	10	&	00BF25	&	Perrier-Bellet	&	2000-08-21	\\
10617	&	VLT	&	NaCo/S27	&	$K_{\rm S}$	&	0.6	&	077.C-0483(A)	&	Melo	&	2006-07-28	\\
10812	&	VLT	&	NaCo/S54	&	NB2.12	&	0.9	&	381.C-0235(A)	&	Kuerster	&	2008-07-04	\\
11048	&	CFHT	&	PUEO/KIR	&	Br$\gamma$	&	2.5	&	01BD03	&	Forveille	&	2001-08-31	\\
11964	&	VLT	&	NaCo/S13	&	NB2.12	&	0.6	&	074.C-0074(A)	&	Udry	&	2004-11-10	\\
12097	&	Subaru	&	IRCS	&	$K$	&	3	&	o08184	&	Dello Russo/Vervack	&	2008-08-05	\\
12781	&	CFHT	&	PUEO/KIR	&	$H_{2}(\nu=1-0)$ 	&	8	&	03BD2	&	Forveille	&	2003-10-15	\\
13389	&	VLT	&	NaCo/S13	&	NB1.64	&	1	&	072.C-0570(A)	&	Beuzit	&	2003-12-09	\\
16536	&	VLT	&	NaCo/S13	&	NB1.64	&	4	&	073.C-0155(A)	&	Beuzit	&	2004-09-23	\\
17609	&	MMT	&	ARIES	&	$K_{{\rm C}~2.09}$	&	1.4	&	UAO-S10/S11	&	De Rosa/Ward-Duong	&	2013-09-18	\\
21088	&	CFHT	&	PUEO/KIR	&	Br$\gamma$	&	4	&	F123	&	Perrier	&	2000-02-18	\\
21556	&	CFHT	&	PUEO/KIR	&	$H_{2}(\nu=1-0)$ 	&	4	&	02BE03	&	Forveille	&	2002-09-18	\\
21932	&	VLT	&	NaCo/S27	&	NB2.17	&	0.4	&	079.C-0216(A)	&	Montagnier	&	2007-09-15	\\
22627	&	VLT	&	NaCo/S13	&	NB2.12	&	0.4	&	70.C-0777(D)	&	Mundt	&	2003-01-23	\\
22738	&	VLT	&	NaCo/S13	&	NB1.64	&	3	&	073.C-0155(A)	&	Beuzit	&	2004-09-23	\\
23452	&	CFHT	&	PUEO/KIR	&	$H_{2}(\nu=1-0)$ 	&	1.5	&	02BE03	&	Forveille	&	2002-09-18	\\
23512	&	CFHT	&	PUEO/KIR	&	$H_{2}(\nu=1-0)$ 	&	15	&	03BD2	&	Forveille	&	2003-10-15	\\
23518	&	MMT	&	ARIES	&	$K_{{\rm C}~2.09}$	&	1.4	&	UAO-S10/S11	&	De Rosa/Ward-Duong	&	2013-09-18	\\
23932	&	VLT	&	NaCo/S27	&	$K_{\rm S}$	&	0.5	&	70.C-0738(A)	&	Beuzit	&	2003-03-16	\\
24186	&	VLT	&	NaCo/S13	&	$H$	&	0.35	&	70.C-0738(A)	&	Beuzit	&	2003-03-18	\\
24284	&	Subaru	&	CIAO	&	$H$	&	10	&	o04203	&	Itoh	&	2004-11-20	\\
25578	&	VLT	&	NaCo/S27	&	$K_{\rm S}$	&	2	&	079.C-0216(A)	&	Montagnier	&	2007-09-15	\\
25878	&	CFHT	&	PUEO/KIR	&	Fe~{\scriptsize II}	&	1	&	03BH10D	&	Liu	&	2004-01-05	\\
26857	&	CFHT	&	PUEO/KIR	&	Fe~{\scriptsize II}	&	3	&	97IIF28	&	Beuzit	&	1997-12-29	\\
28368	&	Subaru	&	HiCIAO	&	$H$	&	1.5	&	o11302	&	Bowler	&	2011-12-27	\\
29052	&	VLT	&	NaCo/S13	&	NB1.64	&	1.5	&	072.C-0570(A)	&	Beuzit	&	2003-12-08	\\
29277	&	CFHT	&	PUEO/KIR	&	Fe~{\scriptsize II}	&	4	&	03AF26	&	Beuzit	&	2003-03-16	\\
29295	&	CFHT	&	PUEO/KIR	&	$H_{2}$	&	0.7	&	97IIF28	&	Beuzit	&	1997-12-29	\\
29316	&	CFHT	&	PUEO/KIR	&	Br$\gamma$	&	10	&	03BD10	&	Forveille	&	2004-01-06	\\
30920	&	VLT	&	NaCo/S13	&	$K_{\rm S}$	&	2	&	077.C-0783(A)	&	Forveille	&	2006-04-11	\\
31292	&	VLT	&	NaCo/S13	&	NB1.64	&	1.5	&	072.C-0570(A)	&	Beuzit	&	2003-12-09	\\
31293	&	VLT	&	NaCo/S13	&	NB1.64	&	1.3	&	072.C-0570(A)	&	Beuzit	&	2003-12-09	\\
31635	&	VLT	&	NaCo/S13	&	NB1.64	&	0.345	&	072.C-0570(A)	&	Beuzit	&	2003-12-09	\\
33142	&	CFHT	&	PUEO/KIR	&	Fe~{\scriptsize II}	&	6	&	05AF19	&	Forveille	&	2005-04-26	\\
33226	&	CFHT	&	PUEO/KIR	&	Fe~{\scriptsize II}	&	1	&	97IIF28	&	Beuzit	&	1997-12-29	\\
33499	&	VLT	&	NaCo/S13	&	NB1.64	&	1	&	072.C-0570(A)	&	Beuzit	&	2003-12-08	\\
34603	&	CFHT	&	PUEO/KIR	&	Fe~{\scriptsize II}	&	2	&	98IF58	&	Beuzit	&	1998-03-07	\\
35191	&	VLT	&	NaCo/S13	&	NB1.26	&	0.75	&	072.C-0570(A)	&	Beuzit	&	2003-12-10	\\
36208	&	VLT	&	NaCo/S13	&	$H$	&	0.345	&	072.C-0570(A)	&	Beuzit	&	2003-12-11	\\
36626	&	CFHT	&	PUEO/KIR	&	Br$\gamma$	&	5	&	F58	&	Perrier-Bellet	&	2000-04-19	\\
36627	&	CFHT	&	PUEO/KIR	&	Br$\gamma$	&	10	&	F58	&	Perrier-Bellet	&	2000-04-19	\\
37217	&	VLT	&	NaCo/S13	&	NB1.64	&	1	&	072.C-0570(A)	&	Beuzit	&	2003-12-08	\\
37288	&	VLT	&	NaCo/S13	&	NB1.64	&	1	&	70.C-0777(E)	&	Mundt	&	2003-02-19	\\
37766	&	VLT	&	NaCo/S13	&	$K_{\rm S}$	&	2.7	&	077.C-0783(A)	&	Forveille	&	2006-04-11	\\
38956	&	CFHT	&	PUEO/KIR	&	$H_{2}$	&	15	&	97IID06	&	Richer/Beuzit	&	1997-12-26	\\
40501	&	VLT	&	NaCo/S13	&	$H$	&	0.8	&	70.C-0738(A)	&	Beuzit	&	2003-03-16	\\
41824	&	CFHT	&	PUEO/KIR	&	Br$\gamma$	&	10	&	07AF12	&	Forveille	&	2007-01-29	\\
45908	&	VLT	&	NaCo/S13	&	NB1.64	&	0.5	&	072.C-0570(A)	&	Beuzit	&	2003-12-09	\\
46655	&	CFHT	&	PUEO/KIR	&	Fe~{\scriptsize II}	&	2	&	03BH10D	&	Liu	&	2004-01-06	\\
46706	&	CFHT	&	PUEO/KIR	&	$H_{2}$	&	2.5	&	04AD8	&	Forveille	&	2004-04-04	\\
46769	&	MMT	&	ARIES	&	$K_{{\rm C}~2.09}$	&	1.4	&	UAO-S2	&	De Rosa	&	2013-05-24	\\
47103	&	VLT	&	NaCo/S13	&	$H$	&	8	&	079.C-0216(A)	&	Montagnier	&	2007-04-10	\\
47425	&	VLT	&	NaCo/S13	&	NB1.26	&	1	&	70.C-0738(A)	&	Beuzit	&	2003-03-17	\\
47513	&	VLT	&	NaCo/S13	&	NB1.64	&	0.5	&	072.C-0570(A)	&	Beuzit	&	2003-12-10	\\
47620	&	MMT	&	ARIES	&	$K_{{\rm C}~2.09}$	&	1.4	&	UAO-S2	&	De Rosa	&	2013-05-24	\\
47780	&	VLT	&	NaCo/S13	&	NB1.75	&	0.5	&	70.C-0738(A)	&	Beuzit	&	2003-03-17	\\
48659	&	VLT	&	NaCo/S13	&	$K_{\rm S}$	&	0.5	&	60.A-9800(J)	&	--$^{a}$	&	2010-02-08	\\
48714	&	CFHT	&	PUEO/KIR	&	$H_{2}$	&	0.5	&	04AD8	&	Forveille	&	2004-04-04	\\
49969	&	VLT	&	NaCo/S13	&	NB2.12	&	0.4	&	70.C-0738(A)	&	Beuzit	&	2003-03-18	\\
49986	&	CFHT	&	PUEO/KIR	&	$J_{\rm cont}$	&	1	&	F58	&	Perrier-Bellet	&	2000-04-18	\\
51007	&	CFHT	&	PUEO/KIR	&	Fe~{\scriptsize II}	&	2	&	03BH10D	&	Liu	&	2004-01-06	\\
51317	&	VLT	&	NaCo/S54	&	NB2.17	&	0.3454	&	079.C-0216(A)	&	Montagnier	&	2007-04-10	\\
53020	&	VLT	&	NaCo/S27	&	$K_{\rm S}$	&	0.5	&	079.C-0216(A)	&	Montagnier	&	2007-05-06	\\
53985	&	CFHT	&	PUEO/KIR	&	Br$\gamma$	&	3.5	&	F58	&	Perrier-Bellet	&	2000-04-19	\\
54035	&	CFHT	&	PUEO/KIR	&	Fe~{\scriptsize II}	&	0.2	&	H1A	&	Roddier	&	2000-04-13	\\
54211	&	CFHT	&	PUEO/KIR	&	K$^{\prime}$	&	0.1	&	00BH12	&	Roddier	&	2000-12-13	\\
54532	&	CFHT	&	PUEO/KIR	&	$H_{2}(\nu=1-0)$ 	&	10	&	05AF19	&	Forveille	&	2005-04-26	\\
55360	&	CFHT	&	PUEO/KIR	&	Fe~{\scriptsize II}	&	1.5	&	98IF58	&	Beuzit	&	1998-03-07	\\
56244	&	VLT	&	NaCo/S13	&	$K_{\rm S}$	&	0.347	&	074.C-0084(B)	&	Neuh$\ddot{a}$user	&	2005-01-07	\\
56528	&	CFHT	&	PUEO/KIR	&	Fe~{\scriptsize II}	&	2	&	99IF59	&	Perrier	&	1999-04-04	\\
57050	&	CFHT	&	PUEO/KIR	&	$H_{2}(\nu=2-1)$ 	&	15	&	05AF19	&	Forveille	&	2005-04-25	\\
57087	&	VLT	&	NaCo/L27	&	L$^{\prime}$	&	0.2	&	081.C-0430(A)	&	Apai	&	2008-04-06	\\
57544	&	CFHT	&	PUEO/KIR	&	$H_{2}$	&	8	&	98IF58	&	Beuzit	&	1998-03-07	\\
57548	&	CFHT	&	PUEO/KIR	&	Fe~{\scriptsize II}	&	1	&	03BH10D	&	Liu	&	2004-01-06	\\
57802	&	CFHT	&	PUEO/KIR	&	Fe~{\scriptsize II}	&	1.5	&	98IF58	&	Beuzit	&	1998-03-08	\\
59406	&	VLT	&	NaCo/S13	&	NB2.17	&	2	&	60.A-9800(J)	&	--	&	2010-02-08	\\
60559	&	CFHT	&	PUEO/KIR	&	Fe~{\scriptsize II}	&	7	&	98IF58	&	Beuzit	&	1998-03-08	\\
60910	&	VLT	&	NaCo/S13	&	$K_{\rm S}$	&	0.3454	&	080.C-0424(A)	&	Vogt	&	2008-02-19	\\
61094	&	VLT	&	NaCo/S27	&	$K_{\rm S}$	&	0.347	&	077.D-0179(A)	&	Neuh$\ddot{a}$user	&	2006-04-22	\\
61629	&	VLT	&	NaCo/S27	&	NB2.17	&	0.4	&	077.C-0483(A)	&	Melo	&	2006-04-29	\\
61874	&	VLT	&	NaCo/S13	&	NB1.64	&	3	&	70.C-0738(A)	&	Beuzit	&	2003-03-14	\\
62452	&	CFHT	&	PUEO/KIR	&	Fe~{\scriptsize II}	&	2	&	03BH10D	&	Liu	&	2004-01-06	\\
62556	&	CFHT	&	PUEO/KIR	&	Fe~{\scriptsize II}	&	5	&	03AF26	&	Beuzit	&	2003-03-17	\\
63510	&	VLT	&	NaCo/S13	&	$K_{\rm S}$	&	8	&	077.C-0783(A)	&	Forveille	&	2006-05-23	\\
65011	&	MMT	&	ARIES	&	$K_{{\rm C}~2.09}$	&	1.4	&	UAO-S2	&	De Rosa	&	2013-05-24	\\
65026	&	CFHT	&	PUEO/KIR	&	$H_{2}(\nu=2-1)$ 	&	8	&	05AF19	&	Forveille	&	2005-04-27	\\
65859	&	CFHT	&	PUEO/KIR	&	$H_{2}$	&	0.5	&	97IIF09	&	Bouvier	&	1998-01-14	\\
66625	&	MMT	&	ARIES	&	$K_{{\rm C}~2.09}$	&	1.4	&	UAO-S2	&	De Rosa	&	2013-05-24	\\
66906	&	CFHT	&	PUEO/KIR	&	Br$\gamma$	&	10	&	F123	&	Perrier	&	2000-02-19	\\
67155	&	VLT	&	NaCo/S13	&	$K_{\rm S}$	&	2	&	077.C-0783(A)	&	Forveille	&	2006-05-23	\\
67164	&	CFHT	&	PUEO/KIR	&	Br$\gamma$	&	15	&	03AF26	&	Beuzit	&	2003-03-21	\\
68469	&	CFHT	&	PUEO/KIR	&	Br$\gamma$	&	3	&	F58	&	Perrier-Bellet	&	2000-04-20	\\
69454	&	VLT	&	NaCo/S13	&	NB1.64	&	1	&	70.C-0738(A)	&	Beuzit	&	2003-03-14	\\
70890	&	VLT	&	NaCo/S13	&	$H$	&	1	&	075.C-0733(A)	&	Beuzit	&	2005-05-01	\\
70975	&	VLT	&	NaCo/S13	&	$K_{\rm S}$	&	10	&	078.C-0441(A)	&	Forveille	&	2007-02-23	\\
71253	&	CFHT	&	PUEO/KIR	&	$H_{2}$	&	10	&	98IF58	&	Beuzit	&	1998-03-08	\\
71898	&	CFHT	&	PUEO/KIR	&	$H_{2}$	&	8	&	02AF43	&	Beuzit	&	2002-06-24	\\
72896	&	VLT	&	NaCo/S13	&	$K_{\rm S}$	&	1.5	&	091.D-0804(A)	&	De Rosa	&	2013-04-19	\\
72944	&	VLT	&	NaCo/S13	&	$K_{\rm S}$	&	0.5	&	076.C-0139(A)	&	Bouy	&	2006-03-01	\\
73470	&	CFHT	&	PUEO/KIR	&	$H_{2}(\nu=2-1)$ 	&	8	&	05AF21	&	Beuzit	&	2005-04-25	\\
74190	&	Subaru	&	IRCS	&	$K$	&	3	&	o06101	&	Imanishi	&	2006-07-20	\\
74995	&	VLT	&	NaCo/S27	&	$K_{\rm S}$	&	5	&	081.C-0600(A)	&	Lagrange	&	2008-06-29	\\
75187	&	CFHT	&	PUEO/KIR	&	Br$\gamma$	&	3	&	02BE03	&	Forveille	&	2002-09-18	\\
76074	&	VLT	&	NaCo/S13	&	NB1.64	&	0.35	&	075.C-0733(A)	&	Beuzit	&	2005-05-02	\\
76832	&	MMT	&	ARIES	&	$K_{{\rm C}~2.09}$	&	1.4	&	UAO-S2	&	De Rosa	&	2013-05-24	\\
76901	&	CFHT	&	PUEO/KIR	&	Br$\gamma$	&	15	&	F58	&	Perrier-Bellet	&	2000-04-19	\\
78353	&	Subaru	&	CIAO	&	$K$	&	30	&	o05104	&	Nakajima	&	2005-07-13	\\
79755	&	CFHT	&	PUEO/KIR	&	$H_{\rm cont}$	&	1	&	01AH25B	&	Baudoz	&	2001-05-03	\\
79762	&	CFHT	&	PUEO/KIR	&	Br$\gamma$	&	10	&	F58	&	Perrier-Bellet	&	2000-04-20	\\
80018	&	VLT	&	NaCo/S13	&	NB1.64	&	1	&	079.C-0216(A)	&	Montagnier	&	2007-05-19	\\
80346	&	CFHT	&	PUEO/KIR	&	Br$\gamma$	&	5	&	07AF12	&	Forveille	&	2007-01-29	\\
80459	&	CFHT	&	PUEO/KIR	&	$J_{\rm cont}$	&	1.5	&	01BF51	&	Gallant	&	2001-08-03	\\
80824	&	CFHT	&	PUEO/KIR	&	K$^{\prime}$	&	0.1	&	01AH25B	&	Baudoz	&	2001-05-01	\\
82809	&	VLT	&	NaCo/S13	&	NB1.64	&	10	&	71.C-0388(A)	&	Beuzit	&	2003-07-20	\\
82817	&	VLT	&	NaCo/S27	&	NB1.64	&	0.345	&	71.D-0465(A)	&	Forveille	&	2003-05-29	\\
83043	&	CFHT	&	PUEO/KIR	&	$H_{2}(\nu=1-0)$ 	&	5	&	05AF19	&	Forveille	&	2005-04-26	\\
83599	&	VLT	&	NaCo/S13	&	NB1.64	&	0.5	&	71.C-0388(A)	&	Beuzit	&	2003-07-19	\\
83762	&	MMT	&	ARIES	&	$K_{{\rm C}~2.09}$	&	1.4	&	UAO-S2	&	De Rosa	&	2013-05-24	\\
83945	&	CFHT	&	PUEO/KIR	&	Br$\gamma$	&	10	&	01AD01	&	Forveille	&	2001-07-07	\\
84099	&	MMT	&	ARIES	&	$K_{{\rm C}~2.09}$	&	1.4	&	UAO-S2	&	De Rosa	&	2013-05-24	\\
84140	&	CFHT	&	PUEO/KIR	&	$H_{2}(\nu=1-0)$ 	&	6	&	04AD8	&	Forveille	&	2004-04-05	\\
84790	&	MMT	&	ARIES	&	$K_{{\rm C}~2.09}$	&	1.4	&	UAO-S2	&	De Rosa	&	2013-05-24	\\
85523	&	VLT	&	NaCo/S27	&	$K_{\rm S}$	&	0.35	&	084.C-0443(A)	&	Lagrange	&	2010-03-20	\\
85665	&	CFHT	&	PUEO/KIR	&	$H_{2}(\nu=1-0)$ 	&	15	&	05AF19	&	Forveille	&	2005-04-26	\\
86057	&	VLT	&	NaCo/S13	&	NB1.64	&	1.8	&	079.C-0216(A)	&	Montagnier	&	2007-05-14	\\
86087	&	MMT	&	ARIES	&	$K_{{\rm C}~2.09}$	&	1.4	&	UAO-S2	&	De Rosa	&	2013-05-24	\\
86162	&	CFHT	&	PUEO/KIR	&	Br$\gamma$	&	1	&	01BF51	&	Gallant	&	2001-08-03	\\
86214	&	VLT	&	NaCo/S13	&	$H$	&	2.5	&	075.C-0733(A)	&	Beuzit	&	2005-05-01	\\
86287	&	CFHT	&	PUEO/KIR	&	Br$\gamma$	&	7	&	F58	&	Perrier-Bellet	&	2000-04-19	\\
86776	&	CFHT	&	PUEO/KIR	&	$H_{2}(\nu=2-1)$ 	&	5	&	02AF43	&	Beuzit	&	2002-07-23	\\
86990	&	VLT	&	NaCo/S13	&	NB1.64	&	1	&	71.C-0388(A)	&	Beuzit	&	2003-07-19	\\
87937	&	VLT	&	NaCo/L27	&	L$^{\prime}$	&	0.2	&	081.C-0430(C)	&	Apai	&	2008-07-03	\\
87938	&	MMT	&	ARIES	&	$K_{{\rm C}~2.09}$	&	1.4	&	UAO-S2	&	De Rosa	&	2013-05-24	\\
88574	&	CFHT	&	PUEO/KIR	&	Br$\gamma$	&	6	&	F58	&	Perrier-Bellet	&	2000-04-20	\\
91699	&	CFHT	&	PUEO/KIR	&	$H_{2}(\nu=2-1)$ 	&	10	&	05AF19	&	Forveille	&	2005-04-26	\\
91768	&	CFHT	&	PUEO/KIR	&	$H_{2}(\nu=1-0)$ 	&	0.5	&	09BC06	&	DeRosa	&	2009-09-01	\\
91772	&	CFHT	&	PUEO/KIR	&	Br$\gamma$	&	1	&	01AH25B	&	Baudoz	&	2001-05-04	\\
92403	&	VLT	&	NaCo/S27	&	IB2.18	&	0.3454	&	091.D-0804(A)	&	De Rosa	&	2013-05-11	\\
92871	&	CFHT	&	PUEO/KIR	&	$H_{2}(\nu=1-0)$ 	&	3	&	05AF19	&	Forveille	&	2005-04-27	\\
93101	&	CFHT	&	PUEO/KIR	&	Pa$\beta$	&	3	&	02BE03	&	Forveille	&	2002-09-18	\\
93873	&	CFHT	&	PUEO/KIR	&	Br$\gamma$	&	5	&	00AD99	&	Forveille	&	2000-08-15	\\
93899	&	CFHT	&	PUEO/KIR	&	K$^{\prime}$	&	10	&	00AD99	&	Forveille	&	2000-08-15	\\
94349	&	VLT	&	NaCo/S13	&	NB1.64	&	7	&	079.C-0216(A)	&	Montagnier	&	2007-05-14	\\
94761	&	CFHT	&	PUEO/KIR	&	Br$\gamma$	&	1.5	&	00BF25	&	Perrier-Bellet	&	2000-08-20	\\
97241	&	CFHT	&	PUEO/KIR	&	Fe~{\scriptsize II}	&	8	&	02BF27	&	Beuzit	&	2002-09-10	\\
97292	&	MMT	&	ARIES	&	$K_{{\rm C}~2.09}$	&	1.4	&	UAO-S10/S11	&	De Rosa/Ward-Duong	&	2013-09-18	\\
99150	&	VLT	&	NaCo/S27	&	$K_{\rm S}$	&	4.7	&	077.C-0483(A)	&	Melo	&	2006-06-09	\\
99701	&	VLT	&	NaCo/S13	&	NB1.64	&	10	&	71.C-0388(A)	&	Beuzit	&	2003-07-19	\\
101180	&	CFHT	&	PUEO/KIR	&	Br$\gamma$	&	4	&	00AD99	&	Forveille	&	2000-08-16	\\
102141	&	VLT	&	NaCo/S13	&	$K_{\rm S}$	&	5	&	083.C-0659(A)	&	Patience	&	2009-06-01	\\
102401	&	MMT	&	ARIES	&	$K_{{\rm C}~2.09}$	&	1.4	&	UAO-S10/S11	&	De Rosa/Ward-Duong	&	2013-09-18	\\
102409	&	VLT	&	NaCo/S13	&	$K_{\rm S}$	&	0.5	&	71.C-0029(A)	&	Mundt	&	2003-07-21	\\
103039	&	CFHT	&	PUEO/KIR	&	Br$\gamma$	&	4	&	01BF50	&	Perrier-Bellet	&	2001-08-06	\\
103096	&	CFHT	&	PUEO/KIR	&	Br$\gamma$	&	1	&	00BF25	&	Perrier-Bellet	&	2000-08-21	\\
103441	&	VLT	&	NaCo/S13	&	NB1.64	&	0.9	&	71.C-0029(A)	&	Mundt	&	2003-07-23	\\
106106	&	CFHT	&	PUEO/KIR	&	Fe~{\scriptsize II}	&	3	&	97IIF28	&	Beuzit	&	1997-12-28	\\
106255	&	VLT	&	NaCo/S13	&	NB1.64	&	1.5	&	075.C-0733(A)	&	Beuzit	&	2005-05-02	\\
106440	&	VLT	&	NaCo/S27	&	$K_{\rm S}$	&	0.35	&	083.C-0599(A)	&	Lagrange	&	2009-09-28	\\
106811	&	Subaru	&	CIAO	&	$K$	&	5	&	o05136	&	Nakajima	&	2005-11-14	\\
108159	&	VLT	&	NaCo/S27	&	$K_{\rm S}$	&	2	&	60.A-9800(J)	&	--	&	2008-10-23	\\
108706	&	CFHT	&	PUEO/KIR	&	Pa$\beta$	&	10	&	02BF27	&	Beuzit	&	2002-09-10	\\
108782	&	CFHT	&	PUEO/KIR	&	Br$\gamma$	&	3	&	01BD02	&	--	&	2001-08-07	\\
109388	&	VLT	&	NaCo/S27	&	$K_{\rm S}$	&	0.5	&	083.C-0151(A)	&	Lagrange	&	2009-08-27	\\
109555	&	CFHT	&	PUEO/KIR	&	Fe~{\scriptsize II}	&	3	&	97IIF28	&	Beuzit	&	1997-12-28	\\
110893	&	CFHT	&	PUEO/KIR	&	Br$\gamma$	&	2	&	01BF50	&	Perrier-Bellet	&	2001-08-04	\\
111313	&	Subaru	&	CIAO	&	$K$	&	10	&	o05104	&	Nakajima	&	2005-07-09	\\
111766	&	VLT	&	NaCo/S27	&	$K_{\rm S}$	&	0.7	&	077.C-0483(A)	&	Melo	&	2006-05-26	\\
111802	&	VLT	&	NaCo/S13	&	NB2.12	&	0.5	&	075.C-0112(A)	&	Udry	&	2005-07-08	\\
112460	&	CFHT	&	PUEO/KIR	&	Fe~{\scriptsize II}	&	1.6	&	97IIF28	&	Beuzit	&	1997-12-28	\\
112774	&	Subaru	&	CIAO	&	$K$	&	5	&	o04104	&	Nakajima	&	2004-09-01	\\
112909	&	MMT	&	ARIES	&	$K_{{\rm C}~2.09}$	&	1.4	&	UAO-S10/S11	&	De Rosa/Ward-Duong	&	2013-09-18	\\
113020	&	CFHT	&	PUEO/KIR	&	Br$\gamma$	&	3	&	01BF50	&	Perrier-Bellet	&	2001-08-04	\\
113229	&	VLT	&	NaCo/S13	&	NB1.64	&	0.345	&	072.C-0570(A)	&	Beuzit	&	2003-12-08	\\
113296	&	CFHT	&	PUEO/KIR	&	Br$\gamma$	&	1	&	00AD99	&	Forveille	&	2000-08-15	\\
114046	&	VLT	&	NaCo/S13	&	NB1.64	&	3.5	&	072.C-0570(A)	&	Beuzit	&	2003-12-09	\\
115332	&	VLT	&	NaCo/S13	&	NB1.64	&	1	&	073.C-0124(A)	&	Udry	&	2004-06-23	\\
115562	&	MMT	&	ARIES	&	$K_{{\rm C}~2.09}$	&	1.4	&	UAO-S2	&	De Rosa	&	2013-05-24	\\
116132	&	CFHT	&	PUEO/KIR	&	Br$\gamma$	&	2	&	00BF25	&	Perrier-Bellet	&	2000-08-21	\\
117473	&	CFHT	&	PUEO/KIR	&	Fe~{\scriptsize II}	&	2	&	98IIF65	&	Perrier	&	1998-09-07	\\
117828	&	VLT	&	NaCo/S13	&	NB1.64	&	0.345	&	072.C-0570(A)	&	Beuzit	&	2003-12-08	\\

\hline
\end{longtable}
\end{center}
\twocolumn

\begin{table}
{\scriptsize
\centering
\caption{\label{table:obssum}Adaptive optics observations summary}
\begin{tabular}{ccccccc}
\hline
Telescope & Instrument &  1st &  2nd   &    Filters & Pixel Scale &  Unique  \\
          &               &  Epoch     &  Epoch      &(bandpass)            & (mas px$^{-1}$)  & Programmes   \\
\hline
CFHT & AOBIR & 94 & 23 & $JHK$ & 38 & 34 \\
VLT & NaCo & 79 & 24 & $JHKL^{\prime}$ & 13, 27, 54 &31 \\
MMT & ARIES & 16 & 10 & $K$ & 20, 40 &  2\\
Subaru & CIAO & 5 & 0 & $HK$ & 22 & 4\\
Subaru & HiCIAO & 1 & 0 & $H$ & 10 & 1\\
Subaru & IRCS & 2 & 0 & $K$ & 21 & 2 \\

\hline
\end{tabular}
}
\end{table}

\subsection{New AO observations}
In addition to the existing archival observations of the sample, 32 dedicated observations -- 16 new first-epoch imaging, and 16 follow-up second epoch imaging -- were obtained. New and follow-up observations for 26 stars were taken in March, May, and September of 2013 using the Arizona Infrared imager and Echelle Spectrograph \citep[ARIES; ][]{aries} at the MMT Observatory. We also obtained second-epoch confirmation imaging for an additional 6 southern targets in our sample from March~2013 through September~2013 with VLT/NaCo (programme ID: 091.D-0804). The details of these observations are listed in Table~\ref{table:obssum}. When possible, follow-up second-epoch images were obtained with similar configurations as the discovery epochs. New observations of targets from the \textsc{MinMs} sample were taken in the $K_{\rm S}$ filter, as the majority of the existing archival observations ($64$~per~cent) were taken in $K_{\rm S}$ band.

\subsection{Archival photographic plates}
In order to extend the AO imaging survey with a wider search for companions at separations from $\sim$$100 - 10,000$~au, we used the SuperCOSMOS Sky Survey Science Archive \citep{hambly} to obtain scans of plates from the UK Schmidt (UKST), ESO Schmidt, and Palomar Oschin Schmidt (POSS) sky surveys for the full 245-star sample. The archival plates provide multiple epochs of observation for each target, taken with the $BVRI$ filters at a pixel scale of 0.67 arcsec px$^{-1}$. With time baselines spanning from 10 to 50 years between the initial and final epochs, and the extremely high proper motions of the targets in our sample (sample median proper motion of 0.63 arcsec yr$^{-1}$), it is possible to search for common proper motion pairs. Mosaics of the plates were made to provide a search radius around each M-dwarf, and the mosaic dimensions correspond to a projected separation of 10,000~au from each primary star.

\section{Data Reduction and Analysis}
\label{sec:Reduction}
\begin{figure*}
\includegraphics[width=1.0\textwidth]{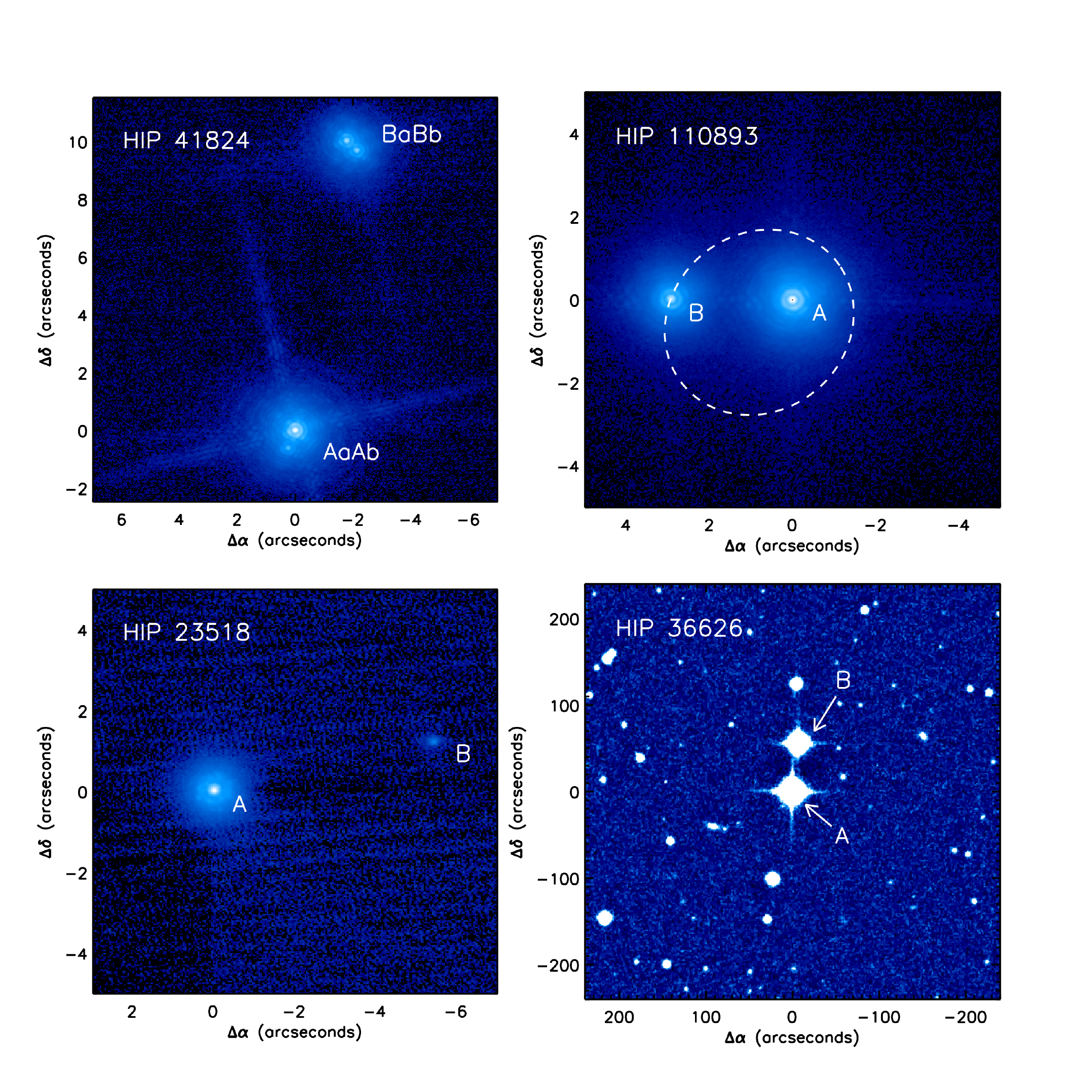}
\caption{Four of the companions identified or recovered within this study. ({\it top left}) A CFHT/AOBIR image of the hierarchical quadruple system HIP~41824 (CU~Cnc), originally discovered by \citet{beuzit2004}, which consists of four M-dwarfs in an $\epsilon$~Lyr-like configuration (e.g. \citealp{2008MNRAS.389..925T}). ({\it top right}) A CFHT/AOBIR image of the HIP 110893 binary system. The measured separation and position angle is consistent with the orbital fit given in \citet{1986A&AS...65..411H}, which is over-plotted (white dashed curve). ({\it bottom left}) An MMT/ARIES image of the newly-discovered binary companion to HIP~23518. With an estimated mass of 0.08~M$_{\odot}$, HIP~23518~B is at the canonical stellar/substellar boundary. ({\it bottom right}) A photographic plate from the Palomar telescope, obtained from the SuperCOSMOS Sky Survey, showing the two components of the wide HIP~36626 (VV~Lyn) binary system \citep{worley62}.}
\label{fig:biggallery}
\end{figure*}

\subsection{Adaptive optics data}
\subsubsection{Image reduction}
For all AO datasets, standardised data reduction techniques of dark subtraction, bad pixel rejection, flat fielding, and sky subtraction were applied. Unsaturated science frames were aligned by fitting a Gaussian to the point spread function (PSF) of the primary in each image to determine the centroid and align on the peak of that fit. After alignment, the individual frames were median combined to form a final reduced science image for each of the \aosamplesize\ M-dwarfs with new or archival AO imaging. The typical combined integration time for a given target was $\sim$60~seconds, with individual exposure times ranging from $0.3-15$ seconds.

\subsubsection{Sensitivity and completeness estimation}
\label{sec:aosens}
Given the heterogeneous origins of the AO datasets, the contrast limits of the observations differ between targets. To assess the sensitivity of the full sample to detecting companions over a range of separations, we generated contrast curves for each target. The ensemble of contrast values was used to estimate the completeness of the study. For each reduced, combined science image, the contrast as a function of separation was measured by first determining the standard deviation of the background level within a five pixel annulus over a range of separations from the primary star, and then calculating the corresponding magnitude difference between the peak value of the primary star PSF and 3$\sigma$ over the background, as tabulated in Table~\ref{table:limits} (Appendix). The maximum angular separation at which a companion could be detected in the image was considered to be the limit where 95~per~cent of the pixels within that radius were within the boundaries of the image field of view.

\subsubsection{Companion detection, photometry and mass estimates}
For each K7-M6 dwarf, the reduced, combined image was carefully visually inspected for candidate stellar companions. Previous comparisons with automated detection procedures have verified the reliability of visual inspection \citep{2009ApJS..181...62M}, and we repeated the inspection multiple times for each target on both individual and combined frames. Examples of detected companions are shown in Figure~\ref{fig:biggallery}. After identification, the fluxes of the candidates and their host stars were measured using aperture photometry in {\sc IDL}. An aperture with radius 2.5 times the average full width at half maximum (FWHM) was chosen in order to measure the total flux associated with the star or candidate in question, with the sky contribution subtracted by defining an annulus of radius 3-9 times the FWHM outside of this aperture. Measurement of the flux ratio between the primary and companion allowed us to derive the magnitude difference of the two objects; any stars unresolved in the 2MASS photometry (with $\Delta m < 4$ and $\rho < 10$ arcsec) were corrected for the individual contributions from each of the stellar components \citep{2011MNRAS.415..854D}. The centroids of the primary target and candidate companion were found with a Gaussian fit, providing an accurate measurement of the pixel separation between the pair of objects. This was converted to an angular separation between the objects using the known pixel scales of the instruments, given in Table~\ref{table:obssum}. We converted angular separation using the \emph{Hipparcos} parallax-based distance measurements into projected separation in au. 

To determine the masses of stellar components in a system, we utilised the low-mass, solar-metallicity evolutionary models of \citet{baraffe98}, assuming a standard field star age of 5~Gyr. Using the absolute magnitudes of the primaries, these models were also applied to the sensitivity curves calculated in Section~\ref{sec:aosens} to determine the completeness of the AO imaging in terms of detectable companion mass.

\subsection{Archival plate analysis}

\subsubsection{Companion detection algorithm}
Detection of co-moving companion candidates within the mosaiced plates was performed by measuring all objects within the plates using Source Extractor \citep{sourceextract}. With these parameters, we derived the proper motion for each object with respect to the stationary background stars in the field. An object was considered a common proper motion (CPM) candidate if its measured proper motion with respect to the background field stars was within 5$\sigma$ of the \emph{Hipparcos}-measured proper motion of the primary. These candidates were then visually inspected by blinking the plates from different epochs to eliminate false positives, such as artefacts from the plate scanning process. As noted in Section~\ref{sec:Sample}, any stars within our \emph{Hipparcos}-selected sample which were themselves companions to earlier spectral types were excluded from our analysis and statistics.

\subsubsection{Sensitivity and completeness estimation}
To determine the sensitivity of the plate images, we calculated radial contrast curves using the same method applied to the AO data, starting at an annulus of five pixels from the centroid position of the primary. The $B$-band images were chosen for the sensitivity estimates as they provided the highest resolution and minimal saturation of the primary. The depth of the plates provided many faint unassociated objects within the field, and the catalogued magnitudes of these unassociated objects corresponded to magnitudes far below the substellar limit for associated objects. Given the depth of the plates, any stellar companions were readily identified by visual inspection. To determine the minimum detection separation for a stellar companion at the bottom of the main sequence, the expected $B$-band magnitude of a 0.08 $M_{\odot}$ star was estimated from models, and the corresponding level of pixel counts in the plate was determined by matching to a star in the PPMXL catalogue \citep{ppmxl} with the same apparent magnitude. 

\subsubsection{Photometry and mass estimates} 
\label{section:mass_estimates}
For each comoving companion identified within the plate images, angular separations and magnitude differences from the primary M-dwarfs were derived from the 2MASS catalogue astrometry and $K_{\rm S}$-band photometry. As in the AO data analysis, the magnitude differences between the components were converted into component masses using 5~Gyr isochrones \citep{baraffe98}. The identified CPM companions were then cross-referenced against binary component identifications in the Washington Double Star catalogue (WDS; \citealp{mason}). Detected companions were also cross-checked against known catalogues of white dwarf/M-dwarf pairs (Silvestri et al. 2001) and white dwarf catalogues (McCook \& Sion, 1999) to remove any contamination from systems with known higher-mass companions.

\subsection{Astrometric confirmation}
\label{subsection:astro_confirmation}


\begin{figure*}
\includegraphics[width=1.0\hsize]{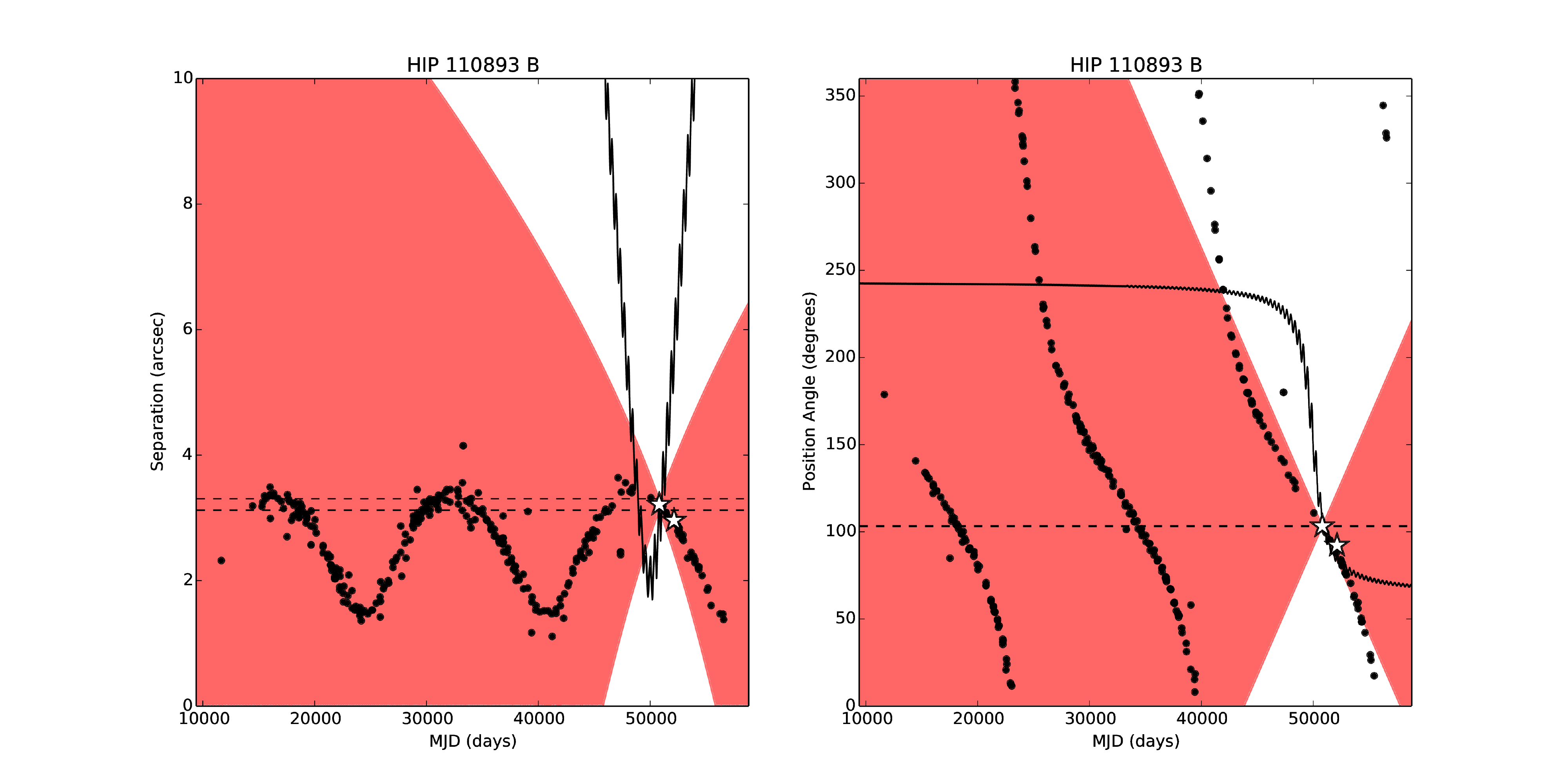}
\caption{Proper motion diagram showing the change in separation (left panel) and position angle (right panel) between the two components of the HIP~110893 binary system. In both diagrams the measurements presented within this study are denoted by large white stars, with the measurements obtained from the WDS catalogue plotted as small circles \citep{mason}. The dashed horizontal lines denote the upper and lower bounds of the separation and position angle measurement within the first epoch. The expected motion of a stationary background object, given the proper motion and parallax of the M3 dwarf target, is enclosed by the solid black lines. Given the maximal change in separation and position angle for circular orbits calculated in Section \ref{subsection:astro_confirmation}, indicative ranges of allowed separations and position angles are denoted by the red shaded region.}
\label{fig:pmd_110893}
\end{figure*}
\begin{figure*}
\includegraphics[width=1.0\hsize]{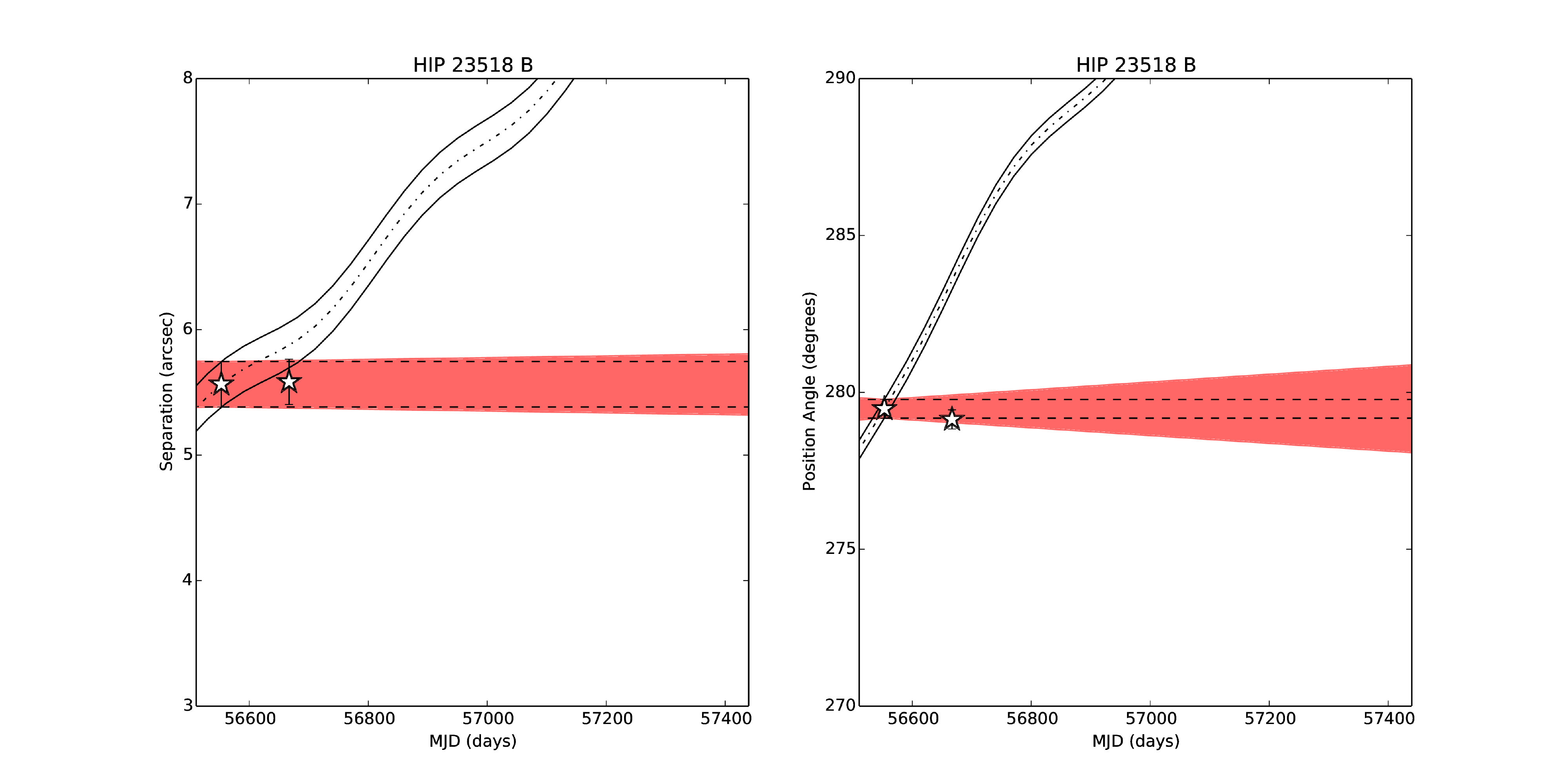}
\caption{Proper motion diagram for the newly-discovered bound companion HIP~23518~B, identified within MMT/ARIES images presented within this study. Symbols, curves, and shading are as with Figure \ref{fig:pmd_110893}.}
\label{fig:pmd_2}
\end{figure*}
\begin{figure*}
\includegraphics[width=1.0\hsize]{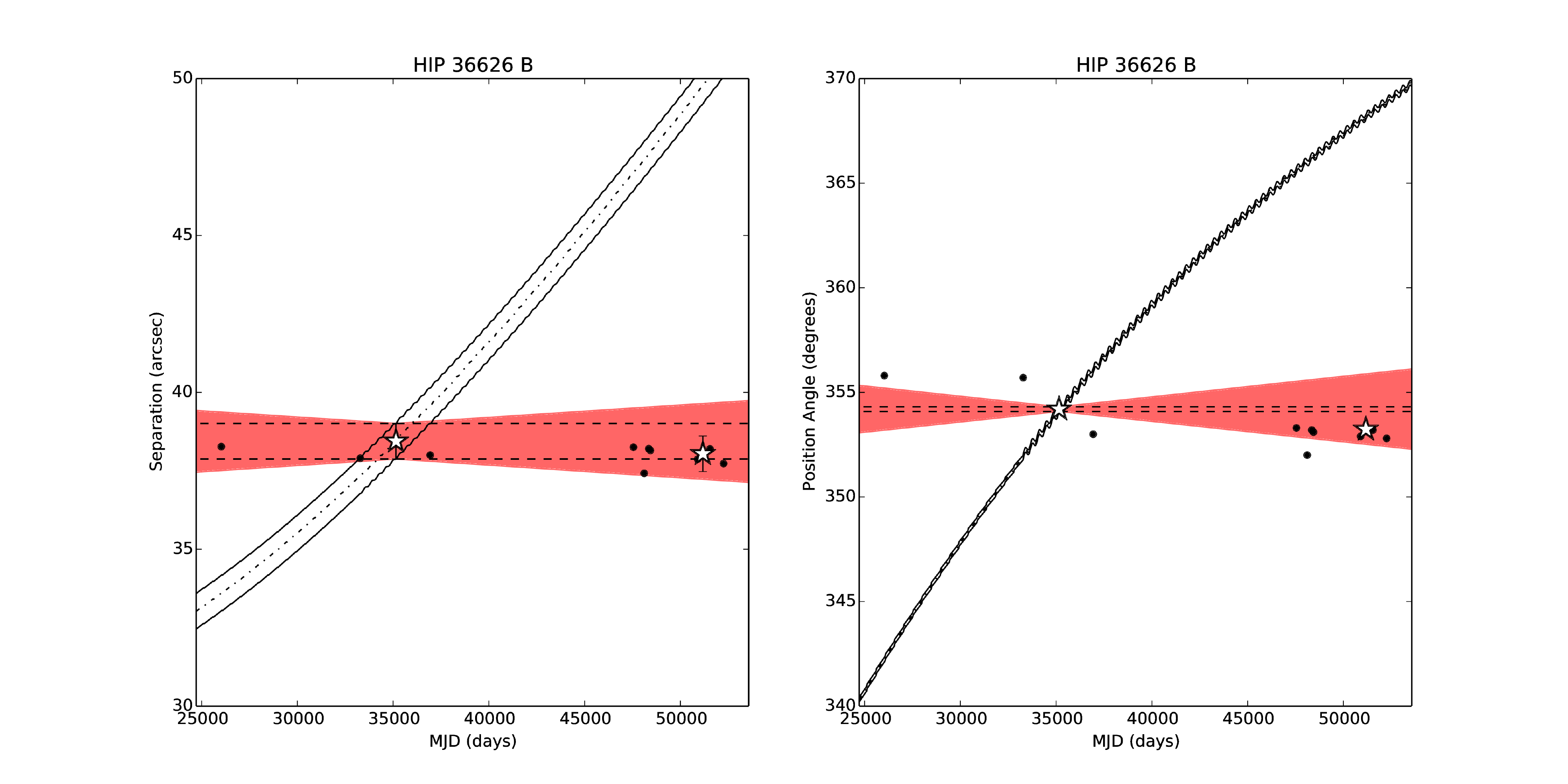}
\caption{Proper motion diagram for the bound companion HIP~36626~B (VV~Lyn~B) identified within the photographic plates. Symbols, curves, and shading are as with Figure \ref{fig:pmd_110893}.}
\label{fig:pmd_3}
\end{figure*}
\begin{figure*}
\includegraphics[width=1.0\hsize]{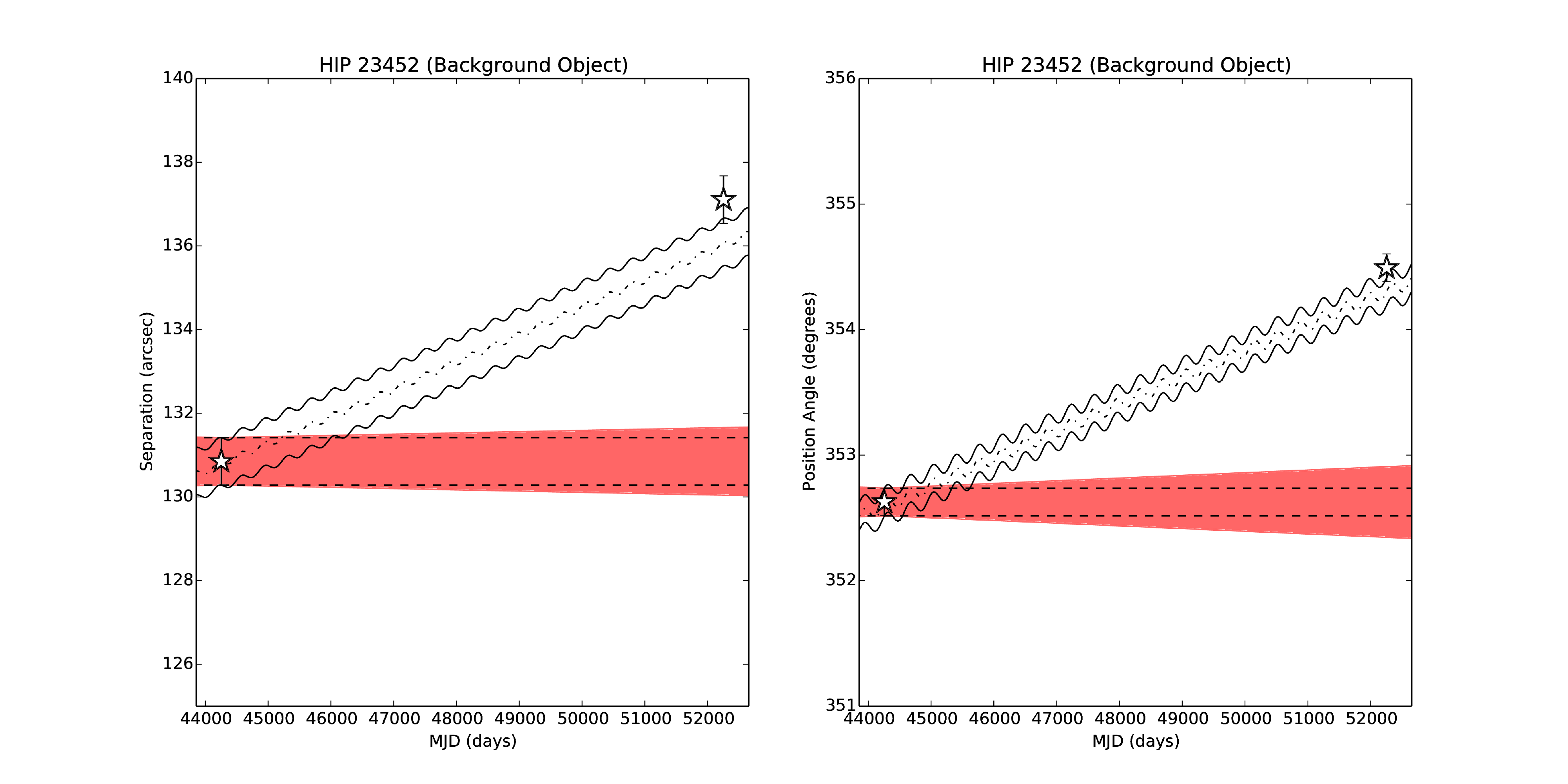}
\caption{Proper motion diagram for a background object within the vicinity of HIP~23452 identified within the photographic plates. Symbols, curves, and shading are as with Figure \ref{fig:pmd_110893}.}
\label{fig:pmd_4}
\end{figure*}


To confirm or reject detected AO and plate CPM candidates, we compared the second-epoch positions with the motion of a background object. The expected motion of a background object was calculated using the \emph{Hipparcos} proper motion and parallax of the primary. In each case, the representative errors in the expected motion of a background object were determined from the first epoch measured uncertainties in position angle and separation. Due to the range of projected separations probed, any resolved companion within the AO images can have measurable orbital motion over the time baseline between the two epochs of observations. In order to estimate a range of possible orbital motions, two scenarios are considered: a face-on, or edge-on circular orbit. For the representative change in the position angle, a face-on circular orbit is used with a semi-major axis equal to the observed projected separation. The period of the orbit is then derived from Kepler's third law, using the masses of the two components given in Table \ref{table:companions}, from which the change in the position angle is calculated.

\begin{figure*}
\includegraphics[width=88mm]{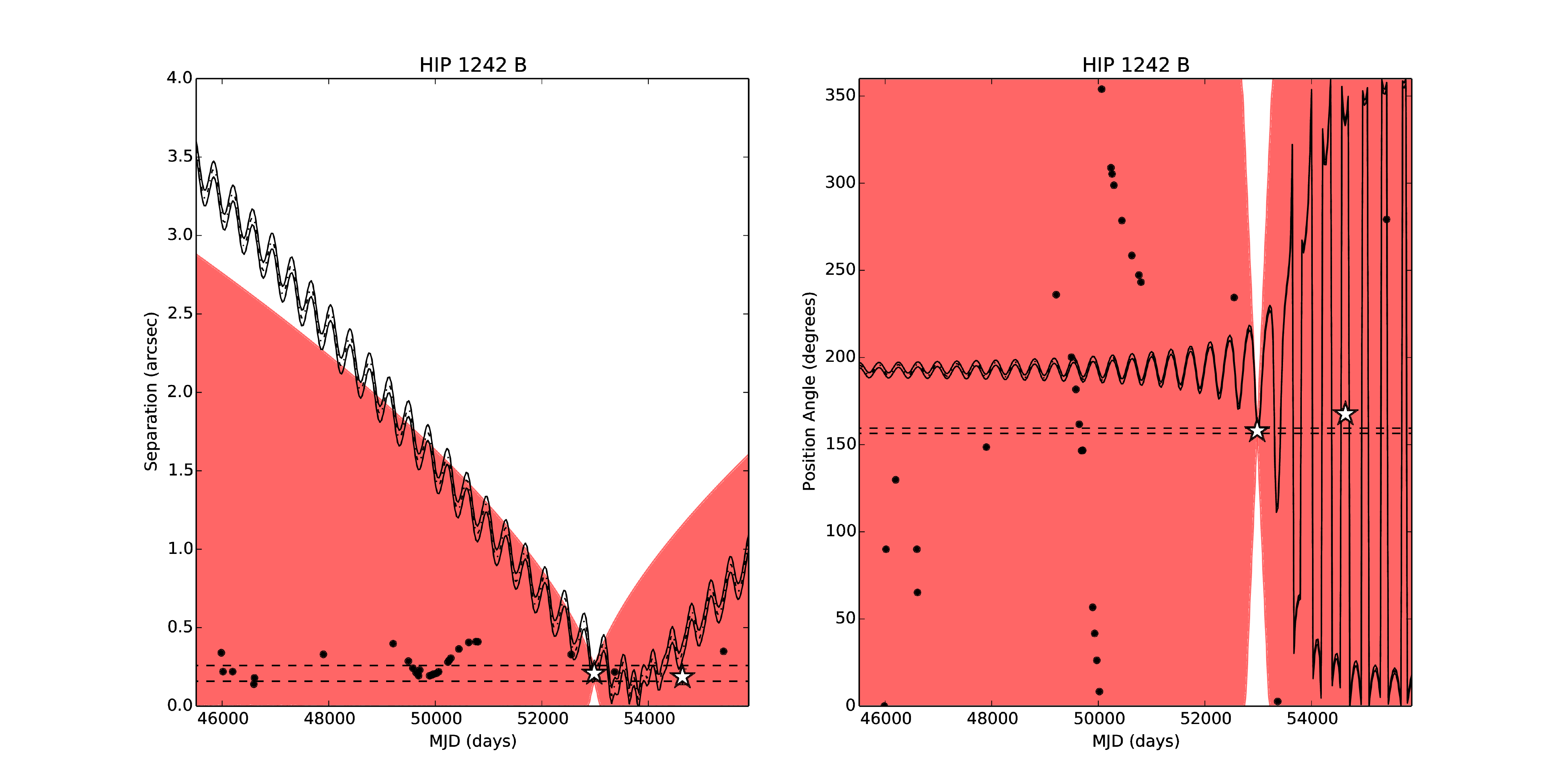}
\includegraphics[width=88mm]{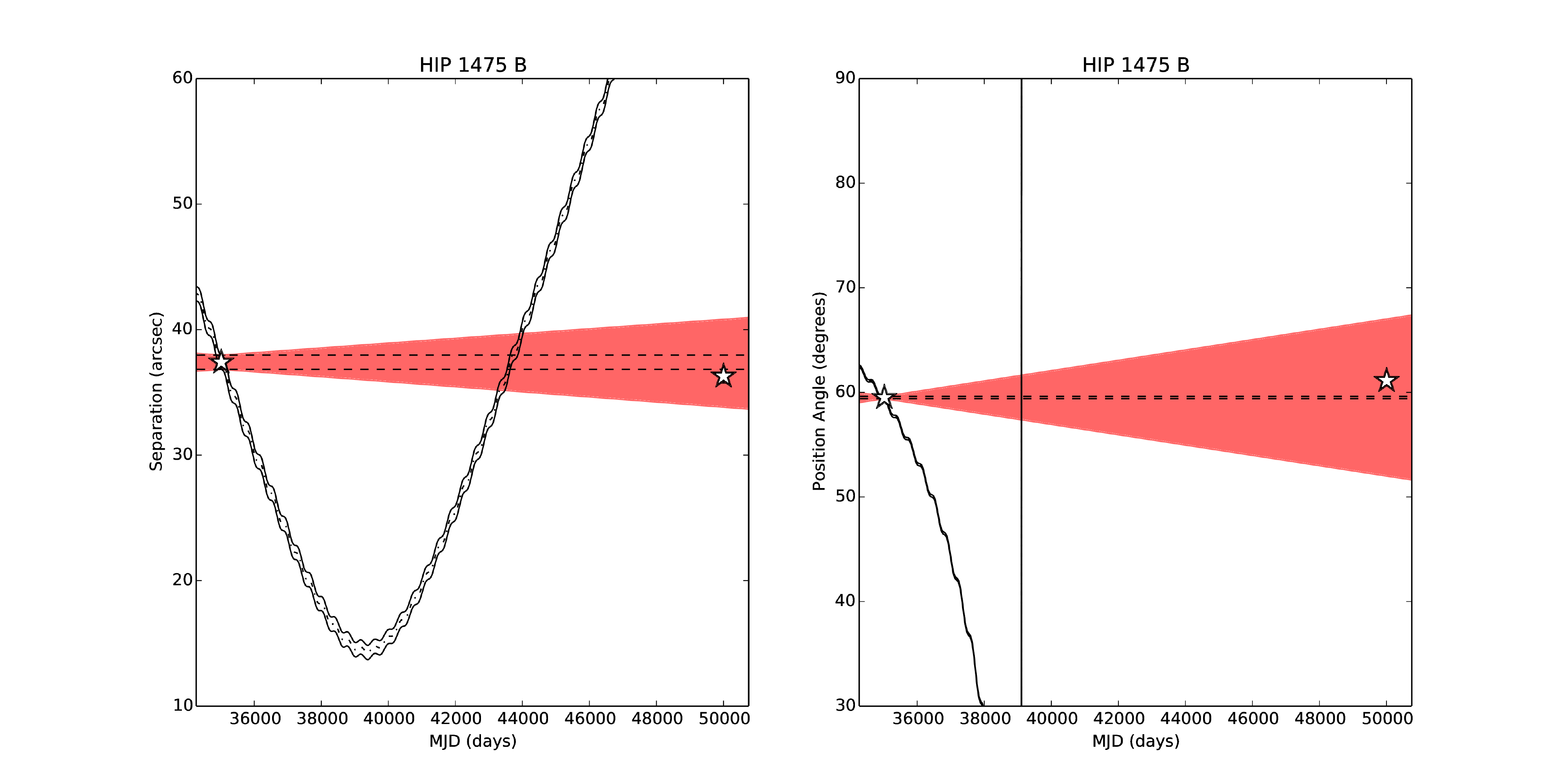}
\\
\includegraphics[width=88mm]{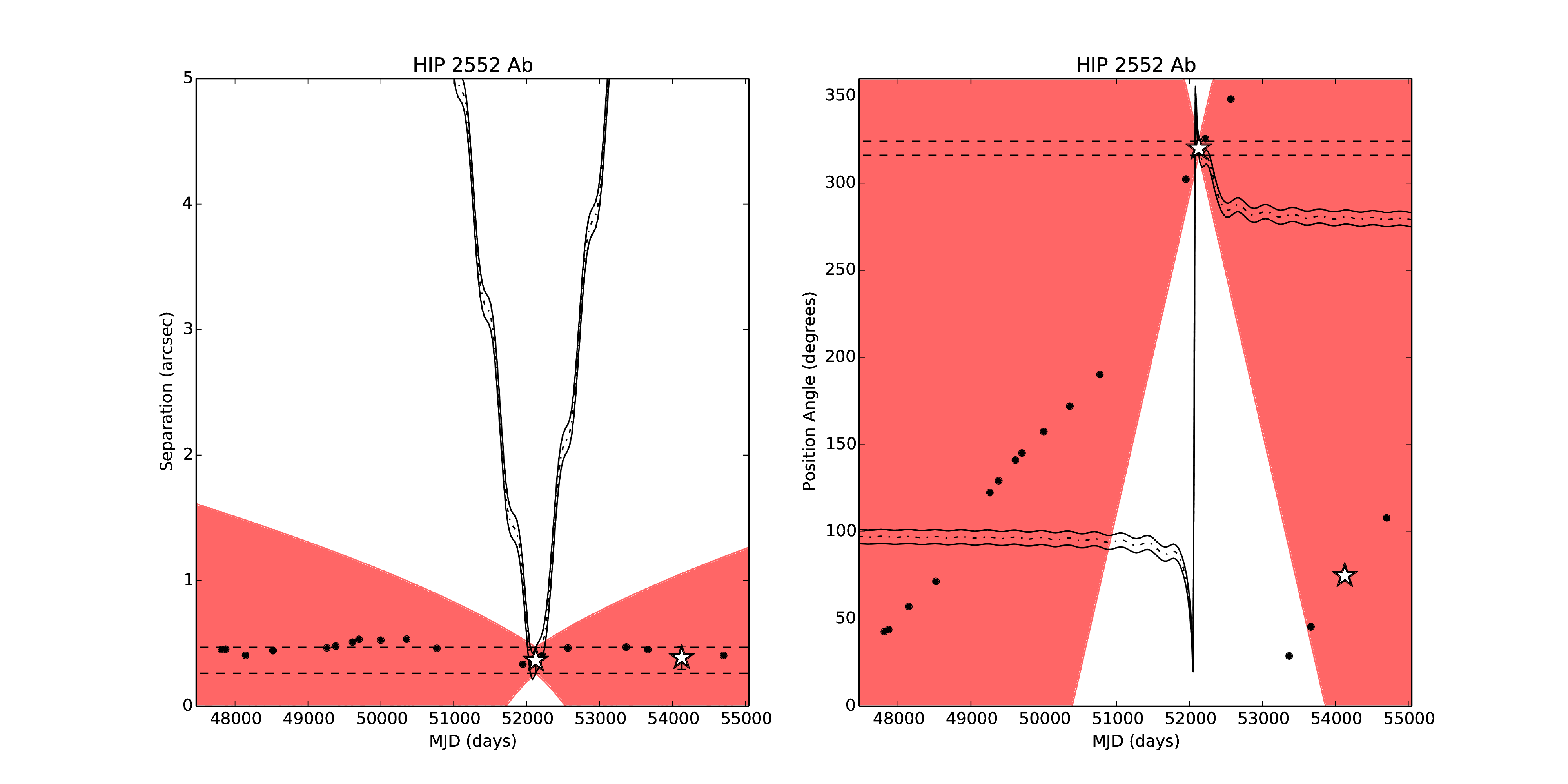}
\includegraphics[width=88mm]{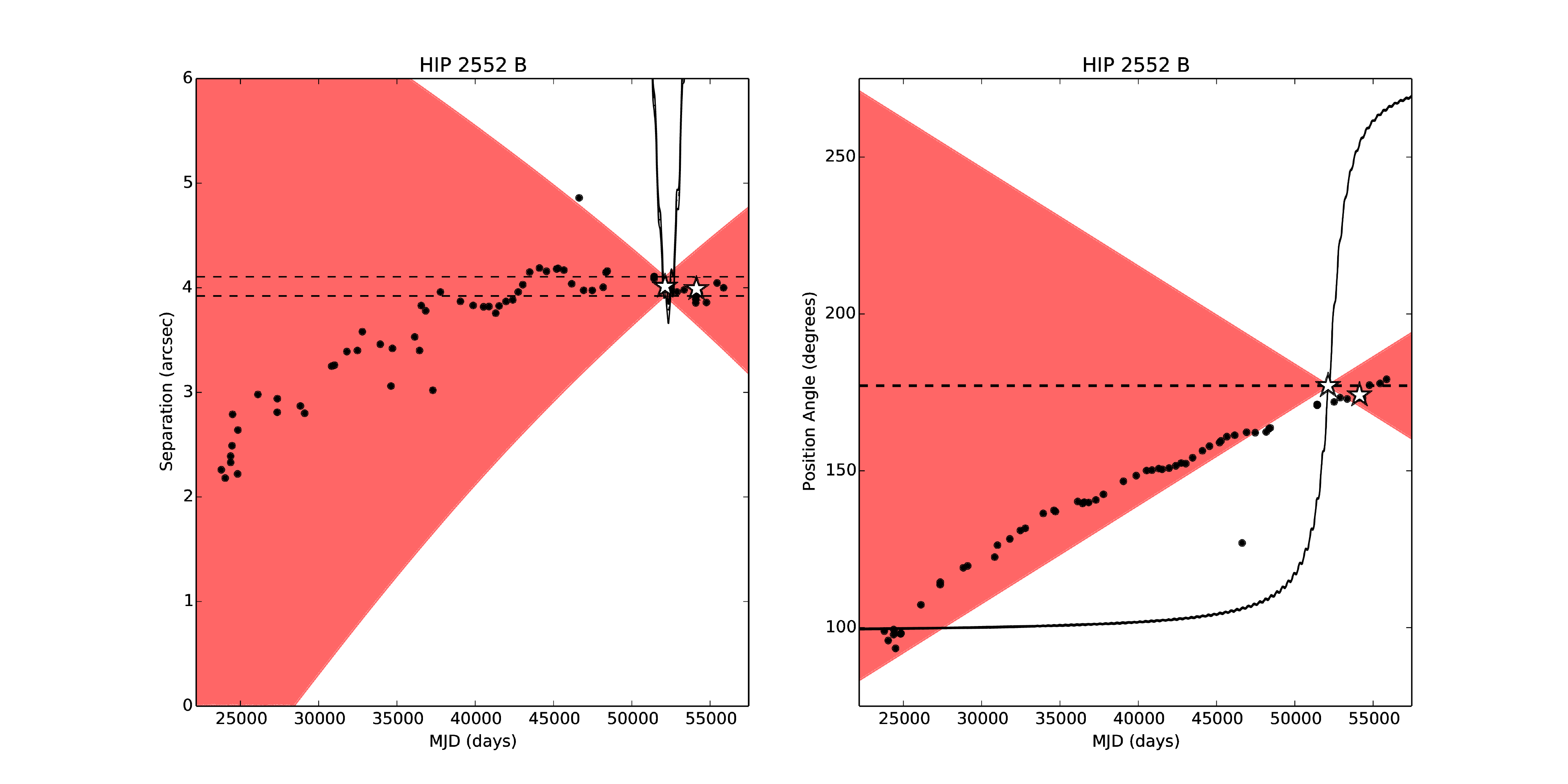}
\\
\includegraphics[width=88mm]{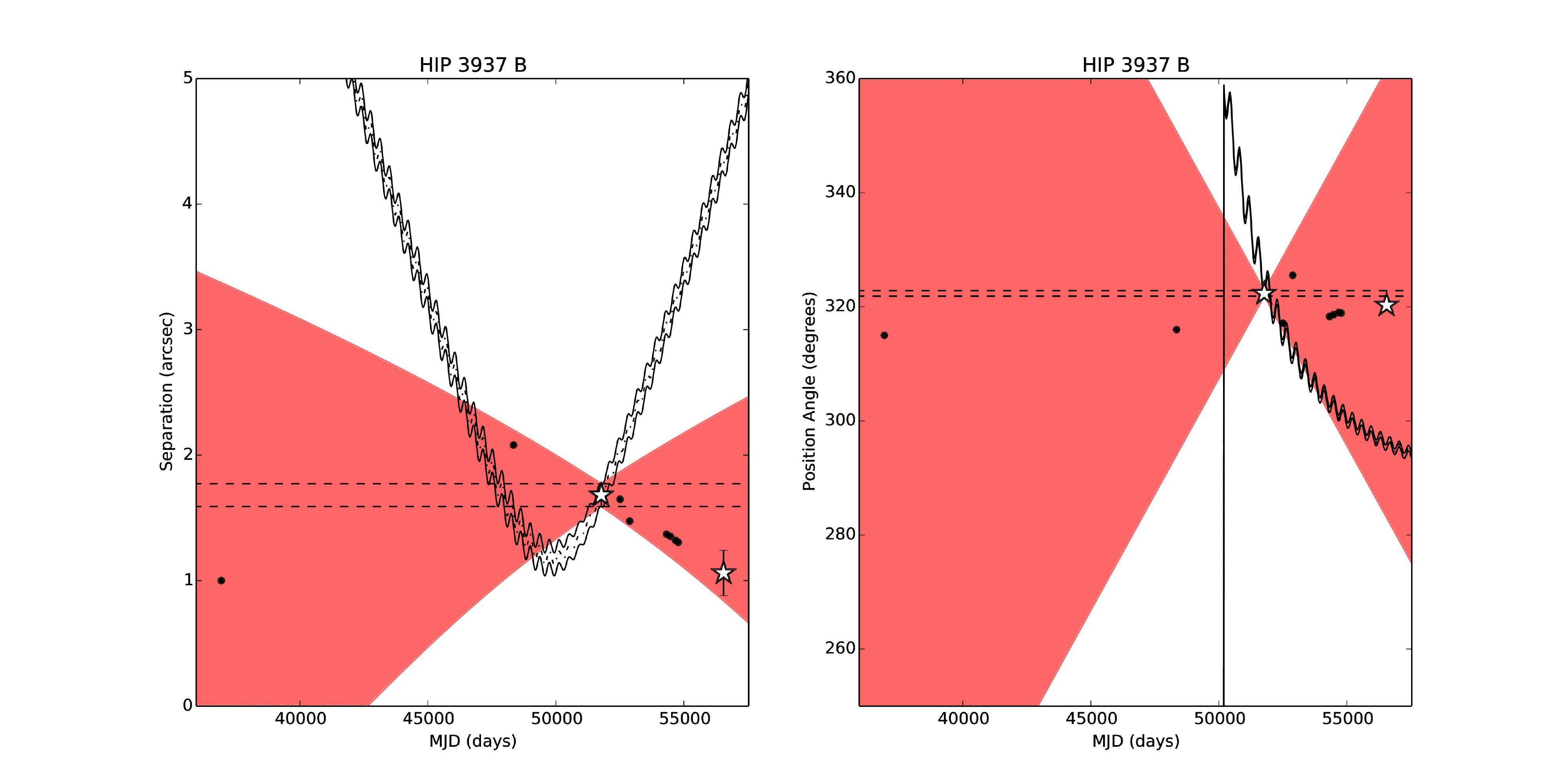}
\includegraphics[width=88mm]{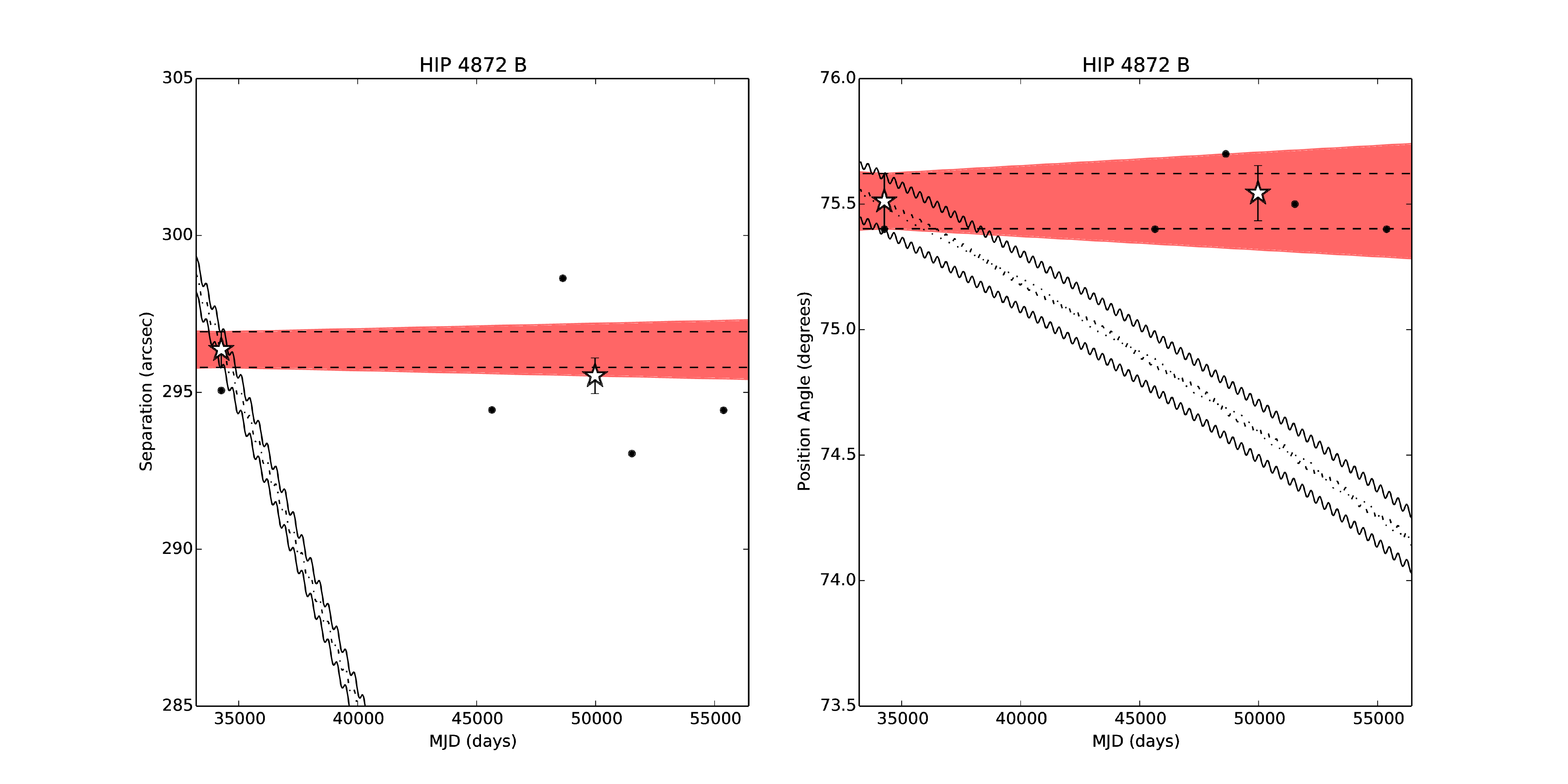}
\\
\includegraphics[width=88mm]{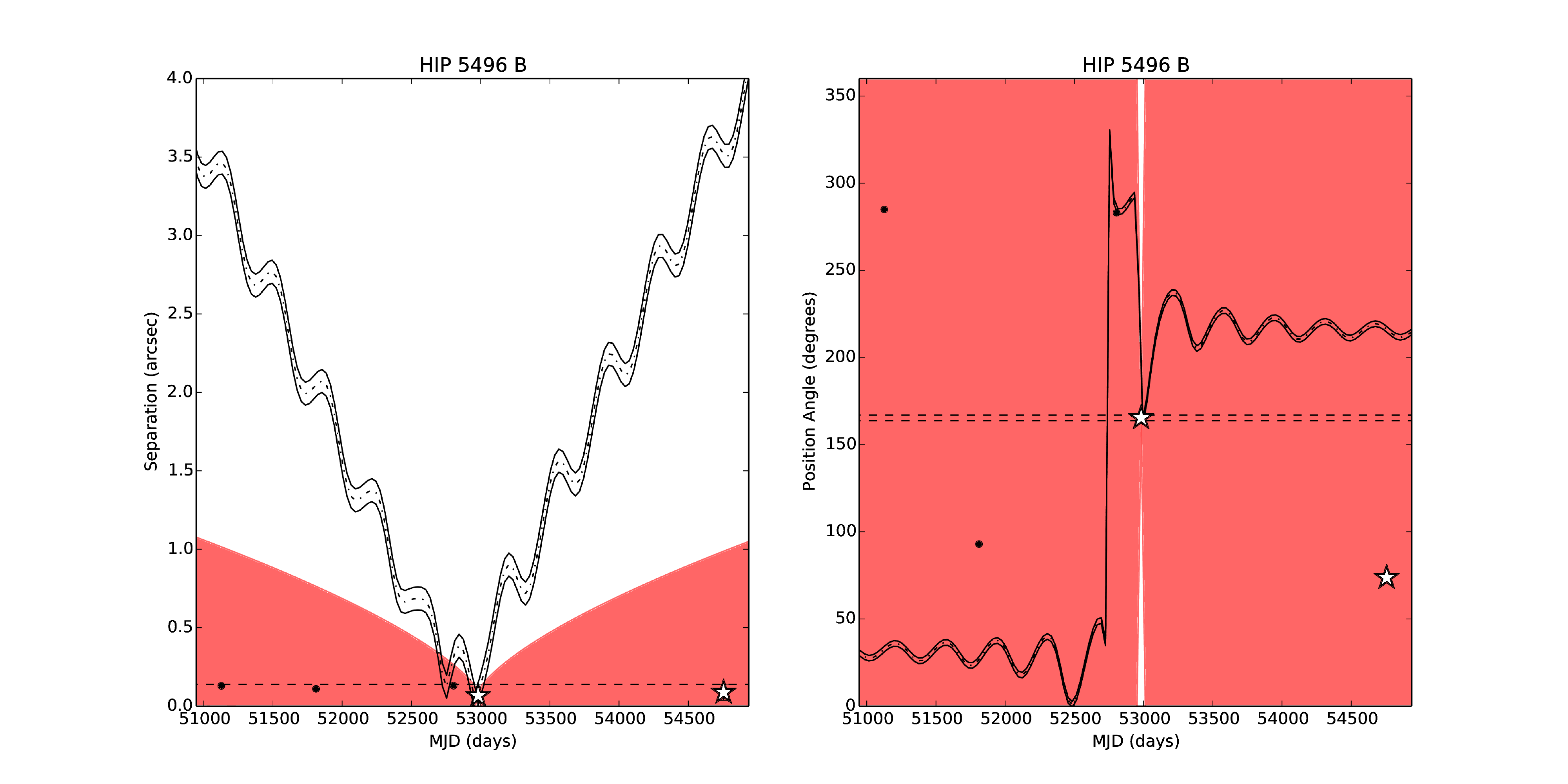}
\includegraphics[width=88mm]{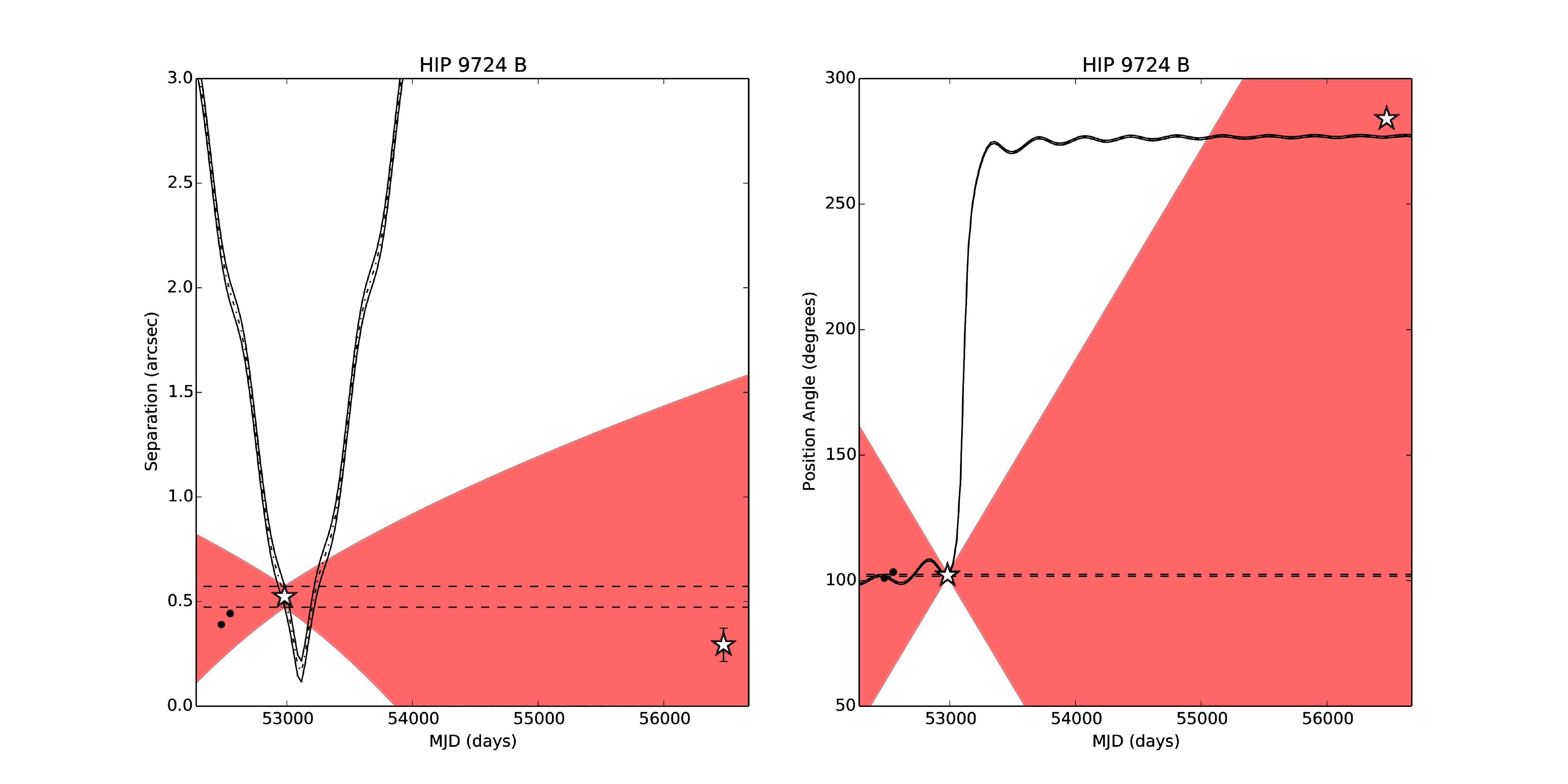}
\\
\includegraphics[width=88mm]{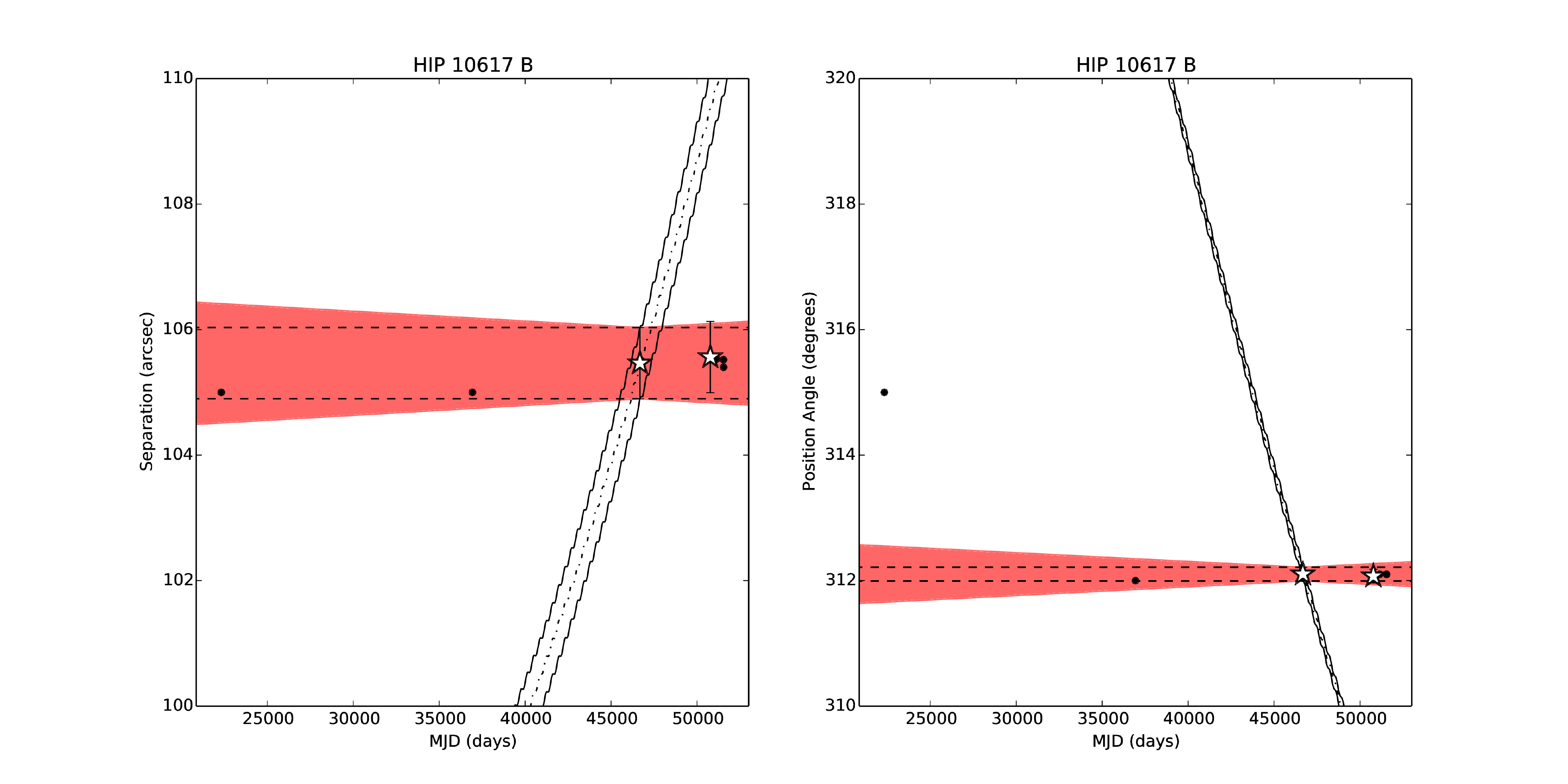}
\includegraphics[width=88mm]{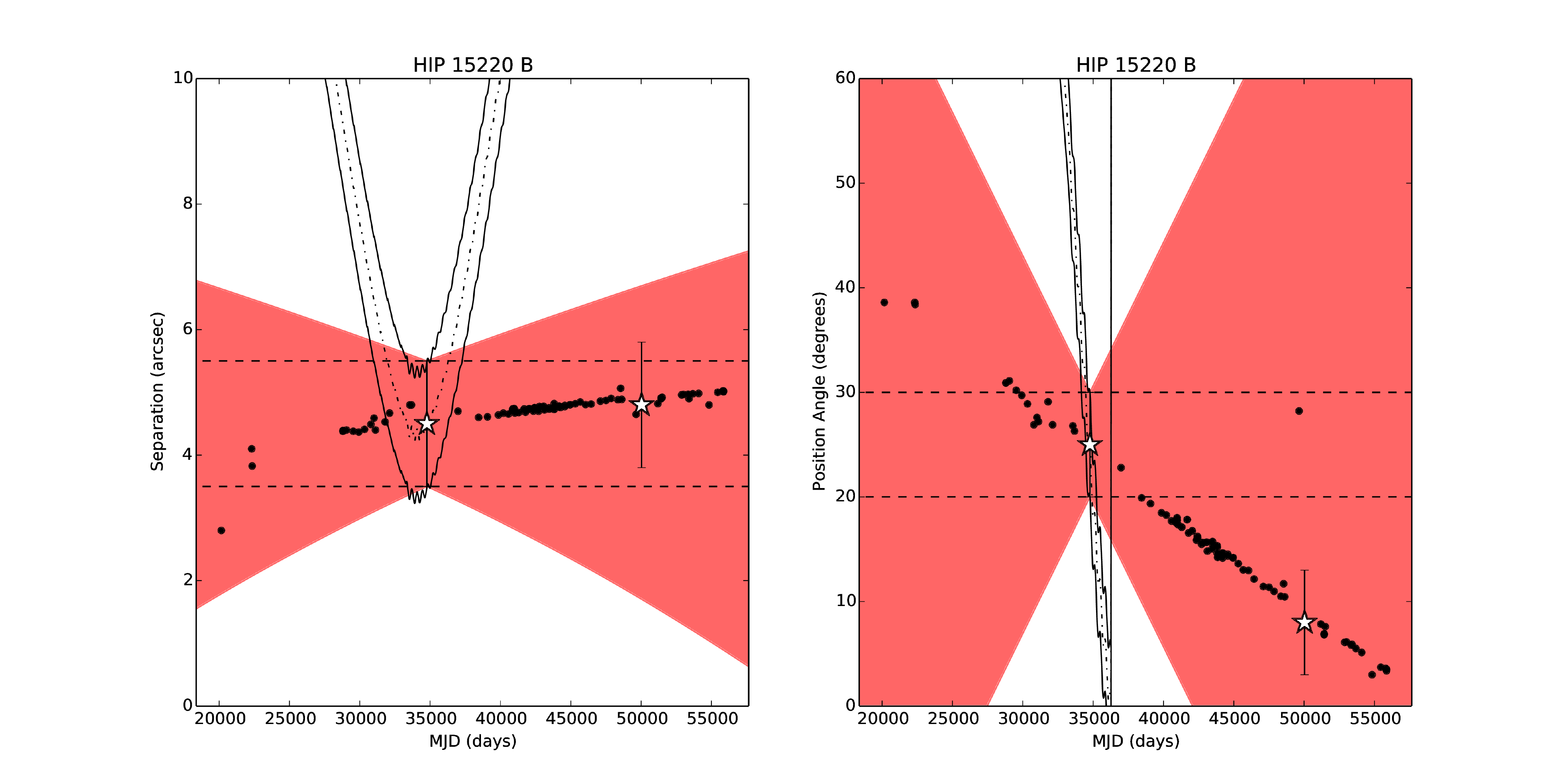}
\caption{Sample proper motion diagrams of the bound companions identified within this study. Symbols, curves, and shading are as with Figure \ref{fig:pmd_110893}. The full set of proper motion diagrams for each of the detected companion candidates in this study is available in the electronic edition of the journal.}
\label{fig:pmd_appendix}
\end{figure*}

In order to estimate the expected change in the separation for a bound component, an edge-on circular orbit is used. As the projected separation does not correspond to a unique value of the semi-major axis ($a$), and the rate of change of the separation depends on both $a$ and the location of the companion within the orbit (characterised by the true anomaly, $\nu$), a large number of orbits were simulated. For a given projected separation ($a_{\rm proj}$), the four unique values of $\nu$ are calculated as 
\begin{equation}
\nu = \pm \cos^{-1}\left(\pm\sqrt{\frac{a_{\rm proj}^2}{a^2}}\right).
\end{equation}
The limiting case for an edge-on circular orbit is when $a = a_{\rm proj}$, with $\nu$ becoming undefined when $a<a_{\rm proj}$. For a given value of $a$, the four values of the true anomaly were increased from $\nu$ to $\nu+2\pi$ over the period determined from Kepler's third law using the masses for each component given in Table \ref{table:companions}. The projected separation at each time step was then calculated as
\begin{equation}
a_{\rm proj} = a |\cos\left(\nu\right)|.
\end{equation}
As the maximal change in the projected separation as a function of time depends on the true semi-major axis, 10,000 simulations were run where the value of $a$ was increased from $a_{\rm proj}$ until the maximum value of $da_{\rm proj}/dt$ was reached (typically between $a = 2-3\times a_{\rm proj}$). These two representative bounds for the change in projected separation and position angle for a bound companion are shown for four examples given in Figure \ref{fig:pmd_110893}, \ref{fig:pmd_2}, \ref{fig:pmd_3}, \ref{fig:pmd_4}, and are shown for all companions which are detected in at least two epochs in Figure \ref{fig:pmd_appendix} (Appendix). The motion of the candidate companion was compared with the expected motion for a background source, and checked for consistency with the expected change for a bound component, within the uncertainties on position angle and separation available from the second epoch measurement. For the the large majority of the historic WDS measurements, no uncertainties are available for additional epochs.

For three of the targets with companions detected at close angular separations ($\rho \la 0.2$~arcsec) --  HIP~36208, HIP~86214, and HIP~114046 -- the candidate is not detected within the second epoch. Given the proper motion of the primary, a background object should have been visible in the field in each of these cases. The lack of detection of the companion in the vicinity of the expected location of a background object is used as evidence for the bound nature of the companion, and they are therefore considered bound for the purposes of this study. Continued monitoring of these targets would be required in order to confirm the physical association of these companion by imaging the companion as it passes through apastron.

\subsubsection{Assessment of binary sample contamination}

We performed two analyses to determine the likelihood of background interlopers contaminating the binary sample. We first determined the background point source density for each of the stars in our sample from the 2MASS catalogue by selecting targets within a 10,000 au radius and within the $K\rm s$-band magnitude limits for potential stellar companions, i.e., fainter than the primary star but brighter than the $K\rm s=10$ substellar limit. We find a median number of 10.5 candidate sources per 10,000~au radius (median of $3.46\times10^{-6}$ sources arcsec$^{-2}$ in a five degree search radius) for the full {\sc MinMs} sample. The nearby distances of stars in our sample lead to the large number of background sources within the full companion search radius. While the maximum separation of an AO-detected binary ($\sim$~35 arcsec) has a very low typical probability of background contamination (often $<<$~1 per cent), the source count method yields significantly higher predictions on the order of a few to 10 per cent for background sources when considering the wider ($> 35$ arcsec) plate-detected candidates.

\begin{figure}
\includegraphics[width=0.50\textwidth]{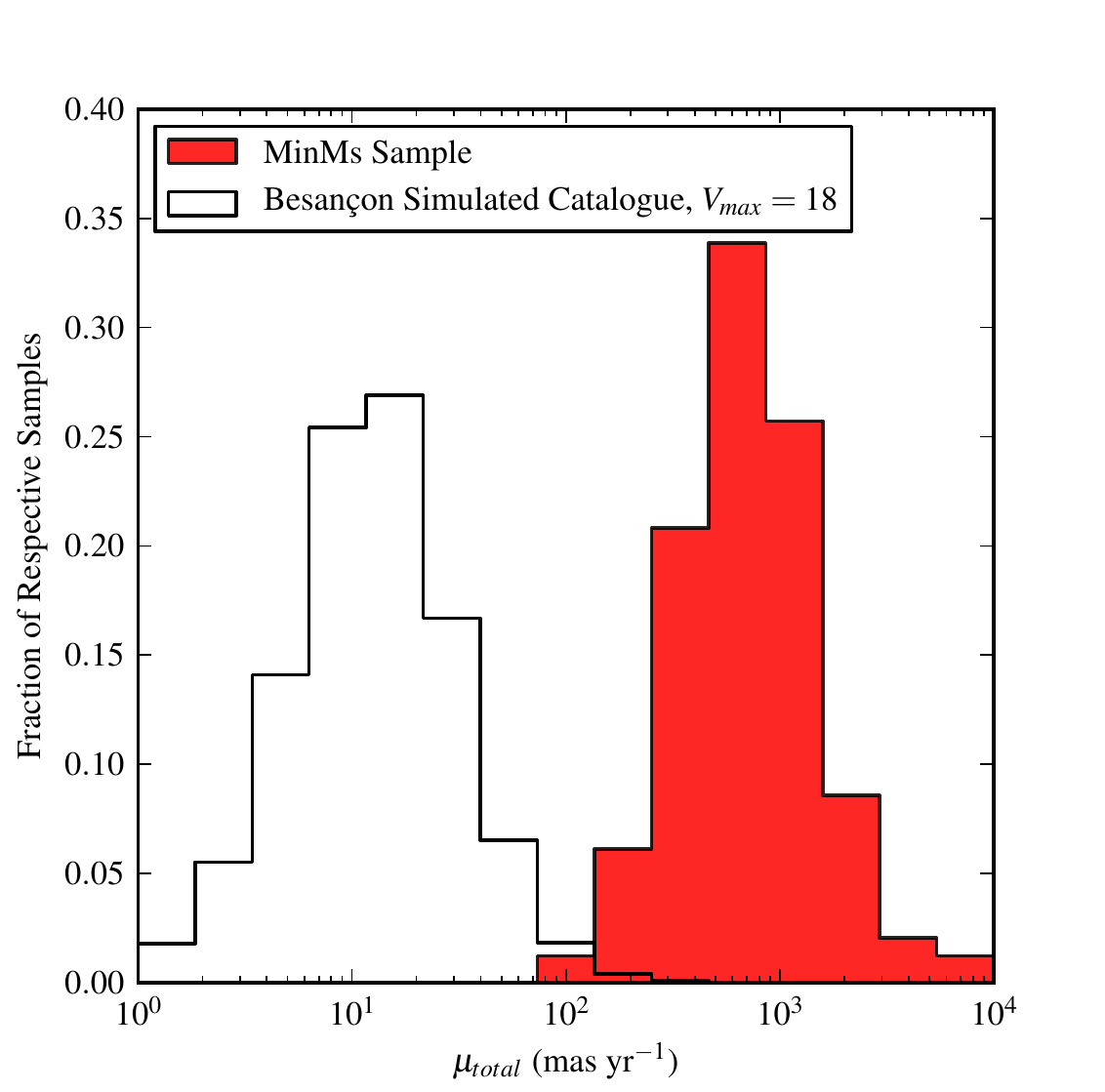}
\caption{The relative proper motion distribution fractions for the {\sc MinMs} sample and a comparable synthetic Besan{\c c}on stellar catalogue \citep{robin2003}, generated with similar sample properties.}
\label{fig:besancon}
\end{figure}

In addition to the background source estimation, we determined the likelihood of unassociated field stars sharing similar 2D proper motions with our sample targets, as such systems would also appear as false positives in the binary statistics. Using the online Besan{\c c}on stellar population synthesis tools \citep{robin2003}, we generated catalogues of over 12,000,000 stars spanning large fields ($0.0 < \alpha < 360$, $-60 < \delta < +60$), with distance and magnitude limits corresponding to the expected companion properties. As seen in Figure~\ref{fig:besancon}, the extremely high proper motion of our sample leads to small overlap with the synthetic population proper motion distribution (even considering only the magnitude of the total proper motion vector and not the constituent 2D vectors, which would further narrow possible matches). For each star in our full sample, we determined the number of proper motion and $K\rm s$-magnitude matches in the synthetic population by selecting on proper motion in both right ascension and declination at various levels of confidence. This yielded the following average numbers of matching synthetic stars over the full sample per target: 1$\sigma$ --  0.08; 3$\sigma$ -- 0.69; 5$\sigma$ -- 1.86; 10$\sigma$ -- 8.23. Of these, each system with a non-zero number of synthetic proper motion matches was selected to check physical association. As shown in Table~\ref{table:besancon}, while these systems may share similar proper motions with some stars in the synthetic sample, the probability of background source contamination from 2MASS in all cases is $<<1$ per cent. The combination of these analyses, particularly when coupled with previous orbital parameter measurements, supports the hypothesis of the physical association of all detected wide systems in our survey.

\begin{table}
{
\centering
\caption{\label{table:besancon} Likelihood of chance proper motion and position matches for low proper motion binaries}
\begin{tabular}{cccccc}
\hline
 HIP    &  \# 5$\sigma$ synth. &  \# 10$\sigma$ synth.  &  $\rho$          &  Source Density & \# Bg. Sources    \\ 
           &   $\mu$ matches       &    $\mu$ matches          &  (asec)   &     (counts asec$^{-2}$) &  (counts)                      \\ 
\hline
3937 & 28 & 127 & 1.68 & 3.15$\times 10^{-3}$ & 2.80$\times 10^{-5}$ \\
15220 & 0 & 1 & 4.5 & 3.21$\times 10^{-5}$ & 2.04$\times 10^{-3}$\\
22738 & 27 & 82 & 7.65 & 3.02$\times 10^{-6}$ & 5.55$\times 10^{-4}$\\
23452  & 0 & 1 & 0.81 & 2.15$\times 10^{-6}$ & 4.44$\times 10^{-6}$\\
28368 & 3 & 11 & 163.15 & 9.54$\times 10^{-6}$ & 7.98$\times 10^{-1}$\\
35191  & 7 & 37 & 0.05 & 7.28$\times 10^{-6}$ & 5.72$\times 10^{-1}$\\
36626  & 1 & 5 & 0.68 & 5.28$\times 10^{-6}$ & 7.67$\times 10^{-6}$\\
41824  & 6 & 39 & 11.04 & 3.42$\times 10^{-6}$ & 1.31$\times 10^{-3}$\\
59406  & 0 & 1 & 84.7 & 3.24$\times 10^{-6}$ & 7.31$\times 10^{-2}$\\
65026  & 200 & 942 & 0.63 & 1.56$\times 10^{-6}$ & 1.95$\times 10^{-6}$\\
65714  & 0 & 1 & 8.86 & 2.87$\times 10^{-6}$ & 7.07$\times 10^{-4}$\\
72944  & 1 & 7 & 4.98 & 1.59$\times 10^{-6}$ & 1.21$\times 10^{-4}$\\
97292 & 0 & 1 & 5.62 & 1.06$\times 10^{-4}$ & 1.05$\times 10^{-8}$\\
102141 & 3 & 6 & 2.82 & 4.21$\times 10^{-6}$ & 1.05$\times 10^{-2}$\\
111802 & 0 & 1 & 25.86 & 1.50$\times 10^{-6}$ & 3.15$\times 10^{-4}$\\

\hline
\end{tabular}
}
\end{table}

\section{Results and Discussion}
\label{sec:Results}

\onecolumn 
\begin{center}
{\scriptsize
\begin{longtable}{l c c c c c c c c c c}
\caption{\label{table:companions} Confirmed companions - measured and derived properties} \\
\hline 
HIP & WDS$^{a}$ & UT  & $\rho$ & $a_{\rm proj}$ & $\theta$ & ${\Delta}m$ & Broadband & $M_{\rm prim}$ & $M_{\rm sec}$ & Epoch \\ 
 & desig. & Date & (arcsec) & (au) & (deg) & (mag) & filter & ($M_{\sun}$) & ($M_{\sun}$) &  \\ 
\hline
\endfirsthead
\caption{continued.} \\
\hline 
HIP & WDS & UT & $\rho$ & $a_{\rm proj}$ & $\theta$ & ${\Delta}m$ & Broadband & $M_{\rm prim}$ & $M_{\rm sec}$ & Epoch \\ 
 & desig. & Date & (arcsec) & (au) & (deg) & (mag) & filter & ($M_{\sun}$) & ($M_{\sun}$)  & \\ 
\hline
\endhead
\hline
\endfoot

\hline
\noalign{\smallskip}
\multicolumn{11}{l}{$^{a}$ WDS catalogue component designation.} \\
\multicolumn{11}{l}{$^{b}$ targets within the WDS catalog, but without an assigned component designation.} \\
\multicolumn{11}{l}{$^{c}$ is a newly confirmed companion without a designation in the WDS catalog ($\star$ denotes a new companion).} \\
\multicolumn{11}{l}{$^{d}$ HIP~91772 is designated as the ``B" component in the WDS catalog, but has the largest estimated mass of the system.} \\
\multicolumn{11}{l}{$^{e}$ denotes a target for which there is no second epoch detection, but instead a non-detection of a background object (Section 4.3).} \\
\multicolumn{11}{l}{$^{f}$ denotes instance where second epoch data were not available, but multiple confirmation epochs were obtained from previous WDS measurements.} \\
\endlastfoot

1242	&	--$^{b}$	&	2003-12-10	&	0.21	$\pm$	0.05	&	1.0	$\pm$	0.3	&	158.0	$\pm$	1.0	&	0.93	&	$H$	&	0.15	&	0.11	&	1	\\
	&		&	2008-06-22	&	0.19	$\pm$	0.01	&	0.9	$\pm$	0.0	&	167.6	$\pm$	3.2	&	0.58	&	$H$	&	0.18	&	0.14	&	2	\\
1475	&	B	&	1954-10-03	&	37.40	$\pm$	0.60	&	134.1	$\pm$	2.0	&	59.5	$\pm$	0.1	&	2.71	&	$K_{S}$	&	0.42	&	0.16	&	1	\\
	&		&	1995-10-13	&	36.28	$\pm$	0.70	&	130.2	$\pm$	2.0	&	61.1	$\pm$	0.2	&	2.71	&	$K_{S}$	&	0.42	&	0.16	&	2	\\
2552	&	Ab	&	2001-08-05	&	0.40	$\pm$	0.10	&	4.0	$\pm$	1.0	&	320.0	$\pm$	4.0	&	1.28	&	$K_{S}$	&	0.40	&	0.21	&	1	\\
	&		&	2007-01-29	&	0.39	$\pm$	0.02	&	3.9	$\pm$	0.2	&	72.5	$\pm$	1.7	&	1.21	&	$K_{S}$	&	0.46	&	0.25	&	2	\\
2552	&	B	&	2001-08-05	&	4.01	$\pm$	0.09	&	40.0	$\pm$	1.0	&	177.1	$\pm$	0.2	&	1.35	&	$K_{S}$	&	0.40	&	0.22	&	1	\\
	&		&	2007-01-29	&	3.99	$\pm$	0.06	&	40.2	$\pm$	0.6	&	171.8	$\pm$	1.1	&	1.35	&	$K_{S}$	&	0.40	&	0.22	&	2	\\
3937	&	--	&	2000-08-19	&	1.68	$\pm$	0.09	&	20.0	$\pm$	2.0	&	322.3	$\pm$	0.5	&	0.61	&	$H$	&	0.26	&	0.19	&	1	\\
	&		&	2013-09-18	&	1.03	$\pm$	0.18	&	12.1	$\pm$	2.1	&	319.8	$\pm$	0.7	&	0.61	&	$K_{S}$	&	0.26	&	0.19	&	2	\\
4872	&	B	&	1952-09-15	&	296.40	$\pm$	0.60	&	2,952.0	$\pm$	45.0	&	75.5	$\pm$	0.3	&	0.61	&	$K_{S}$	&	0.56	&	0.2	&	1	\\
	&		&	1995-09-14	&	295.50	$\pm$	0.60	&	2,943.5	$\pm$	45.0	&	75.5	$\pm$	0.3	&	0.61	&	$K_{S}$	&	0.56	&	0.2	&	2	\\
5496	&	--	&	2003-12-09	&	0.07	$\pm$	0.07	&	0.5 $^{+0.6}_{-0.5}$			&	165.0	$\pm$	2.0	&	0.44	&	$H$	&	0.43	&	0.36	&	1	\\
	&		&	2008-10-17	&	0.09	$\pm$	0.01	&	0.7	$\pm$	0.1	&	73.7	$\pm$	1.8	&	0.26	&	$H$	&	0.52	&	0.45	&	2	\\
9724	&	--	&	2003-12-09	&	0.52	$\pm$	0.05	&	4.8	$\pm$	0.5	&	102.1	$\pm$	0.4	&	3.25	&	$H$	&	0.47	&	0.11	&	1	\\
	&		&	2013-07-03	&	0.62	$\pm$	0.01	&	5.7	$\pm$	0.5	&	102.1	$\pm$	0.4	&	3.25	&	$H$	&	0.47	&	0.11	&	2	\\
10617	&	--	&	1986-09-03	&	105.50	$\pm$	0.60	&	1,511.0	$\pm$	68.0	&	312.1	$\pm$	0.1	&	2.11	&	$K_{S}$	&	0.40	&	0.27	&	1	\\
	&		&	1997-11-22	&	105.60	$\pm$	0.60	&	1,512.1	$\pm$	68.0	&	312.1	$\pm$	0.1	&	2.11	&	$K_{S}$	&	0.40	&	0.27	&	2	\\
15220	&	B	&	1954-01-28	&	6.75	$\pm$	1.00	&	97.2	$\pm$	15.0	&	12.9	$\pm$	5.0	&	0.07	&	$K_{S}$	&	0.50	&	0.49	&	1	\\
	&		&	1995-11-12	&	5.65	$\pm$	1.00	&	81.2	$\pm$	15.0	&	6.0	$\pm$	5.0	&	0.07	&	$K_{S}$	&	0.50	&	0.49	&	2	\\
22738	&	--	&	2004-09-23	&	7.65	$\pm$	0.05	&	85.0	$\pm$	2.0	&	314.7	$\pm$	0.1	&	0.72	&	$H$	&	0.36	&	0.25	&	1	\\
	&		&	2012-08-24	&	7.47	$\pm$	0.56	&	83.1	$\pm$	0.2	&	314.7	$\pm$	0.1	&	0.71	&	$H$	&	0.35	&	0.25	&	2	\\
23452	&	B	&	2002-09-18	&	0.81	$\pm$	0.09	&	6.9	$\pm$	0.8	&	285.4	$\pm$	0.7	&	1.30	&	$K_{S}$	&	0.59	&	0.37	&	1	\\
	&		&	2014-01-13	&	0.89	$\pm$	0.18	&	7.6	$\pm$	1.6	&	335.4	$\pm$	4.8	&	1.30	&	$K_{S}$	&	0.59	&	0.37	&	2	\\
23518	&	$\star$$^{c}$	&	2013-09-18	&	5.60	$\pm$	0.20	&	76.0	$\pm$	3.0	&	279.5	$\pm$	0.3	&	4.81	&	$K_{S}$	&	0.54	&	0.08	&	1	\\
	&		&	2014-01-10	&	5.58	$\pm$	0.20	&	76.1	$\pm$	2.4	&	279.1	$\pm$	0.3	&	4.81	&	$K_{S}$	&	0.54	&	0.08	&	2	\\
23932	&	--	&	2003-03-16	&	0.06	$\pm$	0.05	&	0.6	$\pm$	0.5	&	30.0	$\pm$	4.0	&	0.10	&	$K_{S}$	&	0.43	&	0.41	&	1	\\
	&		&	2011-02-16	&	0.07	$\pm$	0.05	&	0.7	$\pm$	0.5	&	44.4	$\pm$	5.5	&	0.10	&	$K_{S}$	&	0.43	&	0.41	&	2	\\
28368	&	--	&	1954-01-05	&	163.10	$\pm$	0.60	&	2,208.0	$\pm$	50.0	&	119.4	$\pm$	0.6	&	0.20	&	$K_{S}$	&	0.53	&	0.22	&	1	\\
	&		&	1996-12-10	&	162.50	$\pm$	0.60	&	2,199.6	$\pm$	50.0	&	119.5	$\pm$	0.6	&	0.20	&	$K_{S}$	&	0.53	&	0.22	&	2	\\
29316	&	B	&	2004-01-07	&	1.72	$\pm$	0.09	&	19.0	$\pm$	1.0	&	30.1	$\pm$	0.2	&	1.50	&	$K_{S}$	&	0.45	&	0.22	&	1	\\
	&		&	2013-09-18	&	0.58	$\pm$	0.18	&	6.3	$\pm$	1.9	&	52.5	$\pm$	1.1	&	1.50	&	$K_{S}$	&	0.45	&	0.22	&	2	\\
30920	&	B	&	2006-04-11	&	1.37	$\pm$	0.05	&	5.6	$\pm$	0.2	&	47.4	$\pm$	0.2	&	1.24	&	$K_{S}$	&	0.21	&	0.1	&	1	\\
	&		&	2009-03-28	&	1.07	$\pm$	0.01	&	4.4	$\pm$	0.0	&	74.6	$\pm$	0.3	&	1.61	&	$K_{S}$	&	0.24	&	0.13	&	2	\\
31293	&	--	&	1977-02-10	&	23.00	$\pm$	0.60	&	207.3	$\pm$	7.0	&	35.3	$\pm$	0.3	&	0.76	&	$K_{S}$	&	0.45	&	0.32	&	1	\\
	&		&	1989-12-31	&	23.22	$\pm$	0.60	&	209.4	$\pm$	7.0	&	34.0	$\pm$	0.3	&	0.76	&	$K_{S}$	&	0.45	&	0.32	&	2	\\
33142	&	--	&	2002-09-12	&	0.30	$\pm$	0.10	&	3.0	$\pm$	1.0	&	250.0	$\pm$	3.0	&	0.47	&	$H$	&	0.32	&	0.25	&	1	\\
	&		&	2005-04-27	&	0.19	$\pm$	0.01	&	1.9	$\pm$	0.1	&	160.5	$\pm$	3.9	&	0.47	&	$H$	&	0.32	&	0.25	&	2	\\
33499	&	--$^{f}$	&	2003-12-08	&	0.83	$\pm$	0.05	&	6.6	$\pm$	0.4	&	282.9	$\pm$	0.3	&	0.01	&	$H$	&	0.24	&	0.24	&	1	\\
35191	&	--	&	2000-02-25	&	0.06	$\pm$	0.05	&	0.7	$\pm$	0.6	&	154.0	$\pm$	6.0	&	0.32	&	$J$	&	0.39	&	0.34	&	1	\\
	&		&	2003-12-08	&	0.06	$\pm$	0.01	&	0.8	$\pm$	0.1	&	334.5	$\pm$	6.0	&	0.32	&	$J$	&	0.39	&	0.34	&	2	\\
36208	&	$\star ^{e}$	&	2003-12-11	&	0.17	$\pm$	0.05	&	0.6	$\pm$	0.2	&	327.0	$\pm$	4.0	&	1.07	&	$H$	&	0.24	&	0.14	&	1	\\
36626	&	Ab	&	2000-04-20	&	0.70	$\pm$	0.10	&	8.0	$\pm$	1.0	&	201.0	$\pm$	4.0	&	1.87	&	$K_{S}$	&	0.51	&	0.21	&	1	\\
	&		&	2013-09-18	&	1.53	$\pm$	0.01	&	18.1	$\pm$	0.1	&	194.4	$\pm$	0.6	&	1.87	&	$K_{S}$	&	0.51	&	0.21	&	2	\\
36626	&	B	&	1955-02-13	&	38.40	$\pm$	0.60	&	456.0	$\pm$	20.0	&	354.2	$\pm$	0.1	&	3.82	&	$K_{S}$	&	0.51	&	0.39	&	1	\\
	&		&	1998-12-28	&	38.04	$\pm$	0.60	&	451.5	$\pm$	20.0	&	353.2	$\pm$	0.1	&	3.82	&	$K_{S}$	&	0.51	&	0.39	&	2	\\
41824	&	Ab	&	1999-02-27	&	0.67	$\pm$	0.09	&	7.0	$\pm$	1.0	&	157.2	$\pm$	0.9	&	1.48	&	$K_{S}$	&	0.38	&	0.11	&	1	\\
	&		&	2007-01-29	&	0.65	$\pm$	0.09	&	7.2	$\pm$	1.0	&	169.9	$\pm$	1.1	&	1.48	&	$K_{S}$	&	0.38	&	0.11	&	2	\\
41824	&	Ba	&	1999-02-27	&	10.17	$\pm$	0.09	&	113.0	$\pm$	10.0	&	348.6	$\pm$	0.1	&	1.74	&	$K_{S}$	&	0.38	&	0.17	&	1	\\
	&		&	2007-01-29	&	10.32	$\pm$	0.09	&	114.2	$\pm$	10.0	&	348.5	$\pm$	0.1	&	1.74	&	$K_{S}$	&	0.38	&	0.17	&	2	\\
41824	&	Bb	&	1999-02-27	&	9.91	$\pm$	0.09	&	110.0	$\pm$	10.0	&	346.2	$\pm$	0.1	&	2.20	&	$K_{S}$	&	0.38	&	0.13	&	1	\\
	&		&	2007-01-29	&	9.58	$\pm$	0.09	&	106.0	$\pm$	10.0	&	346.0	$\pm$	0.1	&	2.20	&	$K_{S}$	&	0.38	&	0.13	&	2	\\
46706	&	--	&	1997-12-28	&	0.70	$\pm$	0.10	&	7.0	$\pm$	1.0	&	239.0	$\pm$	1.0	&	0.09	&	$K_{S}$	&	0.42	&	0.41	&	1	\\
	&		&	2004-04-05	&	0.52	$\pm$	0.01	&	5.2	$\pm$	0.0	&	49.8	$\pm$	0.3	&	0.09	&	$K_{S}$	&	0.54	&	0.53	&	2	\\
47620	&	B	&	1955-01-26	&	89.10	$\pm$	0.60	&	1,096.0	$\pm$	25.0	&	77.1	$\pm$	0.1	&	0.60	&	$K_{S}$	&	0.52	&	0.46	&	1	\\
	&		&	2000-01-13	&	88.80	$\pm$	0.60	&	1,091.8	$\pm$	25.0	&	77.5	$\pm$	0.1	&	0.60	&	$K_{S}$	&	0.52	&	0.46	&	2	\\
49969	&	--	&	2000-02-19	&	0.20	$\pm$	0.10	&	2.0	$\pm$	1.0	&	253.0	$\pm$	8.0	&	0.81	&	$H$	&	0.43	&	0.29	&	1	\\
	&		&	2003-03-18	&	0.16	$\pm$	0.01	&	1.9	$\pm$	0.0	&	61.8	$\pm$	1.4	&	0.51	&	$H$	&	0.49	&	0.36	&	2	\\
54211	&	B	&	1955-03-19	&	30.20	$\pm$	0.60	&	147.0	$\pm$	3.0	&	131.1	$\pm$	0.2	&	4.40	&	$K_{S}$	&	0.40	&	0.1	&	1	\\
	&		&	2000-03-14	&	31.91	$\pm$	0.60	&	154.7	$\pm$	3.0	&	126.6	$\pm$	0.2	&	4.40	&	$K_{S}$	&	0.40	&	0.1	&	2	\\
59406	&	--	&	1954-04-02	&	84.70	$\pm$	0.60	&	1,066.0	$\pm$	32.0	&	120.9	$\pm$	0.1	&	1.58	&	$K_{S}$	&	0.36	&	0.26	&	1	\\
	&		&	1994-05-18	&	85.12	$\pm$	0.60	&	1,071.7	$\pm$	32.0	&	120.9	$\pm$	0.1	&	1.58	&	$K_{S}$	&	0.36	&	0.26	&	2	\\
60910	&	--$^{f}$	&	2008-02-19	&	0.07	$\pm$	0.05	&	0.9	$\pm$	0.7	&	114.0	$\pm$	5.0	&	0.70	&	$K_{S}$	&	0.32	&	0.22	&	1	\\
62556	&	--	&	1997-12-27	&	0.20	$\pm$	0.10	&	2.0	$\pm$	1.0	&	240.0	$\pm$	7.0	&	0.55	&	$H$	&	0.37	&	0.28	&	1	\\
	&		&	2003-03-17	&	0.32	$\pm$	0.04	&	3.3	$\pm$	0.4	&	202.8	$\pm$	4.3	&	0.55	&	$H$	&	0.37	&	0.28	&	2	\\
63510	&	B	&	2005-05-01	&	0.28	$\pm$	0.05	&	3.2	$\pm$	0.6	&	357.0	$\pm$	1.0	&	3.26	&	$K_{S}$	&	0.58	&	0.12	&	1	\\
	&		&	2006-05-23	&	0.24	$\pm$	0.01	&	2.8	$\pm$	0.1	&	307.4	$\pm$	4.5	&	3.71	&	$K_{S}$	&	0.58	&	0.12	&	2	\\
65011	&	B	&	1950-05-15	&	19.60	$\pm$	0.20	&	260.0	$\pm$	9.0	&	123.1	$\pm$	0.3	&	0.66	&	$K_{S}$	&	0.63	&	0.3	&	1	\\
	&		&	1990-05-06	&	17.80	$\pm$	0.20	&	236.1	$\pm$	9.0	&	128.9	$\pm$	0.3	&	0.66	&	$K_{S}$	&	0.63	&	0.3	&	2	\\
65026	&	B	&	2000-04-20	&	0.63	$\pm$	0.09	&	7.0	$\pm$	1.0	&	107.0	$\pm$	6.0	&	0.85	&	$K_{S}$	&	0.67	&	0.52	&	1	\\
	&		&	2005-04-27	&	1.26	$\pm$	0.09	&	13.5	$\pm$	1.0	&	94.2	$\pm$	0.2	&	0.85	&	$K_{S}$	&	0.67	&	0.52	&	2	\\
65714	&	--	&	1956-04-08	&	8.90	$\pm$	0.60	&	123.0	$\pm$	9.0	&	45.7	$\pm$	2.0	&	1.36	&	$K_{S}$	&	0.48	&	0.22	&	1	\\
	&		&	1996-04-21	&	7.65	$\pm$	0.60	&	106.2	$\pm$	9.0	&	51.5	$\pm$	2.0	&	1.36	&	$K_{S}$	&	0.48	&	0.22	&	2	\\
71898	&	--	&	2002-06-25	&	2.88	$\pm$	0.09	&	31.0	$\pm$	1.0	&	109.9	$\pm$	0.2	&	3.94	&	$K_{S}$	&	0.40	&	0.08	&	1	\\
	&		&	2004-04-25	&	2.77	$\pm$	0.01	&	29.7	$\pm$	0.1	&	105.8	$\pm$	0.2	&	3.94	&	$K_{S}$	&	0.40	&	0.08	&	2	\\
72896	&	--	&	2001-08-06	&	1.00	$\pm$	0.10	&	11.0	$\pm$	1.0	&	116.0	$\pm$	1.0	&	0.77	&	$K_{S}$	&	0.29	&	0.2	&	1	\\
	&		&	2013-04-19	&	1.02	$\pm$	0.01	&	10.3	$\pm$	0.0	&	93.7	$\pm$	0.0	&	0.77	&	$K_{S}$	&	0.37	&	0.25	&	2	\\
72944	&	Ba	&	2005-06-04	&	4.92	$\pm$	0.05	&	47.5	$\pm$	0.9	&	32.5	$\pm$	0.2	&	2.66	&	$K_{S}$	&	0.49	&	0.09	&	1	\\
	&		&	2006-03-01	&	4.91	$\pm$	0.01	&	47.4	$\pm$	0.1	&	33.4	$\pm$	0.1	&	2.66	&	$K_{S}$	&	0.49	&	0.09	&	2	\\
72944	&	Bb	&	2005-06-04	&	4.98	$\pm$	0.05	&	48.1	$\pm$	0.9	&	32.0	$\pm$	0.2	&	4.80	&	$K_{S}$	&	0.49	&	0.08	&	1	\\
	&		&	2006-03-01	&	5.01	$\pm$	0.01	&	48.4	$\pm$	0.1	&	32.4	$\pm$	0.1	&	4.80	&	$K_{S}$	&	0.49	&	0.08	&	2	\\
73470	&	--	&	2003-03-17	&	1.36	$\pm$	0.09	&	16.0	$\pm$	1.0	&	179.2	$\pm$	0.2	&	2.26	&	$K_{S}$	&	0.60	&	0.24	&	1	\\
	&		&	2005-04-25	&	1.54	$\pm$	0.01	&	18.1	$\pm$	0.1	&	173.0	$\pm$	0.1	&	2.26	&	$K_{S}$	&	0.62	&	0.25	&	2	\\
79755	&	--	&	1955-04-23	&	64.00	$\pm$	0.60	&	684.0	$\pm$	9.0	&	12.7	$\pm$	0.1	&	0.99	&	$K_{S}$	&	0.66	&	0.47	&	1	\\
	&		&	1994-06-18	&	64.87	$\pm$	0.60	&	693.2	$\pm$	9.0	&	13.5	$\pm$	0.1	&	0.99	&	$K_{S}$	&	0.66	&	0.47	&	2	\\
80018	&	--	&	2007-05-19	&	4.92	$\pm$	0.05	&	41.0	$\pm$	1.0	&	226.9	$\pm$	0.2	&	3.02	&	$H$	&	0.38	&	0.1	&	1	\\
	&		&	2013-03-29	&	4.43	$\pm$	0.02	&	37.0	$\pm$	0.1	&	226.0	$\pm$	0.1	&	3.02	&	$H$	&	0.38	&	0.1	&	2	\\
82817	&	B	&	2003-05-29	&	0.11	$\pm$	0.05	&	0.7	$\pm$	0.3	&	275.0	$\pm$	2.0	&	0.28	&	$H$	&	0.43	&	0.38	&	1	\\
	&		&	2005-07-21	&	0.22	$\pm$	0.01	&	1.4	$\pm$	0.1	&	173.8	$\pm$	0.1	&	0.28	&	$H$	&	0.43	&	0.38	&	2	\\
82817	&	C	&	1954-06-01	&	73.10	$\pm$	0.60	&	453.0	$\pm$	16.0	&	313.8	$\pm$	0.1	&	0.28	&	$H$	&	0.43	&	0.2	&	1	\\
	&		&	1996-04-25	&	72.75	$\pm$	0.60	&	450.7	$\pm$	16.0	&	313.3	$\pm$	0.1	&	0.28	&	$H$	&	0.43	&	0.2	&	2	\\
82817	&	F	&	1954-06-01	&	232.50	$\pm$	0.60	&	1,440.0	$\pm$	50.0	&	155.4	$\pm$	0.4	&	4.58	&	$K_{S}$	&	0.43	&	0.09	&	1	\\
	&		&	1996-04-25	&	232.01	$\pm$	0.60	&	1,437.4	$\pm$	50.0	&	155.4	$\pm$	0.4	&	4.58	&	$K_{S}$	&	0.43	&	0.09	&	2	\\
84140	&	B	&	1999-04-04	&	1.11	$\pm$	0.09	&	6.6	$\pm$	0.6	&	225.0	$\pm$	0.5	&	0.01	&	$K_{S}$	&	0.34	&	0.34	&	1	\\
	&		&	2004-04-05	&	0.25	$\pm$	0.09	&	1.5	$\pm$	0.5	&	166.6	$\pm$	5.5	&	0.01	&	$K_{S}$	&	0.47	&	0.47	&	2	\\
84794	&	--	&	1951-07-05	&	15.80	$\pm$	0.60	&	183.0	$\pm$	11.0	&	267.3	$\pm$	0.2	&	1.10	&	$K_{S}$	&	0.44	&	0.29	&	1	\\
	&		&	1996-08-09	&	16.81	$\pm$	0.60	&	194.9	$\pm$	11.0	&	268.7	$\pm$	0.2	&	1.10	&	$K_{S}$	&	0.44	&	0.29	&	2	\\
86057	&	--	&	2005-05-01	&	3.76	$\pm$	0.05	&	37.0	$\pm$	1.0	&	325.4	$\pm$	0.3	&	1.93	&	$H$	&	0.45	&	0.18	&	1	\\
	&		&	2007-05-14	&	3.94	$\pm$	0.05	&	38.3	$\pm$	1.0	&	323.7	$\pm$	0.3	&	1.93	&	$H$	&	0.45	&	0.18	&	2	\\
86214	&	$\star ^{e}$	&	2005-05-01	&	0.17	$\pm$	0.05	&	0.9	$\pm$	0.3	&	190.0	$\pm$	5.0	&	1.01	&	$H$	&	0.23	&	0.14	&	1	\\
91772	&	A$^{d}$	&	2001-05-04	&	12.53	$\pm$	0.09	&	44.7	$\pm$	0.5	&	170.1	$\pm$	0.1	&	0.60	&	$K_{S}$	&	0.26	&	0.19	&	1	\\
	&		&	2009-09-01	&	11.98	$\pm$	0.04	&	42.8	$\pm$	0.1	&	219.4	$\pm$	0.2	&	0.60	&	$K_{S}$	&	0.26	&	0.19	&	2	\\
93899	&	--	&	1951-07-13	&	116.60	$\pm$	0.60	&	1,020.0	$\pm$	21.0	&	290.2	$\pm$	0.5	&	1.01	&	$K_{S}$	&	0.32	&	0.32	&	1	\\
	&		&	1994-06-12	&	115.40	$\pm$	0.60	&	1,010.1	$\pm$	21.0	&	291.0	$\pm$	0.5	&	1.01	&	$K_{S}$	&	0.32	&	0.32	&	2	\\
94349	&	--	&	2003-07-19	&	0.08	$\pm$	0.05	&	0.8	$\pm$	0.6	&	58.0	$\pm$	6.0	&	0.80	&	$H$	&	0.34	&	0.23	&	1	\\
	&		&	2007-05-14	&	0.16	$\pm$	0.01	&	1.7	$\pm$	0.0	&	319.7	$\pm$	1.8	&	0.80	&	$H$	&	0.41	&	0.28	&	2	\\
94761	&	B	&	1950-08-12	&	72.90	$\pm$	0.60	&	428.0	$\pm$	4.0	&	149.5	$\pm$	0.5	&	4.64	&	$K_{S}$	&	0.49	&	0.09	&	1	\\
	&		&	1995-08-16	&	75.36	$\pm$	0.60	&	442.4	$\pm$	4.0	&	151.6	$\pm$	0.5	&	4.64	&	$K_{S}$	&	0.49	&	0.09	&	2	\\
97292	&	B$^{f}$	&	2013-09-18	&	5.60	$\pm$	0.20	&	76.0	$\pm$	3.0	&	140.1	$\pm$	0.3	&	0.71	&	$K_{S}$	&	0.49	&	0.37	&	1	\\
99150	&	--	&	1976-05-30	&	41.60	$\pm$	0.60	&	619.0	$\pm$	47.0	&	111.0	$\pm$	0.5	&	1.70	&	$K_{S}$	&	0.37	&	0.33	&	1	\\
	&		&	1996-09-17	&	41.60	$\pm$	0.60	&	621.0	$\pm$	47.0	&	109.4	$\pm$	0.5	&	1.70	&	$K_{S}$	&	0.37	&	0.33	&	2	\\
102141	&	C	&	2003-07-21	&	2.82	$\pm$	0.05	&	30.0	$\pm$	1.0	&	171.2	$\pm$	0.2	&	0.03	&	$K_{S}$	&	0.54	&	0.53	&	1	\\
	&		&	2009-06-01	&	2.51	$\pm$	0.01	&	26.8	$\pm$	0.0	&	159.4	$\pm$	0.6	&	0.03	&	$K_{S}$	&	0.66	&	0.65	&	2	\\
106255	&	--	&	2003-12-10	&	0.19	$\pm$	0.05	&	1.5	$\pm$	0.4	&	162.0	$\pm$	1.0	&	1.12	&	$H$	&	0.27	&	0.15	&	1	\\
	&		&	2005-05-02	&	0.16	$\pm$	0.01	&	1.3	$\pm$	0.0	&	128.2	$\pm$	0.5	&	1.12	&	$H$	&	0.31	&	0.17	&	2	\\
110893	&	--	&	1997-12-28	&	3.21	$\pm$	0.09	&	12.9	$\pm$	0.4	&	103.3	$\pm$	0.2	&	0.98	&	$K_{S}$	&	0.27	&	0.17	&	1	\\
	&		&	2001-08-04	&	2.96	$\pm$	0.01	&	11.8	$\pm$	0.1	&	86.9	$\pm$	0.3	&	0.98	&	$K_{S}$	&	0.32	&	0.2	&	2	\\
111766	&	--	&	2006-05-26	&	0.78	$\pm$	0.05	&	10.0	$\pm$	1.0	&	175.5	$\pm$	0.2	&	0.24	&	$K_{S}$	&	0.38	&	0.34	&	1	\\
	&		&	2013-05-08	&	0.86	$\pm$	0.01	&	11.4	$\pm$	0.1	&	123.0	$\pm$	0.3	&	0.24	&	$K_{S}$	&	0.38	&	0.34	&	2	\\
111802	&	--	&	1984-10-15	&	25.90	$\pm$	0.60	&	225.0	$\pm$	6.0	&	351.6	$\pm$	0.1	&	0.30	&	$K_{S}$	&	0.60	&	0.32	&	1	\\
	&		&	1999-08-11	&	28.07	$\pm$	0.60	&	244.1	$\pm$	6.0	&	352.2	$\pm$	0.1	&	0.30	&	$K_{S}$	&	0.60	&	0.32	&	2	\\
114046	&	$\star ^{e}$	&	2003-12-09	&	0.07	$\pm$	0.05	&	0.2	$\pm$	0.2	&	267.0	$\pm$	3.0	&	1.21	&	$H$	&	0.44	&	0.25	&	1	\\
116132	&	B	&	1997-12-25	&	5.18	$\pm$	0.09	&	32.0	$\pm$	0.7	&	93.8	$\pm$	0.3	&	1.37	&	$K_{S}$	&	0.35	&	0.17	&	1	\\
	&		&	2000-08-21	&	5.28	$\pm$	0.01	&	32.7	$\pm$	0.0	&	94.0	$\pm$	0.1	&	1.37	&	$K_{S}$	&	0.39	&	0.2	&	2	\\

\hline
\end{longtable}}
\end{center}
\twocolumn


\subsection{Detected co-moving companions and survey completeness}
\label{sec:detections}
Based on the multi-epoch AO and wide-field imaging data analysis, a total of 47 AO-detected and 20 wide-field-detected physically associated companions were identified in the \textsc{MinMs} sample (41 systems with one or more companions detected in the close AO data, and 17 systems with one or more companions detected in the wide plate data, for a total of 58 multiple systems and 187 single stars). Due to some overlap in search space between the two types of data, two different companions were detected independently by the two techniques, resulting in 65 unique co-moving companions to the {\sc MinMs} sample. Among the AO-detected companions, four are newly identified within this study. The measured astrometry and relative photometry for each of the resolved companions are reported in Table~\ref{table:companions}, along with the inferred masses of each component derived from an evolutionary model \citep{baraffe98}, as described in Section~\ref{section:mass_estimates}. The 65 companions are distributed in 53 binary systems, three triple systems, and two quadruple systems.

\begin{figure}
\includegraphics[width=0.50\textwidth]{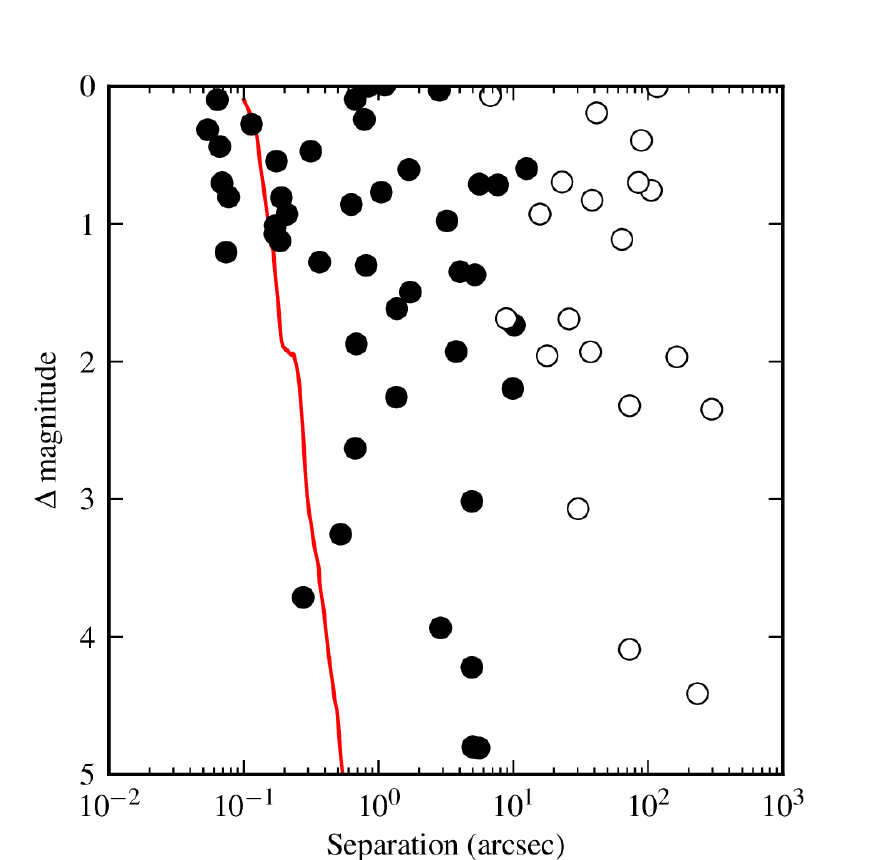}
\caption{The separations and magnitude differences for the confirmed stellar companions in our survey. Companions shown were detected in either the AO data (in one or more of the $JHK{\rm s}$ filters, filled circles), the plate data (only $K_{\rm S}$ magnitude differences from 2MASS cross-matching shown, open circles), or in a few cases, with both imaging techniques. The red line represents the median 3$\sigma$ sensitivity of the sample, and continues beyond the canonical substellar limit. The survey is sensitive to companions above, and to the right of, this sensitivity limit.}
\label{fig:deltamag_sep}
\end{figure}

The observed angular separations and magnitude differences for all of the companions are plotted in Figure~\ref{fig:deltamag_sep}, and the 3$\sigma$ median sensitivity of the sample is also shown. The angular separations range from 0.05~arcsec to 4.94~arcmin. The magnitude differences measured in one of the infrared filters ranged from $\Delta m=0.0-5.0$. Only stellar companions are considered for this study, so the maximum magnitude difference needed for sensitivity to an M9 companion ($M_{K} \sim 10$~mag) to an M0 primary ($M_{K} \sim 4.5$~mag) is only 5.5~mag at $K_{\rm S}$-band. The limited dynamic range required to reach the bottom of the main sequence, combined with the sensitive observations, results in a very uniform completeness to stellar companions. Detection limits for all targets are reported in Table~\ref{table:limits} (Appendix).

\begin{figure}
\includegraphics[width=0.50\textwidth]{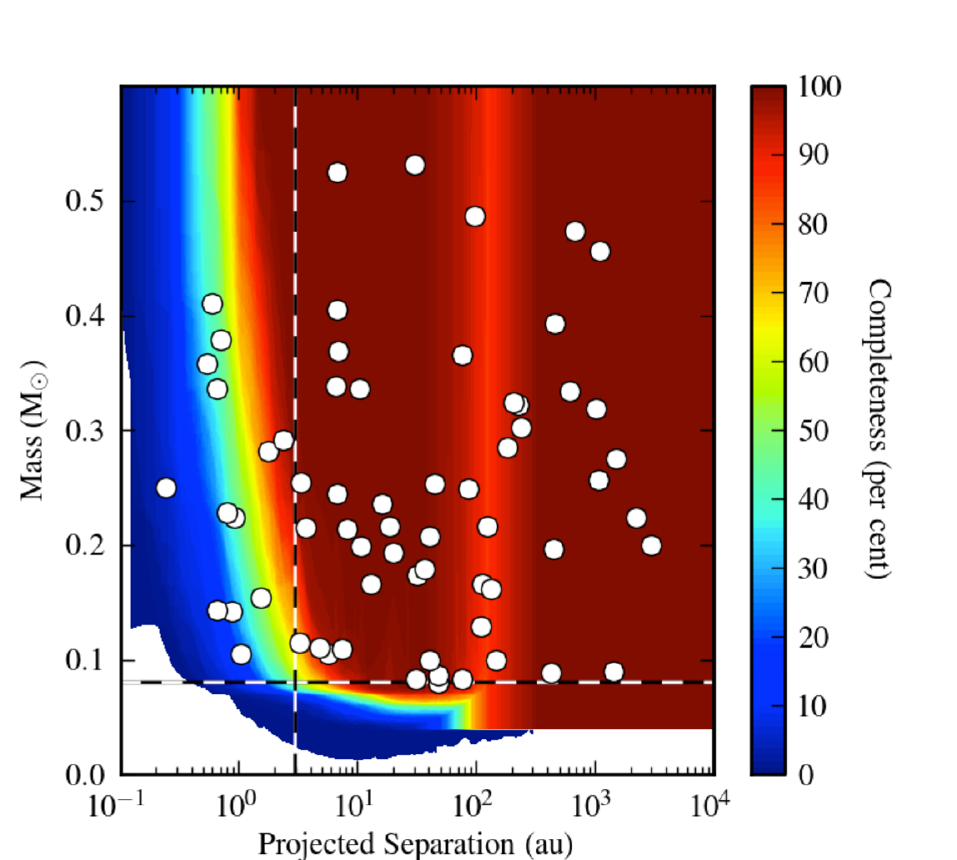}
\caption{The completeness of the {\sc MinMs} survey in terms of derived physical parameters. The horizontal black and white dashed line designates the substellar limit of 0.08~M$_{\odot}$, while the vertical dashed line at 3~au indicates the minimum interior separation to which the AO subsample is 85~per cent complete. These lines define the minimum mass and separation thresholds used to assess population statistics for a subsample minimizing detection biases. Stellar companions are over-plotted, and span the parameter range of our survey.}
\label{fig:masscontour}
\end{figure}

\begin{figure}
\includegraphics[width=0.50\textwidth]{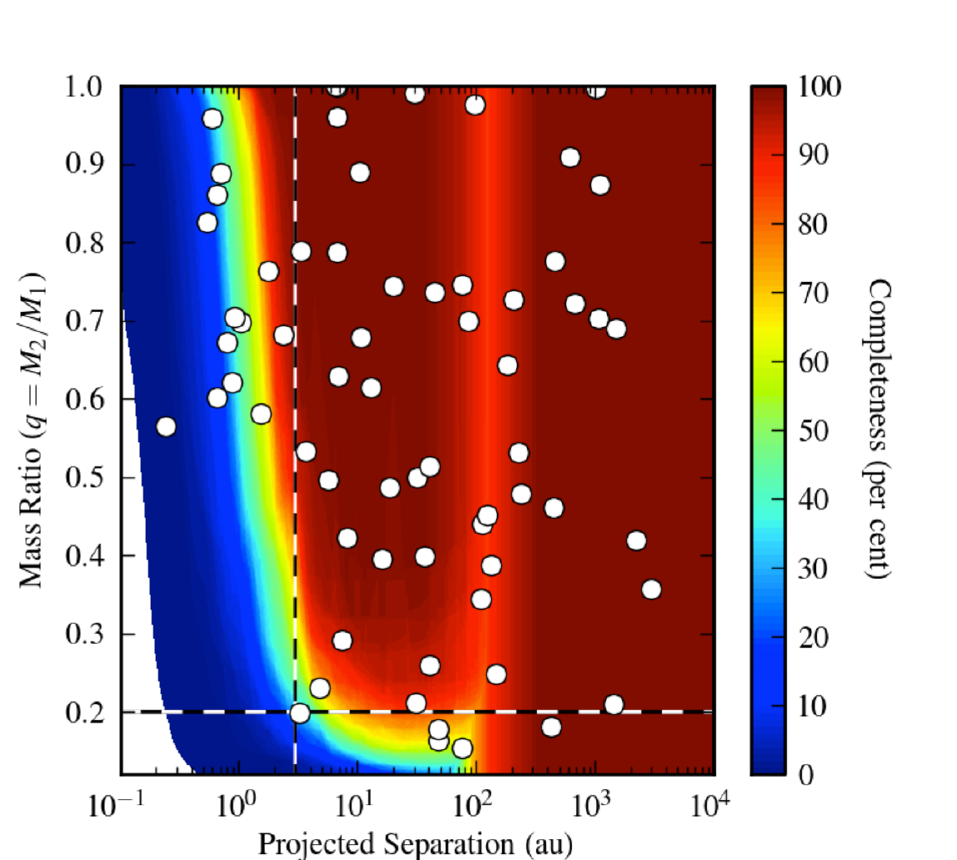}
\caption{The completeness of the {\sc MinMs} survey, shown in terms of the mass ratio of the companion to the primary star. The horizontal dashed line at $q = 0.2$ designates the mass ratio limit for uniform completeness, while the vertical dashed line indicates the minimum separation limit of 3~au used to assess population statistics.}
\label{fig:massratiocontour}
\end{figure}

The observational properties are transformed into physical properties of mass and projected separation in Figure~\ref{fig:masscontour}. Wider than a few au, the completeness is very high (typically $>95$~per~cent), as shown in Figure~\ref{fig:masscontour}. The detected companions are also included in the figure, and co-moving objects down to the stellar/substellar limit are identified in the {\sc MinMs} study. The vertical and horizontal dashed lines define the boundaries of the region considered for constructing the distributions of companion separation, secondary mass, and total system mass. The closer pairs ($<3$~au) are not included in the population statistics, since the interpretation of the data would be dependent on a large correction factor. A similar plot in terms of mass ratio and projected separation is given in Figure~\ref{fig:massratiocontour}, and the dashed lines delineate the systems included in determining the mass ratio distribution of the {\sc MinMs} sample. Although the sensitivity is very uniform to a companion mass limit of 0.08~M$_{\sun}$, the range of target masses makes the lower mass ratio limit of $q \ge 0.2$ for uniform completeness in the mass ratio distribution.

\subsection{Comparison of sample properties and binary detections}
\label{sec:malm}
To determine if any selection biases impacted the study, particularly given the magnitude-limited nature of our survey, we analysed binary occurrence against attributes such as distance, magnitude, and metallicity. To test whether observed binaries were consistent with being drawn randomly from the survey sample, all systems were rank-ordered by a given attribute and the binaries were plotted to create a cumulative distribution function, which could then be compared against a random distribution. This was performed using a method similar to calculating the Gini coefficient, a widely-used statistic in economics measuring wealth distribution within societies \citep{gini}. For each system attribute, the area underneath the binary distribution, normalized to 0.5, was calculated. The error on the area was estimated with a Monte Carlo simulation of 10,000 random distributions, each with the same number of binaries in the same number of total targets. In this study, the 58 multiple systems within 245 total systems leads us to an expected area of 0.5 $\pm$ 0.03.

\begin{figure}
\includegraphics[width=0.50\textwidth]{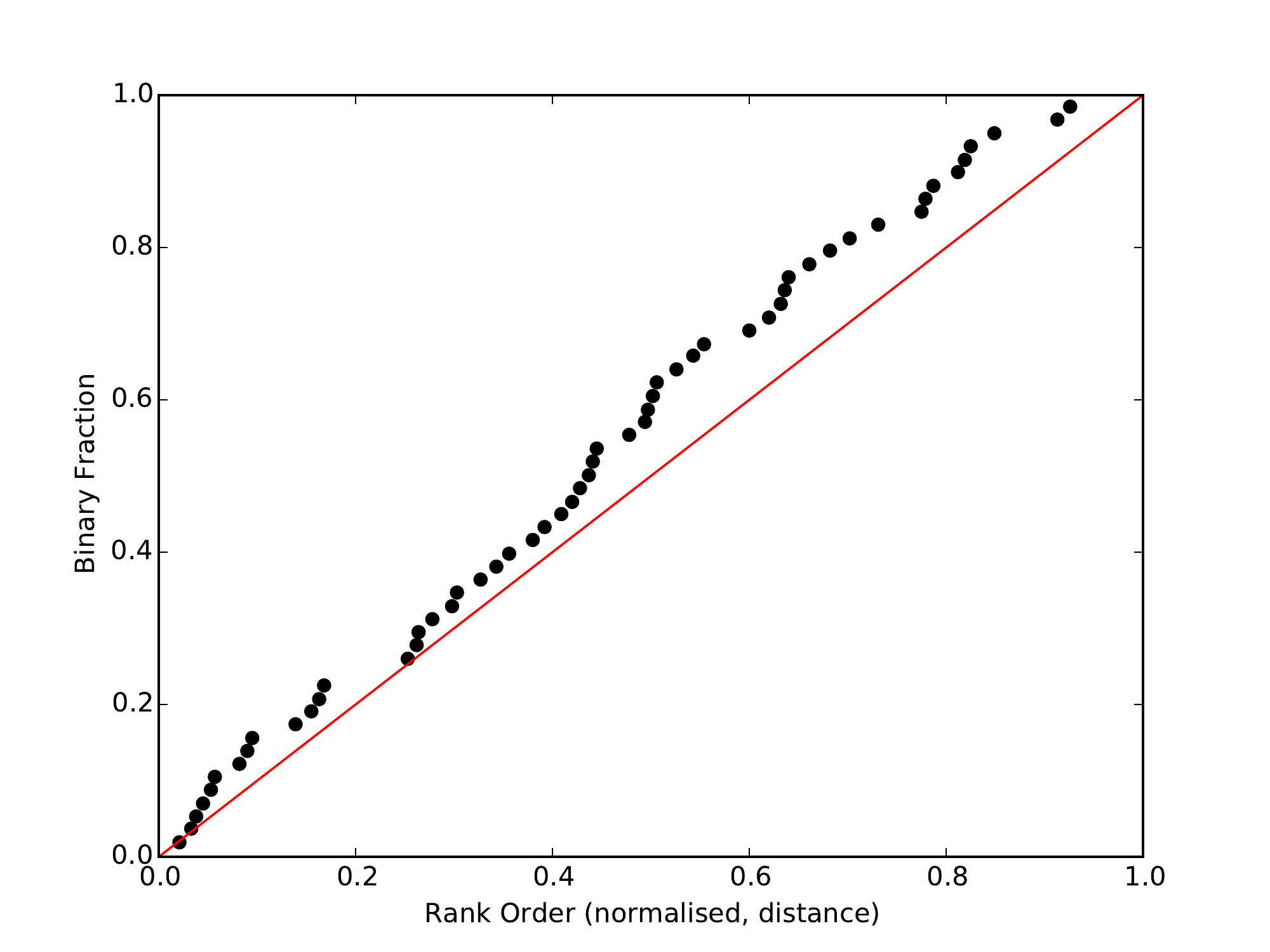}
\caption{Binary fraction of the stars within our sample at a given distance versus normalized rank-ordered distance of the full sample. The solid red line denotes the expected trend for a random distribution, with integrated area normalised to 0.5 $\pm$ 0.03 (errors drawn from Monte Carlo simulations of 10,000 random distributions). The black points show the binary fraction as a function of rank-ordered distance; with an integrated area of 0.54, the binary occurrence is consistent with being drawn from a random distribution, implying an unbiased observation of binaries in the \textsc{MinMs} sample with respect to system distance.}
\label{fig:dist_gini}
\end{figure}

\begin{figure}
\includegraphics[width=0.50\textwidth]{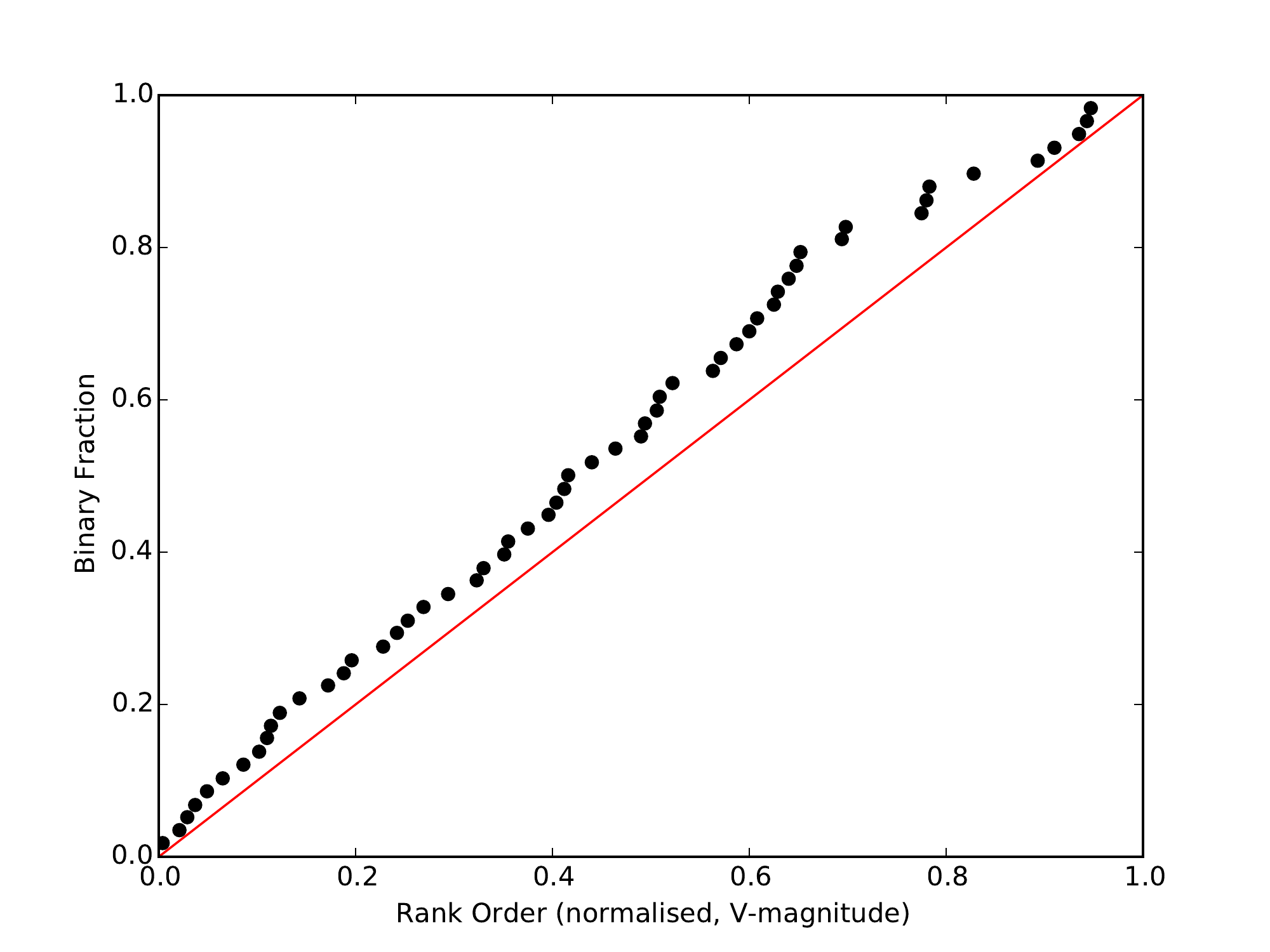}
\caption{Binary fraction of the stars within our sample with a given $V$-band magnitude versus normalized rank-ordered magnitude of the full sample. Colors and symbols are as in the preceding figure for distance distribution. The integrated binary distribution with respect to magnitude is also 0.54, and consistent with being drawn randomly, implying an unbiased observation of binaries in the \textsc{MinMs} sample with respect to system brightness.}
\label{fig:mag_gini}
\end{figure}

The comparisons for the systems in our survey are shown in Figures~\ref{fig:dist_gini} for the distances of the systems, and Figure~\ref{fig:mag_gini}, for the magnitude distribution of the target sample. For the distance distribution, we find the area beneath the binary distribution to be 0.54, slightly over 1$\sigma$ away from the expected area value. This may be seen in a slight overabundance of binaries near rank order 0.5, corresponding to systems at $\sim$8 pc, but is still consistent with a random distribution, and so the sample does not appear to be significantly biased toward systems at closer or further distances. Similarly, for the $V$-band magnitude distribution in Figure~\ref{fig:mag_gini}, the area was also 0.54, and is again consistent with a random underlying distribution. Similar comparisons were performed for the binaries within only the 196-star AO subsample, with area values of 0.49 for the distance distribution and 0.52 for the magnitude distribution, which are also consistent with selection from a random distribution.

Using literature values for the 124 stars within our sample with previous metallicity measurements (Table~\ref{table:sample}), subsamples of the binaries and single stars were also compared with the same technique. With 20 observed multiple systems within the 124 star subsample with metallicity measurements, the expected integrated area of the binary distribution function is $0.5 \pm 0.06$. With an observed value of 0.45, the survey also appears consistent with a random distribution with respect to stellar metallicity, and suggests no dependence of binarity upon the system metallicity. With very few measurements of extremely metal-poor or metal-rich stars within our sample, no binaries were detected at extreme metallicities, underscoring the utility of additional sample metallicity and age estimations in determining robust statistics for various population subsamples.

\subsection{Projected separation distribution}
\begin{figure}
\includegraphics[width=0.50\textwidth]{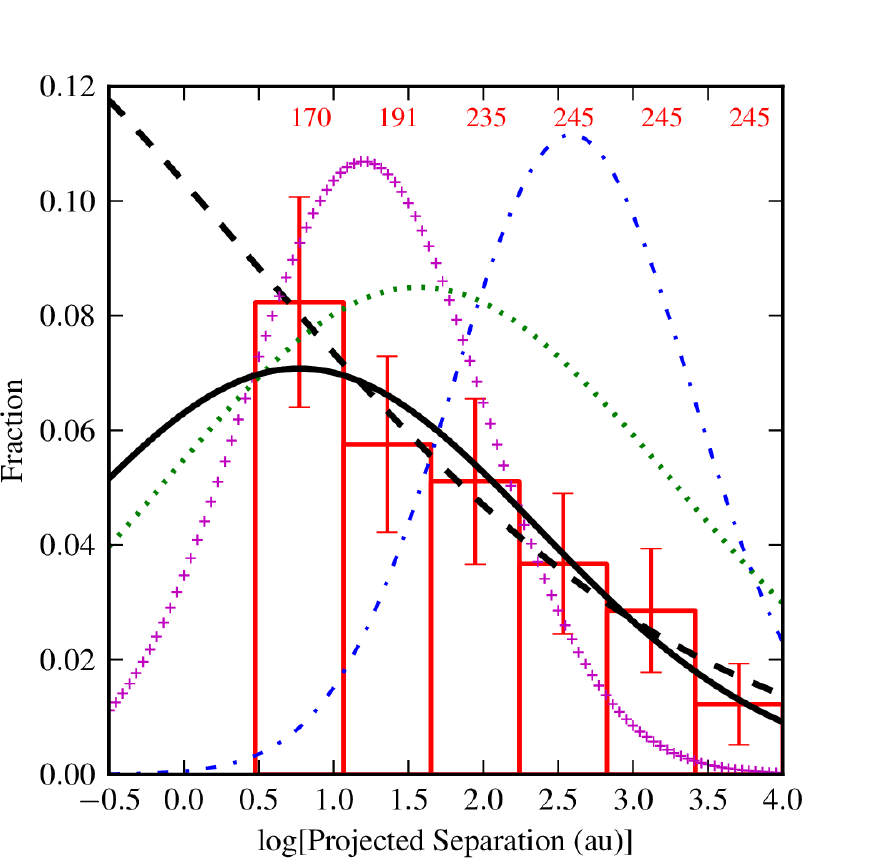}
\caption{The distribution of projected separations for the companions resolved between 3 and 10,000~au, corrected for incompleteness (red histogram). Assuming a log-normal distribution, a Gaussian was fit to the observed distribution. Allowing all parameters to vary freely, the distribution is best fit by a Gaussian with a mean of $\mu_{\log(a)} = -0.66$ and width of $\sigma_{\log(a)} = 1.86$ (dashed black curve). A restricted fit was also computed by fixing the mean of the distribution to the centre of the first bin of the distribution ($\mu_{\log(a)}=0.77$), with a best-fitting Gaussian of width $\sigma_{\log(a)} = 1.34$ (solid black curve). These fits represent the two limiting cases of the true distribution, given the observed distribution within the restricted separation range. For comparison, the distribution of companion separations for A-star primaries (blue dot-dash curve, \citealp{2014MNRAS.437.1216D}), solar-type primaries (green dotted curve, \citealp{raghavan}), and for M-dwarf primaries in a previous lucky-imaging survey (magenta crosses, \citealp{janson}).}
\label{fig:separationdist}
\end{figure}

The separation range of 3~au defined from the survey completeness to the outer limit of 10,000~au, spans 3.5 orders of magnitude. The distribution of projected separations was constructed over six equally sized bins from $\log\left(a_{\rm proj}\right)$ of 0.5 to 4.0, and is shown in Figure~\ref{fig:separationdist}. For each bin, the number of targets in the sample that was sensitive to 95~per~cent of the bin range is indicated above each bin, and the companion frequency per bin was determined by dividing the number of resolved companions within the bin by the subset of targets complete to 95 per cent of the bin. The error bars were calculated from Poisson statistics. 

Over the range of the survey data, the distribution rises monotonically to the smaller separations covered in the 3~au to 10~au bin. The best-fit log-normal distribution is plotted, and due to the unconstrained location of the peak of the distribution, a best-fitting $\mu_{\log(a)}=-0.66$ and $\sigma_{\log(a)} = 1.86$ is found. Fixing the peak to be at the centre of the first bin of the observed distribution ($\mu_{\log (a)}=0.77$), a significantly narrower distribution is fit ($\sigma_{\log(a)}=1.34$).  Compared to a previous survey of M-stars at a larger range of distances with less sensitivity to wider companions \citep{janson}, the peak of the restricted fit to the {\sc MinMs} distribution is lower, but significantly wider (Figure \ref{fig:separationdist}). Two additional comparison curves are shown for the distributions of companions to higher mass primaries. The solar-type distribution peak occurs at a somewhat wider separation \citep{raghavan} and the fit is consistently higher than the corresponding value of the M-star distribution with the exception of the bin for the smallest separations. The fit to the projected separation distribution of A-star systems \citep{2014MNRAS.437.1216D} is substantially different, with a markedly larger value for the peak and an overall higher normalisation. Given the larger distances of the A-star sample ($D\leq 75$~pc), the inner limit for data used to construct the A-star distribution is located at the third bin of the {\sc MinMs} data (log$~a_{\rm proj} = 1.6$). Considering the fits to the M-dwarf, solar-type and A-star surveys, the {\sc MinMs} results provide further evidence to support the trend of a wider typical system separation as a function of host star mass that has been noted in previous studies. The {\sc MinMs} study refines the population statistics for low-mass stars with the large sample and comprehensive mass and separation sensitivity.

\subsection{Mass ratio distribution}
\begin{figure}
\includegraphics[width=0.50\textwidth]{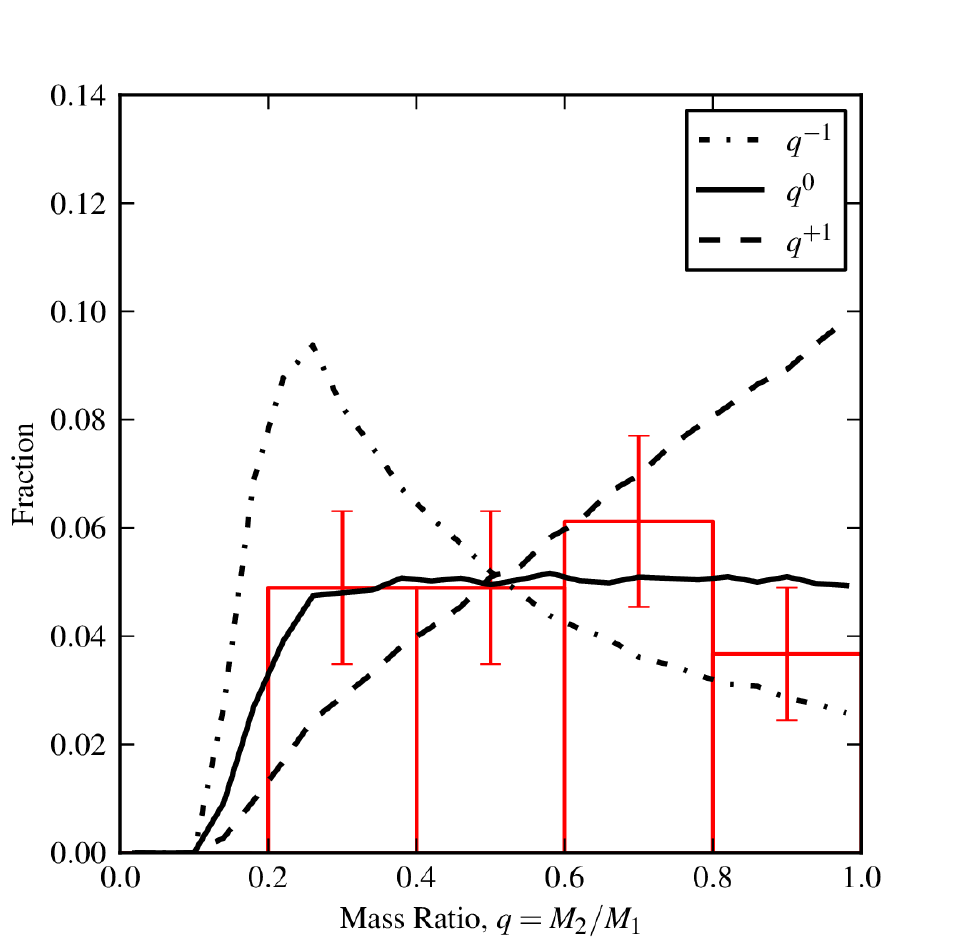}
\caption{The distribution of companion mass ratios for the companions resolved in the {\sc MinMs} study. Only those companions with projected separations between $30-10,000$~au and mass ratios of $q\ge0.2$ are included to minimize observational biases. The predicted shape of the mass ratio distribution for companions drawn from a distribution rising to more equal-mass companions ($f(q)\propto q^1$, dashed curve), a flat distribution ($f(q)\propto q^0$, solid curve), and a falling distribution ($f(q)\propto q^{-1}$, dot-dashed curve) are plotted for reference. The observed mass ratio distribution is most consistent with the flat distribution.}
\label{fig:massratiodist}
\end{figure}

Based on the survey completeness shown in Figure~\ref{fig:massratiocontour}, the systems included in the mass ratio distribution have a mass ratio of $q\ge 0.2$. The shape of the observed mass ratio distribution is consistent with a flat distribution, as shown in Figure~\ref{fig:massratiodist}. Because the target masses are close to the stellar limit, all targets are not sensitive to the full range of mass ratios ($q \ge 0.2$). The median mass of a star in the sample is $\sim$0.44~M$_{\sun}$ and the mass ratio limit for the median star corresponds to a secondary mass limit of 0.09~M$_{\sun}$. For the lowest-mass stars in the sample, the bottom of the main sequence occurs at a mass ratio greater than $q > 0.2$. As the lowest-mass primaries in our sample have masses in the $0.12-0.18~M_{\sun}$ range, corresponding to a companion at the stellar/substellar boundary with $q\sim0.6$, the two lowest mass ratio bins are affected by this potential bias.

The mass ratio values were sorted by the separation of the pair and split into two sets by a range of dividing separations, and the inner and outer distributions were compared to each other with a Kolmogorov-Smirnov (KS) test. The inner and outer distributions were not found to be significantly different, regardless of the dividing separation used. This result is different from what was found for both G-star \citep{raghavan} and A-star binaries \citep{2014MNRAS.437.1216D}. Since there was no difference in the mass ratio distributions, the full set of companions was used to investigate the shape of the true underlying distribution. In order to estimate the expected shape of the mass ratio distribution, taking into account the distribution of primary masses, we performed a series of Monte Carlo simulations to construct comparison mass ratio distributions from pairing the {\sc MinMs} primaries with secondaries of different companion mass ratio functions (Figure \ref{fig:massratiodist}). 

We have assumed that the full mass range of the \textsc{MinMs} primaries can be described as a simple population with a single underlying mass ratio distribution. The pairing simulations were based on a population of $1 \times 10^{5}$ primary stars with a distribution of masses matching the primaries in the \textsc{MinMs} survey. For each primary, a companion mass was drawn randomly from rising, falling, or uniform mass ratio distributions over the full stellar companion mass range of $0.08M_{\odot}$ to the given primary mass, producing $1 \times 10^{5}$ companions. We then divided the masses of the secondaries by the masses of the primaries to obtain the corresponding mass ratios. From this analysis, only those stars with $M_{\odot} < 0.4$ were biased against detecting mass ratios of $q<0.2$ (making up 35 per cent of the sample) and given that the sample stars fall off quickly toward lower-mass primaries, only 10 per cent of primaries $< 0.3 M_{\odot}$ were similarly biased, resulting in a bias in half of the $q = 0.2 - 0.4$ bin. In this way, the rising, falling and uniform distributions are compared with similar completeness to the mass ratio distribution of the observed companions. This is similar to previous analyses for higher-mass primaries, and statistical analyses of companion mass ratio distributions \citep{reg2011, reg2013}. In light of our implicit assumption that the $q$-distribution is the same for the full range of \textsc{MinMs} primaries, the observed mass ratio distribution is thus consistent with companions drawn from a distribution flat in mass ratio ($f(q)\propto q^0$).

\subsection{Companion and total system mass distributions}
\begin{figure}
\includegraphics[width=0.50\textwidth]{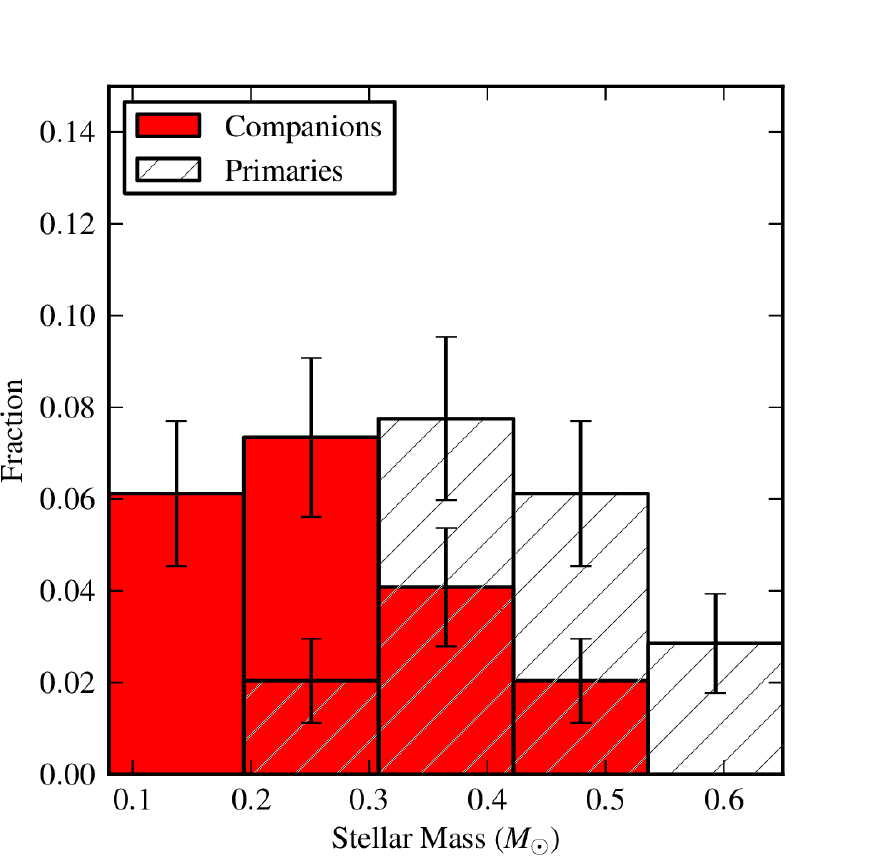}
\caption{The distribution of primary masses and companion masses for all stellar companions and hosts found within the {\sc MinMs} study, shown in fractions of the total sample. A KS test performed on the single star masses against the masses of primaries in binaries showed no preference toward stellar mass correlating with presence of companions over the K7-M6 spectral type range of our sample.}
\label{fig:primcompmass}
\end{figure}

\begin{figure}
\includegraphics[width=0.50\textwidth]{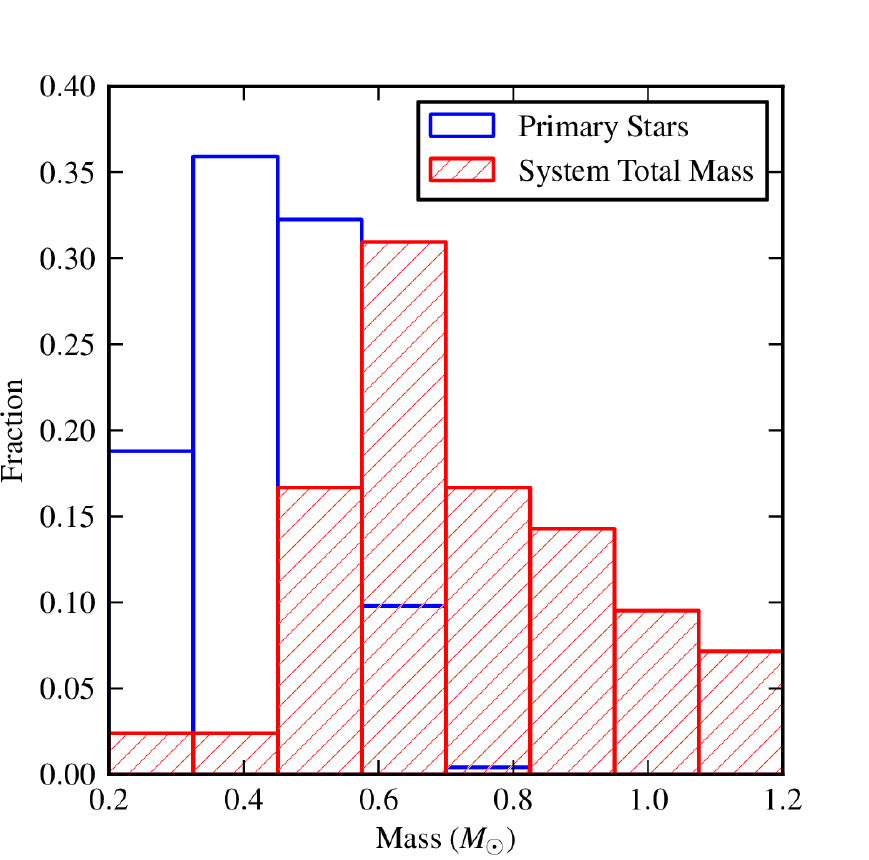}
\caption{The distribution of total system mass, calculated as the sum of all components within a given multiple system, including all companions with separations of $a_{\rm proj} \ge 3$~au and $q\ge 0.2$, normalised by the number of such companions (red hatched histogram). Plotted for reference is the distribution of primary masses given in Figure \ref{fig:prim_mass_dist}, normalised by the overall sample size (blue histogram).}
\label{fig:total_mass}
\end{figure}

Given the high level of completeness to the bottom of the main sequence, the distributions of primary and companion mass were constructed for all pairs with $>$3~au projected separation. Unlike the flat mass ratio distribution, the companion mass distribution shown in Figure~\ref{fig:primcompmass} rises continuously with decreasing mass. Similarly, the total system mass distribution in Figure~\ref{fig:total_mass} rises to smaller masses. For comparison, a KS test was performed to compare the mass distribution of single stars and that of primaries in binary systems; with a \emph{p}-value of 0.34, no intrinsic difference was found between the two populations, indicating no preference for massive primaries having more companions (or vice versa). For triple and quadruple systems, all companions and the primary are summed to determine the system mass included in Figure~\ref{fig:primcompmass}. For comparison with the total system distribution, the target star mass distribution is also plotted in Figure~\ref{fig:total_mass}.

\subsection{Frequency of companions}
There are two quantities that define the multiplicity of the sample. The multiplicity fraction (MF) quantifies the number of multiple systems within the sample:
\begin{equation}
MF = \frac{b + t + q + ...}{s + b + t + q + ...},
\end{equation}
where $s$ is the number of singles, $b$ the number of binaries, $t$ the number of triples, $q$ the number of quadruples, and so forth (e.g. \citealp{Reipurth1993,Goodwin2004}). The companion star fraction (CSF) quantifies the total number of companions within the sample as the following:
\begin{equation}
CSF = \frac{b + 2t + 3q + ...}{s + b + t + q + ...},
\end{equation}
\citep{2002ApJ...581..654P}. As described in Section~\ref{sec:detections}, AO data exist for a subsample of 196 stars within the 245-star MinMs sample. Therefore, the multiplicity fraction can be calculated for separate subsamples, accounting for selection effects over different ranges of projected separation. Of the AO data, which forms a subsample searching projected separations in the 1-100 AU range, we find 41 binaries of the 196 stars with AO imaging, corresponding to a MF of $21 \pm 3$~per~cent. For the plate data, we find 17 wide binaries covering the 100-10000 au range within the full 245 star sample with plate imaging, corresponding to a MF for the wide subsample of $7 \pm 2$~per~cent. This is consistent with the tighter semi-major axis distribution and lower multiplicity seen in comparison to higher-mass primaries.

\begin{figure}
\includegraphics[width=0.50\textwidth]{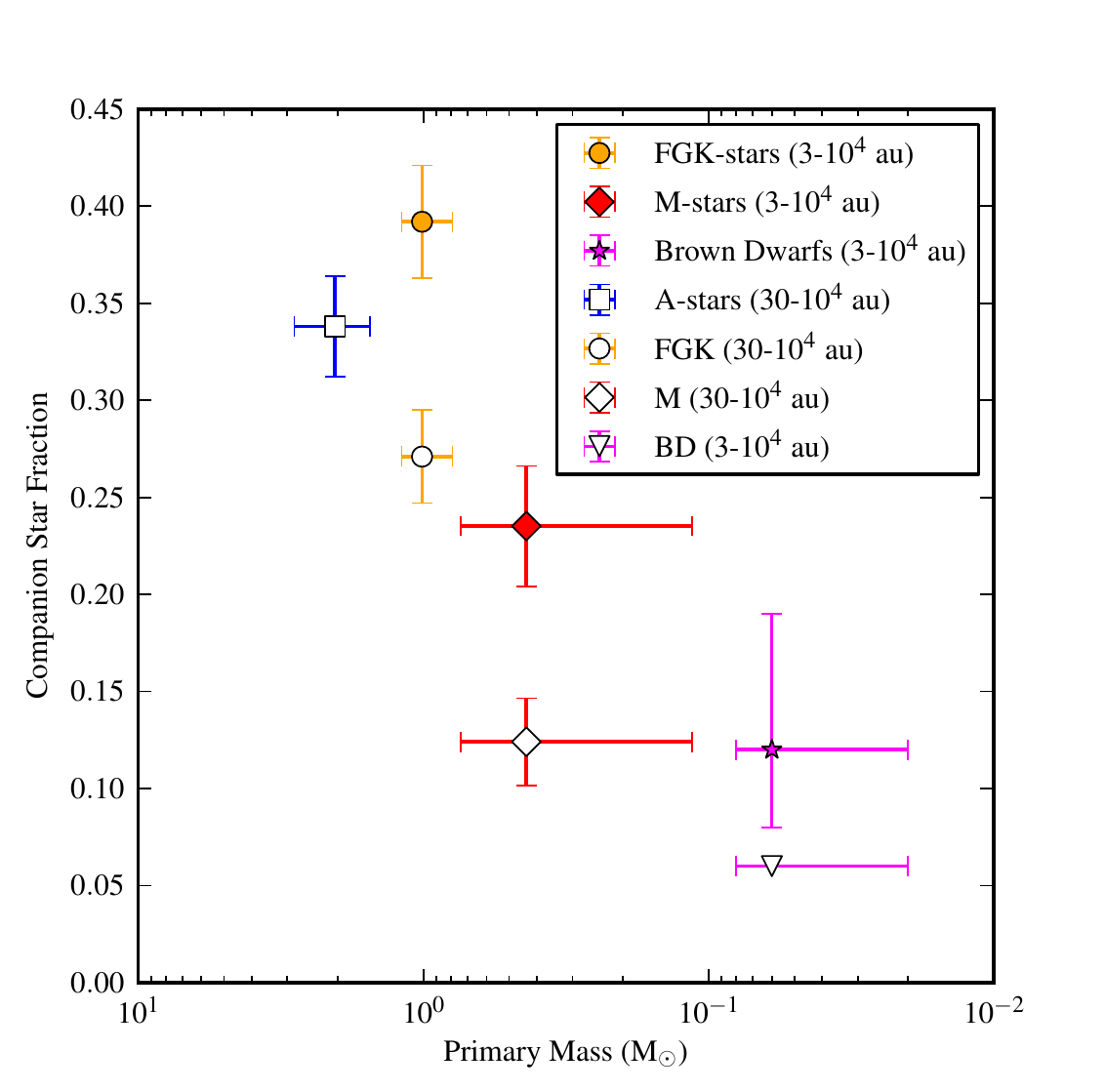}
\caption{The companion star fractions measured over the full $3-10,000$~au (filled symbol) and restricted $30-10,000$~au (open symbol) separation ranges. Comparison values for both ranges, from left to right, are shown for solar-type stars (yellow circles, \citealp{raghavan}), and M-dwarfs (red diamonds, this study). Over the $3-10,000$~au range, a measured value exists for brown dwarf primaries (pink star, \citealp{2006ApJS..166..585B}). Over the restricted $30-10,000$~au range, additional comparison values are available for field A-type stars (blue square, \citealp{2014MNRAS.437.1216D}), as well as an upper limit to the CSF$_{30-10,000{\rm au}}$ for brown dwarf primaries (pink downward triangle, \citealp{2007AJ....133..971A}), demonstrating the decreasing trend of multiplicity with later spectral type.}
\label{fig:CSF_3-10000au}
\end{figure}

The CSF over $3 - 10,000$~au is calculated by summing the fraction in each separation bin of Figure \ref{fig:separationdist}, since each bin is already corrected for incompleteness. The resulting CSF$_{\rm 3-10,000au}$ value is $23.5\pm3.2$~per~cent, as shown in Figure~\ref{fig:CSF_3-10000au}. For closer separations, the survey is not fully sensitive to the bottom of the main sequence, but a lower limit on the total CSF can be estimated by combining the bound companions resolved within this study with those reported within the WDS catalogue and the Ninth catalogue of Spectroscopic Binary Orbits (SB9; \citealp{Pourbaix:2004}). The resulting CSF $_{\rm total}$ is $34.7^{+2.9}_{-3.2}$~per~cent. (Similarly, the lower limit on the total multiplicity of the sample was estimated as MF$_{\rm total}=28.6^{+2.7}_{-3.1}$~per~cent.)

To place the {\sc MinMs} results in a broader context, the CSF is compared with samples with different primary star masses. Two separation ranges are considered: the full $3-10,000$~au range and the more restrictive $30-10,000$~au range. The $3-10,000$~au range requires samples of nearby field objects to reach the smaller projected separations, so the {\sc MinMs} value is compared with field Solar-type stars \citep{raghavan} and field brown dwarfs \citep{2006ApJS..166..585B}. As shown in Figure~\ref{fig:CSF_3-10000au}, the CSF calculated over both separation ranges shows a decline with primary mass. The solar-type star study was sensitive to the bottom of the main sequence as is the case for the {\sc MinMs} survey, while the surveys of brown dwarfs were typically sensitive to $q\ga 0.5$. Over the restricted $30-10,000$~au range, for which a CSF$_{\rm 30-10,000au}=12.4\pm2.3$~per~cent was measured for the M-dwarf primaries within this study, it is possible to include the results from the more massive A-stars \citep{2014MNRAS.437.1216D}. Due to the paucity of wide companions to brown dwarfs, there is only an upper limit on the brown dwarf companion star fraction over this separation range \citep{2007AJ....133..971A}. 

\subsection{Higher order multiple systems}

Since the {\sc MinMs} companions include a large range of separations from a minimum of 0.2~au to a maximum of 2,952~au, and $\sim$200 stars have both AO and wide field imaging, it is possible to detect higher order multiple systems within the survey. A total of three triple and two quadruple systems were identified. All the triple systems are in hierarchical arrangements and the ratio of inner to outer projected separations ranges from 347:1 to 27:1. The threshold for stability of multiple systems is considered to be a separation ratio for the outer to inner pair of 5:1 \citep{2006epbm.book.....E}, suggesting that all the triple systems are likely stable. For the two quadruples, one system consists of two widely separated close pairs, while the other system is composed of a close pair and two wider and lower mass stars. For a full accounting of the frequency of higher order multiples, radial velocity measurements of the sample with AO and wide field imaging are required.

\begin{table}
{
\centering
\caption{\label{table:sbs}Summary of known spectroscopic binaries in the {\sc MinMs} sample}
\begin{tabular}{ccccc}
\hline
HIP & Reference &  SB Type &  Period   &    AO Resolved? \\ 
        &                     &            &  (days)   &                                     \\ 
\hline
9724  & N02 & 1 & 6818 & Yes \\
11964  & E59 & 1 & 1.962 & No \\
25953 & MB89 & 2 & - & No \\
34603 & TP86 & 1 & 10.428 & No \\
38082 & MB89 & 2 & - & No \\
65011 & T97 & 1 & 200.26 & No \\
76901 & N02 & 1 & 62.628 & No \\
80346 & N02 & 1 & 1366.1 & No \\
82809 & M01 & 1 & 2.9655 & No \\
82817  & M01 & 1 & 2.9655 & No \\
111802  & HM65 & 1 & 4.0832 & No \\

\hline
\end{tabular}
}
\end{table}

To determine whether any spectroscopic systems were previously known within the sample, we cross-referenced the SB9 catalogue and the 70 star M-dwarf spectroscopic survey by \citep[hereafter MB89]{marcybenitz} against our sample. Among the 50 \textsc{MinMs} stars within the MB89 study, only two spectroscopic binaries are known. From the cross-reference with SB9, an additional 10 spectroscopic binaries were identified within our sample from the following surveys, listed in Table~\ref{table:sbs}: \citealp{2002ApJS..141..503N} (N02), \citealp{1959MNRAS.119..526E} (E59), \citealp{1986AJ.....92.1424T} (TP86), \citealp{1997A&AS..121...71T} (T97), \citealp{2001MNRAS.325..343M} (M01), and \citealp{1965ApJ...141..649H} (HM65). Of these spectroscopic systems, one was resolved in archival AO imaging (HIP~9724). As the completeness of the SB9 catalogue is unknown since non-detections are not reported, we do not include these binaries in our analysis. The incompleteness of the spectroscopic results underscores the need for a comprehensive spectroscopic companion survey of nearby M-dwarfs, such as the CARMENES survey \citep{carmenes}. By building upon the current results, the {\sc MinMs} sample represents an ideal target set for developing a comprehensive understanding of the frequency and properties of higher-order multiple systems with low-mass primaries.

\section{Summary and Conclusions}
\label{sec:conclusions}

With a combination of multi-epoch adaptive optics and wide-field imaging, we have conducted a companion search program targeting 245~late K and early M-dwarfs which are within 15~pc with \emph{Hipparcos} parallax uncertainties of $<10$~per~cent, $M_{V} > 8$, and which are not companions to earlier spectral type primaries. Companions with projected separations as small as 0.2~au were resolved, and beginning at a separation of 3~au and continuing to 10,000~au, the observations were sensitive to the companions down to the stellar/substellar limit. Within this complete range, a total of 65 co-moving companions were detected, 47 with AO observations and 18 from wide-field imaging data, of which four are newly resolved within this study.

Over the complete $3-10,000$~au separation range, the companion star fraction is $23.5\pm3.2$~per~cent. With the large sample size, the uncertainties are reduced by a factor of $\sim$2 relative to previous studies, which makes it possible to determine that the M-dwarf CSF is distinctly lower than the solar-type value \citep{raghavan} and higher than the corresponding brown dwarf fraction \citep{2006ApJS..166..585B}. By considering the separation range of $30-10,000$~au to enable a comparison with the more distant and more massive A-stars \citep{2014MNRAS.437.1216D}, a CSF$_{30-10,000au}=12.4\pm2.3$ was measured, and a systematic decline in CSF as a function of primary mass is observed.

Dividing the observed range of companion separations into six bins, the separation distribution is seen to rise continuously toward the smallest separations to which the survey is sensitive. When compared to the results of a previous lucky imaging survey of M-dwarfs \citep{janson}, the {\sc MinMs} distribution appears to have a wider spread, likely due to the enhanced coverage that enabled a direct measurement of the widest systems. The mass ratio distribution is flat across the $q=0.2-1.0$ range, similar to that observed for companions to solar-type stars \citep{raghavan}, but different from the rise toward lower mass ratios seen for higher-mass stars \citep{2014MNRAS.437.1216D}. 

The {\sc MinMs} study can serve as a benchmark dataset for comparisons with the companion star properties of more distant populations of young low-mass stars. Given the existing large coverage, the {\sc MinMs} sample is also an ideal set to pursue radial velocity searches for the closest companions to complete the separation distribution, and to provide the first complete accounting of high order multiplicity of M-dwarfs. 


\section*{Acknowledgments}
The authors wish to thank the anonymous referee for providing a thorough review and helpful comments which improved this manuscript. KWD is supported by the National Science Foundation Graduate Research Fellowship under Grant No. DGE-1311230. RJDR acknowledges support from Science and Technology Facilities Council grants ST/H002707/1 and ST/K005588/1. RJP acknowledges support from the Royal Astronomical Society in the form of a research fellowship. We are grateful to the MMT Observatory staff for their excellent support of these observations. We also thank B. Svoboda for valuable programming discussions. This research has made use of the SIMBAD database, operated at CDS, Strasbourg, France. This research makes use of data products from the Two Micron All Sky Survey, which is a joint project of the University of Massachusetts and the Infrared Processing and Analysis Center/California Institute of Technology, funded by the National Aeronautics and Space Administration and the National Science Foundation. This research has made use of the Washington Double Star catalogue maintained at the US Naval Observatory. This research used the facilities of the Canadian Astronomy Data Centre operated by the National Research Council of Canada with the support of the Canadian Space Agency. Results are based on data obtained from the ESO Science Archive Facility. This research has made use of data obtained from the SuperCOSMOS Science Archive, prepared and hosted by the Wide Field Astronomy Unit, Institute for Astronomy, University of Edinburgh, which is funded by the UK Science and Technology Facilities Council.

\bibliographystyle{mn2e}
\setlength{\bibhang}{2.0em}
\setlength\labelwidth{0.0em}
\bibliography{mstarsv13_arxiv}

\begin{thebibliography}{78}
\expandafter\ifx\csname natexlab\endcsname\relax\def\natexlab#1{#1}\fi

\bibitem[{{Allen} {et~al}\mbox{.}(2007){Allen}, {Koerner}, {McElwain}, {Cruz},
  \& {Reid}}]{2007AJ....133..971A}
{Allen} P.~R., {Koerner} D.~W., {McElwain} M.~W., {Cruz} K.~L., {Reid} I.~N.,
  2007, \aj, 133, 971

\bibitem[{{Andrews} {et~al}\mbox{.}(2009){Andrews}, {Wilner}, {Hughes}, {Qi},
  \& {Dullemond}}]{Andrews:2009}
{Andrews} S.~M., {Wilner} D.~J., {Hughes} A.~M., {Qi} C., {Dullemond} C.~P.,
  2009, \apj, 700, 1502

\bibitem[{{Artymowicz} \& {Lubow}(1994)}]{Artymowicz1994}
{Artymowicz} P., {Lubow} S.~H., 1994, \apj, 421, 651

\bibitem[{{Baraffe} {et~al}\mbox{.}(1998){Baraffe}, {Chabrier}, {Allard}, \&
  {Hauschildt}}]{baraffe98}
{Baraffe} I., {Chabrier} G., {Allard} F., {Hauschildt} P.~H., 1998, \aap, 337,
  403

\bibitem[{{Bastian}, {Covey} \& {Meyer}(2010){Bastian}, {Covey}, \&
  {Meyer}}]{Bastian2010}
{Bastian} N., {Covey} K.~R., {Meyer} M.~R., 2010, \araa, 48, 339

\bibitem[{{Bergfors} {et~al}\mbox{.}(2010){Bergfors}, {Brandner}, {Janson},
  {Daemgen}, {Geissler}, {Henning}, {Hippler}, {Hormuth}, {Joergens}, \&
  {K{\"o}hler}}]{Bergfors}
{Bergfors} C. {et~al.}, 2010, \aap, 520, A54

\bibitem[{{Bertin} \& {Arnouts}(1996)}]{sourceextract}
{Bertin} E., {Arnouts} S., 1996, 117, 393

\bibitem[{{Beuzit} {et~al}\mbox{.}(2004){Beuzit}, {S{\'e}gransan}, {Forveille},
  {Udry}, {Delfosse}, {Mayor}, {Perrier}, {Hainaut}, {Roddier}, {Roddier}, \&
  {Mart{\'{\i}}n}}]{beuzit2004}
{Beuzit} J.-L. {et~al.}, 2004, \aap, 425, 997

\bibitem[{{Browning} {et~al}\mbox{.}(2010){Browning}, {Basri}, {Marcy}, {West},
  \& {Zhang}}]{browning}
{Browning} M.~K., {Basri} G., {Marcy} G.~W., {West} A.~A., {Zhang} J., 2010,
  \aj, 139, 504

\bibitem[{{Burgasser} {et~al}\mbox{.}(2006){Burgasser}, {Kirkpatrick}, {Cruz},
  {Reid}, {Leggett}, {Liebert}, {Burrows}, \& {Brown}}]{2006ApJS..166..585B}
{Burgasser} A.~J., {Kirkpatrick} J.~D., {Cruz} K.~L., {Reid} I.~N., {Leggett}
  S.~K., {Liebert} J., {Burrows} A., {Brown} M.~E., 2006, \apjs, 166, 585

\bibitem[{{Cutri} {et~al}\mbox{.}(2003){Cutri}, {Skrutskie}, {van Dyk},
  {Beichman}, {Carpenter}, {Chester}, {Cambresy}, {Evans}, {Fowler}, {Gizis},
  {Howard}, {Huchra}, {Jarrett}, {Kopan}, {Kirkpatrick}, {Light}, {Marsh},
  {McCallon}, {Schneider}, {Stiening}, {Sykes}, {Weinberg}, {Wheaton},
  {Wheelock}, \& {Zacarias}}]{cutri}
{Cutri} R.~M. {et~al.}, 2003, {2MASS All Sky Catalog of point sources.}

\bibitem[{{De Rosa} {et~al}\mbox{.}(2011){De Rosa}, {Bulger}, {Patience},
  {Leland}, {Macintosh}, {Schneider}, {Song}, {Marois}, {Graham}, {Bessell}, \&
  {Doyon}}]{2011MNRAS.415..854D}
{De Rosa} R.~J. {et~al.}, 2011, \mnras, 415, 854

\bibitem[{{De Rosa} {et~al}\mbox{.}(2014){De Rosa}, {Patience}, {Wilson},
  {Schneider}, {Wiktorowicz}, {Vigan}, {Marois}, {Song}, {Macintosh}, {Graham},
  {Doyon}, {Bessell}, {Thomas}, \& {Lai}}]{2014MNRAS.437.1216D}
{De Rosa} R.~J. {et~al.}, 2014, \mnras, 437, 1216

\bibitem[{{Doyon} {et~al}\mbox{.}(1998){Doyon}, {Nadeau}, {Vallee}, {Starr},
  {Cuillandre}, {Beuzit}, {Beigbeder}, \& {Brau-Nogue}}]{doyon}
{Doyon} R., {Nadeau} D., {Vallee} P., {Starr} B.~M., {Cuillandre} J.~C.,
  {Beuzit} J.-L., {Beigbeder} F., {Brau-Nogue} S., 1998, in Society of
  Photo-Optical Instrumentation Engineers (SPIE) Conference Series, Vol. 3354,
  Society of Photo-Optical Instrumentation Engineers (SPIE) Conference Series,
  {Fowler} A.~M., ed., pp. 760--768

\bibitem[{{Eggleton}(2006)}]{2006epbm.book.....E}
{Eggleton} P., 2006, {Evolutionary Processes in Binary and Multiple Stars}

\bibitem[{{ESA}(1997)}]{esaref}
{ESA}, 1997, VizieR Online Data Catalog, 1239, 0

\bibitem[{{Evans}(1959)}]{1959MNRAS.119..526E}
{Evans} D.~S., 1959, \mnras, 119, 526

\bibitem[{{Fischer} \& {Marcy}(1992)}]{fischermarcy}
{Fischer} D.~A., {Marcy} G.~W., 1992, 396, 178

\bibitem[{{Gaidos} \& {Mann}(2014)}]{gaidosmann2014}
{Gaidos} E., {Mann} A.~W., 2014, \apj, 791, 54

\bibitem[{{Gini}(1955)}]{gini}
{Gini} C., 1955, Memorie di metodologica statistica, 1, 156

\bibitem[{{Goodwin}(2010)}]{Goodwin2010}
{Goodwin} S.~P., 2010, Royal Society of London Philosophical Transactions
  Series A, 368, 851

\bibitem[{{Goodwin}, {Whitworth} \& {Ward-Thompson}(2004){Goodwin},
  {Whitworth}, \& {Ward-Thompson}}]{Goodwin2004}
{Goodwin} S.~P., {Whitworth} A.~P., {Ward-Thompson} D., 2004, \aap, 423, 169

\bibitem[{{Gray} {et~al}\mbox{.}(2006){Gray}, {Corbally}, {Garrison},
  {McFadden}, {Bubar}, {McGahee}, {O'Donoghue}, \& {Knox}}]{gray2006}
{Gray} R.~O., {Corbally} C.~J., {Garrison} R.~F., {McFadden} M.~T., {Bubar}
  E.~J., {McGahee} C.~E., {O'Donoghue} A.~A., {Knox} E.~R., 2006, \aj, 132, 161

\bibitem[{{Hambly} {et~al}\mbox{.}(2001){Hambly}, {MacGillivray}, {Read},
  {Tritton}, {Thomson}, {Kelly}, {Morgan}, {Smith}, {Driver}, {Williamson},
  {Parker}, {Hawkins}, {Williams}, \& {Lawrence}}]{hambly}
{Hambly} N.~C. {et~al.}, 2001, 326, 1279

\bibitem[{{Hawley}, {Gizis} \& {Reid}(1996){Hawley}, {Gizis}, \&
  {Reid}}]{pmsu2}
{Hawley} S.~L., {Gizis} J.~E., {Reid} I.~N., 1996, 112, 2799

\bibitem[{{Heintz}(1986)}]{1986A&AS...65..411H}
{Heintz} W.~D., 1986, 65, 411

\bibitem[{{Henry} {et~al}\mbox{.}(2006){Henry}, {Jao}, {Subasavage},
  {Beaulieu}, {Ianna}, {Costa}, \& {M{\'e}ndez}}]{Henry2006}
{Henry} T.~J., {Jao} W.-C., {Subasavage} J.~P., {Beaulieu} T.~D., {Ianna}
  P.~A., {Costa} E., {M{\'e}ndez} R.~A., 2006, \aj, 132, 2360

\bibitem[{{Herbig} \& {Moorhead}(1965)}]{1965ApJ...141..649H}
{Herbig} G.~H., {Moorhead} J.~M., 1965, \apj, 141, 649

\bibitem[{{Hodapp} {et~al}\mbox{.}(2008){Hodapp}, {Suzuki}, {Tamura}, {Abe},
  {Suto}, {Kandori}, {Morino}, {Nishimura}, {Takami}, {Guyon}, {Jacobson},
  {Stahlberger}, {Yamada}, {Shelton}, {Hashimoto}, {Tavrov}, {Nishikawa},
  {Ukita}, {Izumiura}, {Hayashi}, {Nakajima}, {Yamada}, \&
  {Usuda}}]{Hodapp2008}
{Hodapp} K.~W. {et~al.}, 2008, in Society of Photo-Optical Instrumentation
  Engineers (SPIE) Conference Series, Vol. 7014, Society of Photo-Optical
  Instrumentation Engineers (SPIE) Conference Series

\bibitem[{{Houk} \& {Cowley}(1975)}]{houk}
{Houk} N., {Cowley} A.~P., 1975, {University of Michigan Catalogue of
  two-dimensional spectral types for the HD stars. Volume I. Declinations -90
  to -53.}

\bibitem[{{Janson} {et~al}\mbox{.}(2012){Janson}, {Hormuth}, {Bergfors},
  {Brandner}, {Hippler}, {Daemgen}, {Kudryavtseva}, {Schmalzl}, {Schnupp}, \&
  {Henning}}]{janson}
{Janson} M. {et~al.}, 2012, \apj, 754, 44

\bibitem[{{Kenyon} \& {Hartmann}(1995)}]{kenyonhartmann}
{Kenyon} S.~J., {Hartmann} L., 1995, 101, 117

\bibitem[{{King} {et~al}\mbox{.}(2012{\natexlab{a}}){King}, {Goodwin},
  {Parker}, \& {Patience}}]{king2}
{King} R.~R., {Goodwin} S.~P., {Parker} R.~J., {Patience} J.,
  2012{\natexlab{a}}, 427, 2636

\bibitem[{{King} {et~al}\mbox{.}(2012{\natexlab{b}}){King}, {Parker},
  {Patience}, \& {Goodwin}}]{king1}
{King} R.~R., {Parker} R.~J., {Patience} J., {Goodwin} S.~P.,
  2012{\natexlab{b}}, 421, 2025

\bibitem[{{Kobayashi} {et~al}\mbox{.}(2000){Kobayashi}, {Tokunaga}, {Terada},
  {Goto}, {Weber}, {Potter}, {Onaka}, {Ching}, {Young}, {Fletcher}, {Neil},
  {Robertson}, {Cook}, {Imanishi}, \& {Warren}}]{Kobayashi2000}
{Kobayashi} N. {et~al.}, 2000, in Society of Photo-Optical Instrumentation
  Engineers (SPIE) Conference Series, Vol. 4008, Optical and IR Telescope
  Instrumentation and Detectors, {Iye} M., {Moorwood} A.~F., eds., pp.
  1056--1066

\bibitem[{{Koen} {et~al}\mbox{.}(2010){Koen}, {Kilkenny}, {van Wyk}, \&
  {Marang}}]{koen}
{Koen} C., {Kilkenny} D., {van Wyk} F., {Marang} F., 2010, \mnras, 403, 1949

\bibitem[{{Kozai}(1962)}]{Kozai1962}
{Kozai} Y., 1962, \aj, 67, 591

\bibitem[{{Lenzen} {et~al}\mbox{.}(2003){Lenzen}, {Hartung}, {Brandner},
  {Finger}, {Hubin}, {Lacombe}, {Lagrange}, {Lehnert}, {Moorwood}, \&
  {Mouillet}}]{lenzen2003}
{Lenzen} R. {et~al.}, 2003, in Society of Photo-Optical Instrumentation
  Engineers (SPIE) Conference Series, Vol. 4841, Society of Photo-Optical
  Instrumentation Engineers (SPIE) Conference Series, {Iye} M., {Moorwood}
  A.~F.~M., eds., pp. 944--952

\bibitem[{{L{\'e}pine} \& {Gaidos}(2011)}]{lepine2011}
{L{\'e}pine} S., {Gaidos} E., 2011, 142, 138

\bibitem[{{Lidov}(1962)}]{Lidov1962}
{Lidov} M.~L., 1962, \planss, 9, 719

\bibitem[{{Luhman} {et~al}\mbox{.}(2010){Luhman}, {Allen}, {Espaillat},
  {Hartmann}, \& {Calvet}}]{Luhman2010}
{Luhman} K.~L., {Allen} P.~R., {Espaillat} C., {Hartmann} L., {Calvet} N.,
  2010, \apjs, 186, 111

\bibitem[{{Luhman} {et~al}\mbox{.}(2003){Luhman}, {Stauffer}, {Muench},
  {Rieke}, {Lada}, {Bouvier}, \& {Lada}}]{Luhman2003}
{Luhman} K.~L., {Stauffer} J.~R., {Muench} A.~A., {Rieke} G.~H., {Lada} E.~A.,
  {Bouvier} J., {Lada} C.~J., 2003, \apj, 593, 1093

\bibitem[{{Marcy} \& {Benitz}(1989)}]{marcybenitz}
{Marcy} G.~W., {Benitz} K.~J., 1989, 344, 441

\bibitem[{{Mason} {et~al}\mbox{.}(2001){Mason}, {Wycoff}, {Hartkopf},
  {Douglass}, \& {Worley}}]{mason}
{Mason} B.~D., {Wycoff} G.~L., {Hartkopf} W.~I., {Douglass} G.~G., {Worley}
  C.~E., 2001, 122, 3466

\bibitem[{{Mazeh} {et~al}\mbox{.}(2001){Mazeh}, {Latham}, {Goldberg}, {Torres},
  {Stefanik}, {Henry}, {Zucker}, {Gnat}, \& {Ofek}}]{2001MNRAS.325..343M}
{Mazeh} T. {et~al.}, 2001, \mnras, 325, 343

\bibitem[{{McCarthy} {et~al}\mbox{.}(1998){McCarthy}, {Burge}, {Angel}, {Ge},
  {Sarlot}, {Fitz-Patrick}, \& {Hinz}}]{aries}
{McCarthy} D.~W., {Burge} J.~H., {Angel} J.~R.~P., {Ge} J., {Sarlot} R.~J.,
  {Fitz-Patrick} B.~C., {Hinz} J.~L., 1998, in Society of Photo-Optical
  Instrumentation Engineers (SPIE) Conference Series, Vol. 3354, Infrared
  Astronomical Instrumentation, {Fowler} A.~M., ed., pp. 750--754

\bibitem[{{Metchev} \& {Hillenbrand}(2009)}]{2009ApJS..181...62M}
{Metchev} S.~A., {Hillenbrand} L.~A., 2009, \apjs, 181, 62

\bibitem[{{Murakawa} {et~al}\mbox{.}(2004){Murakawa}, {Suto}, {Tamura},
  {Kaifu}, {Takami}, {Takato}, {Oya}, {Hayano}, {Gaessler}, \&
  {Kamata}}]{Murakawa2004}
{Murakawa} K. {et~al.}, 2004, pasj, 56, 509

\bibitem[{{Murakawa} {et~al}\mbox{.}(2003){Murakawa}, {Suto}, {Tamura},
  {Takami}, {Takato}, {Hayashi}, {Doi}, {Kaifu}, {Hayano}, {Gaessler}, \&
  {Kamata}}]{Murakawa2003}
{Murakawa} K. {et~al.}, 2003, in Society of Photo-Optical Instrumentation
  Engineers (SPIE) Conference Series, Vol. 4841, Instrument Design and
  Performance for Optical/Infrared Ground-based Telescopes, {Iye} M.,
  {Moorwood} A.~F.~M., eds., pp. 881--888

\bibitem[{{Neves} {et~al}\mbox{.}(2014){Neves}, {Bonfils}, {Santos},
  {Delfosse}, {Forveille}, {Allard}, \& {Udry}}]{neves2014}
{Neves} V., {Bonfils} X., {Santos} N.~C., {Delfosse} X., {Forveille} T.,
  {Allard} F., {Udry} S., 2014, \aap, 568, A121

\bibitem[{{Newton} {et~al}\mbox{.}(2014){Newton}, {Charbonneau}, {Irwin},
  {Berta-Thompson}, {Rojas-Ayala}, {Covey}, \& {Lloyd}}]{newton2014}
{Newton} E.~R., {Charbonneau} D., {Irwin} J., {Berta-Thompson} Z.~K.,
  {Rojas-Ayala} B., {Covey} K., {Lloyd} J.~P., 2014, \aj, 147, 20

\bibitem[{{Nidever} {et~al}\mbox{.}(2002){Nidever}, {Marcy}, {Butler},
  {Fischer}, \& {Vogt}}]{2002ApJS..141..503N}
{Nidever} D.~L., {Marcy} G.~W., {Butler} R.~P., {Fischer} D.~A., {Vogt} S.~S.,
  2002, \apjs, 141, 503

\bibitem[{{Parker} \& {Goodwin}(2009)}]{parkergoodwin}
{Parker} R.~J., {Goodwin} S.~P., 2009, \mnras, 397, 1041

\bibitem[{{Parker} \& {Meyer}(2014)}]{Parker2014}
{Parker} R.~J., {Meyer} M.~R., 2014, ArXiv e-prints

\bibitem[{{Parker} \& {Quanz}(2013)}]{parkerquanz}
{Parker} R.~J., {Quanz} S.~P., 2013, \mnras, 436, 650

\bibitem[{{Patience} {et~al}\mbox{.}(2002){Patience}, {White}, {Ghez},
  {McCabe}, {McLean}, {Larkin}, {Prato}, {Kim}, {Lloyd}, {Liu}, {Graham},
  {Macintosh}, {Gavel}, {Max}, {Bauman}, {Olivier}, {Wizinowich}, \&
  {Acton}}]{2002ApJ...581..654P}
{Patience} J. {et~al.}, 2002, \apj, 581, 654

\bibitem[{{Perryman} {et~al}\mbox{.}(1997){Perryman}, {Lindegren},
  {Kovalevsky}, {Hoeg}, {Bastian}, {Bernacca}, {Cr{\'e}z{\'e}}, {Donati},
  {Grenon}, {Grewing}, {van Leeuwen}, {van der Marel}, {Mignard}, {Murray}, {Le
  Poole}, {Schrijver}, {Turon}, {Arenou}, {Froeschl{\'e}}, \&
  {Petersen}}]{oldhip}
{Perryman} M.~A.~C. {et~al.}, 1997, 323, L49

\bibitem[{{Pourbaix} {et~al}\mbox{.}(2004){Pourbaix}, {Tokovinin}, {Batten},
  {Fekel}, {Hartkopf}, {Levato}, {Morrell}, {Torres}, \&
  {Udry}}]{Pourbaix:2004}
{Pourbaix} D. {et~al.}, 2004, \aap, 424, 727

\bibitem[{{Poveda} {et~al}\mbox{.}(2009){Poveda}, {Allen}, {Costero},
  {Echevarr{\'{\i}}a}, \& {Hern{\'a}ndez-Alc{\'a}ntara}}]{poveda}
{Poveda} A., {Allen} C., {Costero} R., {Echevarr{\'{\i}}a} J.,
  {Hern{\'a}ndez-Alc{\'a}ntara} A., 2009, \apj, 706, 343

\bibitem[{{Quirrenbach} {et~al}\mbox{.}(2010){Quirrenbach}, {Amado}, {Mandel},
  {Caballero}, {Mundt}, {Ribas}, {Reiners}, {Abril}, {Aceituno}, {Afonso},
  {Barrado y Navascues}, {Bean}, {B{\'e}jar}, {Becerril}, {B{\"o}hm},
  {C{\'a}rdenas}, {Claret}, {Colom{\'e}}, {Costillo}, {Dreizler},
  {Fern{\'a}ndez}, {Francisco}, {Galad{\'{\i}}}, {Garrido}, {Gonz{\'a}lez
  Hern{\'a}ndez}, {Gu{\`a}rdia}, {Guenther}, {Guti{\'e}rrez-Soto}, {Joergens},
  {Hatzes}, {Helmling}, {Henning}, {Herrero}, {K{\"u}rster}, {Laun}, {Lenzen},
  {Mall}, {Martin}, {Mart{\'{\i}}n-Ruiz}, {Mirabet}, {Montes}, {Morales},
  {Morales Mu{\~n}oz}, {Moya}, {Naranjo}, {Rabaza}, {Ram{\'o}n}, {Rebolo},
  {Reffert}, {Rodler}, {Rodr{\'{\i}}guez}, {Rodr{\'{\i}}guez Trinidad},
  {Rohloff}, {S{\'a}nchez Carrasco}, {Schmidt}, {Seifert}, {Setiawan},
  {Solano}, {Stahl}, {Storz}, {Su{\'a}rez}, {Thiele}, {Wagner}, {Wiedemann},
  {Zapatero Osorio}, {del Burgo}, {S{\'a}nchez-Blanco}, \& {Xu}}]{carmenes}
{Quirrenbach} A. {et~al.}, 2010, in Society of Photo-Optical Instrumentation
  Engineers (SPIE) Conference Series, Vol. 7735, Society of Photo-Optical
  Instrumentation Engineers (SPIE) Conference Series

\bibitem[{{Raghavan} {et~al}\mbox{.}(2010){Raghavan}, {McAlister}, {Henry},
  {Latham}, {Marcy}, {Mason}, {Gies}, {White}, \& {ten Brummelaar}}]{raghavan}
{Raghavan} D. {et~al.}, 2010, 190, 1

\bibitem[{{Reggiani} \& {Meyer}(2013)}]{reg2013}
{Reggiani} M., {Meyer} M.~R., 2013, \aap, 553, A124

\bibitem[{{Reggiani} \& {Meyer}(2011)}]{reg2011}
{Reggiani} M.~M., {Meyer} M.~R., 2011, \apj, 738, 60

\bibitem[{{Reid} \& {Gizis}(1997)}]{Reid1997}
{Reid} I.~N., {Gizis} J.~E., 1997, \aj, 113, 2246

\bibitem[{{Reid}, {Hawley} \& {Gizis}(1995){Reid}, {Hawley}, \&
  {Gizis}}]{pmsu1}
{Reid} I.~N., {Hawley} S.~L., {Gizis} J.~E., 1995, 110, 1838

\bibitem[{{Reipurth} \& {Zinnecker}(1993)}]{Reipurth1993}
{Reipurth} B., {Zinnecker} H., 1993, \aap, 278, 81

\bibitem[{{Robin} {et~al}\mbox{.}(2003){Robin}, {Reyl{\'e}}, {Derri{\`e}re}, \&
  {Picaud}}]{robin2003}
{Robin} A.~C., {Reyl{\'e}} C., {Derri{\`e}re} S., {Picaud} S., 2003, \aap, 409,
  523

\bibitem[{{Roeser}, {Demleitner} \& {Schilbach}(2010){Roeser}, {Demleitner}, \&
  {Schilbach}}]{ppmxl}
{Roeser} S., {Demleitner} M., {Schilbach} E., 2010, \aj, 139, 2440

\bibitem[{{Rojas-Ayala} {et~al}\mbox{.}(2012){Rojas-Ayala}, {Covey},
  {Muirhead}, \& {Lloyd}}]{rojasayala}
{Rojas-Ayala} B., {Covey} K.~R., {Muirhead} P.~S., {Lloyd} J.~P., 2012, \apj,
  748, 93

\bibitem[{{Rousset} {et~al}\mbox{.}(2003){Rousset}, {Lacombe}, {Puget},
  {Hubin}, {Gendron}, {Fusco}, {Arsenault}, {Charton}, {Feautrier}, {Gigan},
  {Kern}, {Lagrange}, {Madec}, {Mouillet}, {Rabaud}, {Rabou}, {Stadler}, \&
  {Zins}}]{rousset2003}
{Rousset} G. {et~al.}, 2003, in Society of Photo-Optical Instrumentation
  Engineers (SPIE) Conference Series, Vol. 4839, Society of Photo-Optical
  Instrumentation Engineers (SPIE) Conference Series, {Wizinowich} P.~L.,
  {Bonaccini} D., eds., pp. 140--149

\bibitem[{{Tokovinin}(2008)}]{2008MNRAS.389..925T}
{Tokovinin} A., 2008, \mnras, 389, 925

\bibitem[{{Tokovinin}(1997)}]{1997A&AS..121...71T}
{Tokovinin} A.~A., 1997, \aap, 121, 71

\bibitem[{{Tokunaga} {et~al}\mbox{.}(1998){Tokunaga}, {Kobayashi}, {Bell},
  {Ching}, {Hodapp}, {Hora}, {Neill}, {Onaka}, {Rayner}, {Robertson}, {Warren},
  {Weber}, \& {Young}}]{Tokunaga1998}
{Tokunaga} A.~T. {et~al.}, 1998, in Society of Photo-Optical Instrumentation
  Engineers (SPIE) Conference Series, Vol. 3354, Infrared Astronomical
  Instrumentation, {Fowler} A.~M., ed., pp. 512--524

\bibitem[{{Tomkin} \& {Pettersen}(1986)}]{1986AJ.....92.1424T}
{Tomkin} J., {Pettersen} B.~R., 1986, \aj, 92, 1424

\bibitem[{{Torres} {et~al}\mbox{.}(2006){Torres}, {Quast}, {da Silva}, {de La
  Reza}, {Melo}, \& {Sterzik}}]{torres}
{Torres} C.~A.~O., {Quast} G.~R., {da Silva} L., {de La Reza} R., {Melo}
  C.~H.~F., {Sterzik} M., 2006, \aap, 460, 695

\bibitem[{{van Leeuwen}(2007)}]{vanleeuwen}
{van Leeuwen} F., 2007, 474, 653

\bibitem[{{Worley}(1962)}]{worley62}
{Worley} C.~E., 1962, \aj, 67, 396

\bibitem[{{Wu} \& {Murray}(2003)}]{Wu2003}
{Wu} Y., {Murray} N., 2003, \apj, 589, 605

\end{thebibliography}

\section*{Appendix}

\onecolumn 
\begin{center}
\begin{longtable}{l l l l l l l l l l c c c}  
\caption{\label{table:limits} Companion detection limits} \\

\hline
\multirow{3}{*}{HIP} & \multicolumn{9}{c}{AO $\Delta$m (mag) at $\rho = (arcsec)$} &  \multicolumn{3}{c}{$\rho$ sensitive to full MS} \\
    & 0.05	&	0.15	&	0.25	&	0.50	&	1.00	&	1.50	&	2.00	&	5.00	&	10.00	& $\rho_{\rm~AO, max}$	& $\rho_{\rm~Plate, min}$ & $\rho_{\rm~Plate, max}$ \\\cline{2-10}
	&		&		&		&		&		&		&		&		&		&	(arcsec)	& (arcsec) & (arcmin) \\
\hline
\endfirsthead
\caption{continued.} \\
\hline 
\multirow{3}{*}{HIP} & \multicolumn{9}{c}{AO $\Delta$m (mag) at $\rho = (arcsec)$} &  \multicolumn{3}{c}{$\rho$ sensitive to full MS} \\
    & 0.05	&	0.15	&	0.25	&	0.50	&	1.00	&	1.50	&	2.00	&	5.00	&	10.00	& $\rho_{\rm~AO, max}$	& $\rho_{\rm~Plate, min}$ & $\rho_{\rm~Plate, max}$ \\\cline{2-10}
	&		&		&		&		&		&		&		&		&		&	(arcsec)	& (arcsec) & (arcmin) \\
\hline
\endhead
\hline
\endfoot

\hline
\noalign{\smallskip}
\endlastfoot

HIP428	&	...	&	0.64	&	1.87	&	4.33	&	6.67	&	8.08	&	8.71	&	9.18	&	9.17	&	34.00	&	5.64	&	14.81	\\
HIP439	&	0.02	&	3.87	&	5.04	&	6.23	&	6.41	&	6.39	&	6.40	&	6.36	&	4.11	&	14.00	&	21.58	&	38.40	\\
HIP1242	&	0.01	&	2.70	&	2.09	&	6.34	&	7.04	&	7.04	&	7.07	&	7.01	&	4.48	&	13.00	&	2.50	&	33.42	\\
HIP1475	&	...	&	0.89	&	1.91	&	4.37	&	6.16	&	7.88	&	8.20	&	8.51	&	8.51	&	35.00	&	12.49	&	46.46	\\
HIP2552	&	...	&	0.63	&	1.49	&	2.36	&	5.14	&	6.76	&	7.75	&	6.47	&	7.90	&	39.00	&	5.37	&	16.56	\\
HIP3937	&	...	&	0.61	&	0.87	&	1.89	&	2.55	&	1.35	&	1.64	&	7.50	&	8.79	&	36.00	&	6.30	&	14.07	\\
HIP4569	&	...	&	...	&	...	&	...	&	...	&	...	&	...	&	...	&	...	&	...	&	5.40	&	13.49	\\
HIP4856	&	...	&	0.69	&	1.94	&	4.28	&	6.90	&	8.02	&	8.63	&	9.52	&	9.34	&	38.00	&	7.93	&	20.24	\\
\hline

\end{longtable}
\end{center}
\twocolumn

\bsp

\label{lastpage}

\end{document}